\documentclass[structabstract]{aa}  
\usepackage{graphicx}
\usepackage{txfonts}
\begin{document}
   \title{Spectroscopic versus Photometric Metallicities\,$\colon$ 
            Milky Way Dwarf Spheroidal Companions as a Test Case}

   \author{S.~Lianou 
              \inst{1} 
              \fnmsep\thanks{Fellow of the Heidelberg Graduate School
                             of Fundamental Physics (HGSFP) and member
                             of the International Max Planck Research
                             School (IMPRS) for Astronomy \& Cosmic
                             Physics at the University of Heidelberg}
           \and
            E. K. Grebel 
               \inst{1}
           \and
            A. Koch 
               \inst{2,3} 
          }

           \institute{Astronomisches Rechen-Institut, Zentrum f\"{u}r
                  Astronomie der Universit\"{a}t Heidelberg,
                  M\"{o}nchhofstrasse 12-14, D-69120 Heidelberg,
                  Germany\\  
                  \email{lianou@ari.uni.heidelberg.de;\,grebel@ari.uni.heidelberg.de}  
           \and
                  University of Leicester, Department of Physics and
                  Astronomy, University Road, LE1 7RH Leicester, UK\\
                  \email{ak326@astro.le.ac.uk}
           \and
                  Landessternwarte, Zentrum f\"{u}r Astronomie der
                  Universit\"{a}t Heidelberg, Koenigstuhl, D-69117
                  Heidelberg, Germany
                  }

   \date{Received March 31, 2011; accepted May 21, 2011}

% \abstract{}{}{}{}{} 
% 5 {} token are mandatory
 
  \abstract
  % context heading (optional)
  % {} leave it empty if necessary  
   {}
  % aims heading (mandatory)
   {The method of deriving photometric metallicities using red giant branch
     stars is applied to resolved stellar populations under the common
     assumption that they mainly consist of single--age old stellar
     populations. We explore the effect of the presence of mixed--age stellar
     populations on deriving photometric metallicities. 
    }
  % methods heading (mandatory)
   {We use photometric data sets for the five Galactic dwarf spheroidals
     Sculptor, Sextans, Carina, Fornax, and Leo\,II in order to derive their
     photometric metallicity distribution functions from their resolved red
     giant branches using isochrones of the Dartmouth Stellar Evolutionary
     Database. We compare the photometric metallicities with published
     spectroscopic metallicities based on the analysis of the near--infrared
     Ca triplet (Ca\,T), both on the metallicity scale of Carretta \& Gratton
     and on the scale defined by the Dartmouth isochrones. In addition, we
     compare the photometric metallicities with published spectroscopic
     metallicities based on spectral synthesis and medium--resolution
     spectroscopy, and on high resolution spectra where available. 
    }
  % results heading (mandatory)
   {The \textit{mean} properties of the spectroscopic and photometric
     metallicity samples are comparable within the intrinsic scatter of each
     method although the mean metallicities of dSphs with pronounced
     intermediate--age population fractions may be underestimated by the
     photometric method by up to a few tenths of dex in [Fe/H]. The
     star-by-star differences of the spectroscopic minus the photometric
     metallicities show a wide range of values along the fiducial
     spectroscopic metallicity range, with the tendency to have systematically
     lower photometric metallicities for those dwarf spheroidals with a higher
     fraction of intermediate--age populations. Such discrepancies persist
     even in the case of the purely old Sculptor dSph, where one would
     na\"ively expect a very good match when comparing with medium or low
     resolution metallicity measurements. Overall, the agreement between Ca\,T
     metallicities and photometric metallicities is very good in the
     metallicity range from $\sim-$2\,dex to $\sim-$1.5\,dex. We find that the
     photometric method is reliable in galaxies that contain small (less than
     15\%) intermediate--age stellar fractions. Therefore, in the presence of
     mixed--age stellar populations, one needs to quantify the fraction of the
     intermediate--age stars in order to assess their effect on determining
     metallicities from photometry alone. Finally, we note that the comparison
     of spectroscopic metallicities of the same stars obtained with different
     methods reveals similarly large discrepancies as the comparison with
     photometric metallicities.}
  % conclusions heading (optional), leave it empty if necessary 
   {} 

   \keywords{Galaxies: dwarf -- 
             Galaxies: stellar content --
             (Galaxies:) Local Group -- 
             Galaxies: abundances}
             
   \maketitle
%
%________________________________________________________________

\section{Introduction}

    There are several techniques one can use to derive the photometric
    metallicities of a stellar system using its resolved old red giant
    branches (RGBs). These include the use of the $(V-I)_{o}$ color of the RGB
    stars at the luminosity corresponding to $M_I\,=\,-3.0$\,mag or
    $M_I\,=\,-3.5$\,mag in conjunction with the empirical relations defined in
    Da Costa \& Armandroff (\cite{sl_dacosta90}; DA90), in Armandroff et
    al.~(\cite{sl_armandroff93}), and in Lee, Freedman \& Madore
    (\cite{sl_lee93}); the use of the fiducial ridge lines or analytic
    functions (Saviane et al.~\cite{sl_saviane00a}) describing the mean locus
    in color--magnitude space of red giants in Galactic globular clusters
    (GCs) with known metal abundances; as well as the use of theoretical
    stellar tracks or isochrones. The latter two techniques serve to either
    bracket the range of the metal abundances or to interpolate between them
    in order to derive the metallicity distribution function. Examples using
    the mean color of the RGB can be found in Mould, Kristian \& Da Costa
    \cite{sl_mould83}, Caldwell et al.~\cite{sl_caldwell98}, Grebel \&
    Guhathakurta \cite{sl_grebel99}, Caldwell \cite{sl_caldwell06}; using GC
    fiducials or analytic fits of GC fiducial loci in Harris, Harris \& Poole
    \cite{sl_harris99}, Sarajedini et al.~\cite{sl_sarajedini02}; using
    interpolation between theoretical tracks or isochrones in Harris \& Harris
    \cite{sl_harris00}, Mouhcine et al.~\cite{sl_mouhcine05}, Crnojevic,
    Grebel \& Koch \cite{sl_crnojevic10}, Bird et al.~\cite{sl_bird10},
    Lianou, Grebel \& Koch \cite{sl_lianou10}. In the case of GCs, dwarf
    spheroidals (dSphs), and the stellar haloes of galaxies, the assumption
    under which these techniques are used is that the red giants represent
    populations of an old age ($\geq10$\,Gyr). For such old populations, a
    star's locus in color--magnitude space is primarily sensitive to
    metallicity, while age spreads only produce a small color spread (Grebel
    \cite{sl_grebel97}; Frayn \& Gilmore \cite{sl_frayn02}). 

    The case of the Local Group (LG) dwarf galaxies has shown that all dwarfs
    studied in detail so far contain a population of old stars (e.g., Grebel
    \cite{sl_grebel01}; Grebel \& Gallagher \cite{sl_grebel04}). Some of these
    systems contain intermediate--age populations as well (from 1\,Gyr up to
    less than 10\,Gyr) in addition to early star formation, thus presenting
    rather complex star formation histories (SFH; Grebel \cite{sl_grebel97};
    Mateo \cite{sl_mateo98}; Tolstoy, Hill \& Tosi \cite{sl_tolstoy09}). In
    particular, this is the case for dwarf irregulars (dIrrs),
    transition--type dwarfs (dIrr/dSphs), dwarf ellipticals (dEs) and the many
    of the more luminous dSphs. Although spectroscopic observations of
    individual stars provide the best means to reveal and break an
    age--metallicity degeneracy along the RGB in systems with complex SFHs, as
    for instance in the case of Carina (Smecker--Hane et
    al.~\cite{sl_smecker-hane94}; Koch et al.~\cite{sl_koch06}), such studies
    are limited to nearby objects within the LG due to the faintness of the
    stars to be targeted and due to crowding. Based on the fact that many of
    the LG dSphs show complex SFHs, the assumption of a single old age for
    their stellar populations does not hold. In dwarf galaxies in more distant
    systems there are clear indications of complex SFHs as well (as traced by,
    e.g., broad RGBs or the presence of luminous asymptotic giant branch
    stars, red clump stars and occasionally even luminous blue main sequences)
    but more detailed information about their SFHs is not available. Moreover,
    in these more distant systems spectroscopy of individual stars along the
    RGB is not feasible with present--day instruments. Thus it is worth
    exploring how the assumption of a single old age affects the
    photometrically derived metallicities of composite populations with a
    range of ages.

    In the present work we perform a comparison of the {\em mean} metallicity
    properties as well as a direct star--by--star comparison between the
    spectroscopically and the photometrically derived metallicities. For
    individual star comparisons, we use the stars in  common to both
    photometric and spectroscopic samples of Galactic dSph companions that
    have been studied in the literature. In order to perform such a
    star--by--star comparison, we use results for five Galactic dSphs, namely
    Carina, Leo\,II, Fornax, Sextans, and Sculptor. The three dSphs Carina,
    Leo\,II, and Fornax have complex star formation and chemical enrichment
    histories with different fractions of intermediate--age stellar
    populations, while Sextans and Sculptor are dominated by old populations.
 
    This paper is structured as follows. In \S2 we present the spectroscopic
    and photometric datasets we use. In \S3 we show our results on the
    comparison of the {\em mean} metallicity properties as well as on the
    star--by--star comparison. In \S4 we discuss our main findings and in \S5
    we present our summary and conclusions.
%__________________________________________________________________

\section{Data}

\subsection{Our dwarf spheroidal galaxy sample}

    The dSph sample was selected such that there are both spectroscopic
    metallicities and photometric results available in the literature. The
    adopted galaxies are the five Galactic dSphs Sculptor, Sextans, Carina,
    Fornax, and Leo\,II, which show a diversity in their SFHs.  

    More specifically, in the case of Sculptor and Sextans the dominant
    population is of an old age (e.g., Hurley--Keller, Mateo \& Grebel
    \cite{sl_hurley-keller99}, Monkiewicz, Mould, Gallagher et
    al.~\cite{sl_monkiewicz99} for Sculptor; Lee et
    al.~\cite{sl_lee03,sl_lee09} for Sextans). Sculptor shows two distinct old
    stellar components in terms of metallicity and kinematics (e.g., Tolstoy
    et al.~\cite{sl_tolstoy04}), as well as a metallicity gradient (Harbeck et
    al.~\cite{sl_harbeck01}). De Boer et al.~(\cite{sl_deboer11}) suggest that
    Sculptor stopped forming stars 7\,Gyr ago. Sextans shows a population
    gradient based on its horizontal branch morphology (Harbeck et
    al.~\cite{sl_harbeck01}), as well as a metallicity gradient, where the
    metal--rich stars are more centrally concentrated and have colder
    kinematics than the metal--poor ones (Battaglia et
    al.~\cite{sl_battaglia11}).

    Carina experienced episodic star formation with at least three distinct
    populations separated by quiescent phases lasting about 4\,Gyr
    (Smecker--Hane et al.~\cite{sl_smecker-hane94}; Smecker--Hane et
    al.~\cite{sl_smecker-hane96}; Mighell \cite{sl_mighell97}; Hurley--Keller,
    Mateo \& Nemec \cite{sl_hurley-keller98}; Monelli et
    al.~\cite{sl_monelli03}). The majority of the stars of Carina formed
    around 7\,Gyr ago (Hurley--Keller et al.~\cite{sl_hurley-keller98}; Rizzi
    et al.~\cite{sl_rizzi03}). Carina shows a mild radial metallicity gradient
    in the sense that the metal-rich population is more centrally concentrated
    (Koch et al.~\cite{sl_koch06}). A similar trend is observed with respect
    to age such that the intermediate--age populations are more centrally
    concentrated (Harbeck et al.~\cite{sl_harbeck01}; Monelli et
    al.~\cite{sl_monelli03}). Leo\,II has both old and intermediate--age
    populations (Aaronson \& Mould \cite{sl_aaronson85}; Lee \cite{sl_lee95};
    Mighell \& Rich \cite{sl_mighell96}; Gullieuszik et
    al.~\cite{sl_gullieuszik08}). It appears that there is no significant
    metallicity gradient present in Leo\,II (Koch et al.~\cite{sl_koch07}). In
    the case of Fornax, the dominant population is of an intermediate--age
    (ca. 3--4 Gyr; Coleman \& de Jong \cite{sl_coleman08}) and it also
    contains old and a young populations (Stetson, Hesser \& Smecker--Hane
    \cite{sl_stetson98}; Saviane, Held \& Bertelli \cite{sl_saviane00b}),
    while showing a strong radial metallicity gradient (Battaglia et
    al.~\cite{sl_battaglia06}).
    
    The global properties of the five dSphs, sorted by
%%%%% TABLE 1 %%%%%%%%%%%%%%%%%%%%%%%%%%%%%%%%%%%%%%%%%%%%%%%%%%%%%%%%%%%%%%
\begin{table}
\begin{minipage}[t]{\columnwidth}
\caption[]{Global properties.}
\label{table1b} 
\centering
\renewcommand{\footnoterule}{}  
\begin{tabular}{l c c c c c c} 

\hline\hline

  Galaxy        &$A_{V}$    &$A_{I}$     &$(m-M)_{O}$        &TRGB \\ 
                &(mag)      &(mag)       &(mag)             &(mag)\\
   (1)          &(2)        &(3)         &(4)               &(5)  \\
                                                                          
\hline

  Sculptor      &$0.245$    &$0.117$     &$19.65\pm0.14$   &$15.70\pm0.10$\\
                                               
  Sextans       &$0.03$     &$0.02$      &$19.90\pm0.06$   &$15.95\pm0.04$\\

  Carina        &$0.109$    &$0.065$     &$20.04\pm0.10$   &$16.10\pm0.10$\\

  Fornax        &$0.186$    &$0.116$     &$20.62\pm0.04$   &$16.75\pm0.02$\\

  Leo\,II       &$0.066$    &$0.041$     &$21.84\pm0.13$   &$17.83\pm0.03$\\

\hline
 
\end{tabular}
\footnotetext[0]{References.-- The extinction is adopted from:
  Bellazzini, Gennari \& Ferraro (\cite{sl_bellazzini05}) for Leo\,II;
  Pont et al.~(\cite{sl_pont04}) for Fornax; Lee et
  al.~(\cite{sl_lee03}) for Sextans, own value for Carina and Sculptor using
  the TRGB method (Lee, Freedman \& Madore \cite{sl_lee93}). The distance
  modulus is adopted from: Bellazzini, Gennari \& Ferraro
  (\cite{sl_bellazzini05}) for Leo\,II; Pont et al.~(\cite{sl_pont04}) for
  Fornax; Lee et al.~(\cite{sl_lee03}) for Sextans, own value for Carina and
  Sculptor using the TRGB method. The \textit{I}--band TRGB is adopted from: Rizzi et
  al.~(\cite{sl_rizzi07a}) for Fornax; Lee et al.~(\cite{sl_lee03}) for
  Sextans; Bellazzini, Gennari \& Ferraro (\cite{sl_bellazzini05}) for
  Leo\,II; own value for Carina and Sculptor.} 
\end{minipage}
\end{table}
%%%%%%%%%%%%%%%%%%%%%%%%%%%%%%%%%%%%%%%%%%%%%%%%%%%%%%%%%%%%%%%%%%%%%%%%%%%%%%%%%%%%
%
    increasing distance modulus, are listed in Table~\ref{table1b}. We show in
    column (1) the galaxy name; in column (2) and (3) the \textit{V}-- and
    \textit{I}--band extinctions, respectively; in column (4) the distance
    modulus; in column (5) the \textit{I}--band magnitude of the tip of the
    RGB (TRGB). 

\subsection{Spectroscopic metallicities and metallicity scales}

    The spectroscopic metallicities of individual stars in dSphs can  be
    inferred either directly by high--resolution measurements of iron
    abundances, [Fe/H], from individual Fe lines, or through low\,/\,medium
    resolution spectroscopic measurements based on different spectral
    indicators. The latter method is the one widely used since it has the
    potential of providing spectra for a large number of stars within a
    reasonable integration time.

 \subsubsection{Ca\,II triplet}

    One way to infer the overall spectroscopic metallicities (strictly
    speaking [M/H], however usually also denoted as [Fe/H]) is from the
    strength of the Ca\,II triplet (Ca\,T) lines at 8498\,\AA, 8542\,\AA\ and
    8662\,\AA. The measured property is the sum of the equivalent widths,
    $\Sigma W$, either of two or of a combination of all three Ca\,T lines
    (e.g., Starkenburg et al.~\cite{sl_starkenburg10}). This is then used to
    derive the reduced equivalent width, $W^{\prime}$, using empirical
    calibrations between the $\Sigma W$ and $(V-V_{HB}$), defined, e.g., in
    Armandroff \& Da Costa (\cite{sl_armandroff91}; hereafter AD91). $V_{HB}$
    is the \textit{V}--band magnitude of the horizontal branch. The
    calibration of $\Sigma W$ as a function of $(V-V_{HB}$) is chosen because
    it removes, to first order, any dependencies on stellar gravity,
    reddening, and distance uncertainties (e.g., AD91). The width $W^{\prime}$
    is then used to derive a metallicity, commonly based on a calibration of
    Galactic GCs with known spectroscopic iron abundances.

    The Galactic GC metallicities are derived in various different ways. Thus,
    several metallicity scales have been defined so far, which include the
    Zinn \& West (\cite{sl_zinn84}; hereafter ZW84), the Carretta \& Gratton
    (\cite{sl_carretta97}; hereafter CG97), the Kraft \& Ivans
    \cite{sl_kraft03}; hereafter KI03) and the Carretta et
    al.~(\cite{sl_carretta09}; hereafter CBG09) metallicity scales. The first
    one uses metallicity sensitive spectrophotometric indices of the
    integrated light of Galactic GCs, while the latter three use high
    resolution spectroscopic measurements of Galactic GC red giants to infer
    their iron abundance from individual \ion{Fe}{I} and \ion{Fe}{II} lines. 

    Commonly used sets of such calibrations are given in AD91 and Da Costa \&
    Armandroff (\cite{sl_dacosta95}) for the ZW84 metallicity scale, and by
    Rutledge, Hesser \& Stetson (\cite{sl_rutledge97}; hereafter R97) for both
    the ZW84 and CG97 metallicity scales. Among these calibrations, the
    definition of the Ca\,T  sum of the equivalent widths, $\Sigma W$, is
    different, depending on how many lines were used and with what
    weight. KI03 provide a similar calibration between their scale of
    \ion{Fe}{II}--based abundances and the reduced widths $W^{\prime}$ of GCs,
    and so do CBG09.

    The Ca\,T method was initially calibrated via Galactic GCs, which are old
    populations and to first order simple stellar populations of a single
    metallicity. They have a different chemical enrichment history than the
    dSphs (AD91; Venn et al.~\cite{sl_venn04}; Koch et
    al.~\cite{sl_koch08a,sl_koch08b}). Subsequently, Cole et
    al.~(\cite{sl_cole04}) extended this method to much younger ages down to
    2.5\,Gyr by including younger open and populous clusters in the Milky Way
    and in the Large Magellanic Cloud, while Carrera et
    al.~(\cite{sl_carrera07}) used Galactic open and globular clusters to
    further extend the method to ages as young as 0.25\,Gyr. The Ca\,T method
    is widely used to derive the metallicities of galaxies that have more
    complex star formation and chemical enrichment histories than those of the
    calibrating Galactic GCs and populous clusters. The implications of the
    different chemical enrichment and star formation history in the dSphs and
    the Galactic GCs for the Ca\,T method have been discussed in Da Costa \&
    Hatzidimitriou (\cite{sl_dacosta98}), Cole et
    al.~(\cite{sl_cole00,sl_cole04}), Pont et al.~(\cite{sl_pont04}), Bosler
    et al.~(\cite{sl_bosler07}), Carrera et al.~(\cite{sl_carrera07}),
    Battaglia et al.~(\cite{sl_battaglia08}) and Koch et
    al.~(\cite{sl_koch08a}). Cole et al.~(\cite{sl_cole04}) have shown that
    the effect of age on the metallicity calibration is negligible as compared
    to the intrinsic scatter of the Ca\,T method for the metallicity ranges on
    the CG97 metallicity scale between $-$2.0 and $-$0.2\,dex in [Fe/H], while
    Carrera et al.~(\cite{sl_carrera07}) extended this metallicity range for
    three metallicity scales between $-$2.2\,dex to $+$0.47\,dex. At lower
    metallicities there is an overestimate of the Ca\,T metallicities as
    compared with metallicities derived from high--resolution measurements
    (e.g., Battaglia et al.~\cite{sl_battaglia08}; Koch et
    al.~\cite{sl_koch08a}). Therefore recent efforts focused on extending the
    scale to even lower metallicities. Starkenburg et
    al.~(\cite{sl_starkenburg10}) recalibrated the empirical relation between
    Ca\,T equivalent width and metallicity, extending the validity range of
    the Ca\,T method to $-$4\,dex. Overall, the method can now be used to
    infer the metallicities of red giants within a metallicity range of
    $-$4\,dex\,$\leq$[Fe/H]$\leq+$0.47\,dex and an age range of
    0.25\,Gyr\,$\leq$age$\leq$13\,Gyr ( Cole et al.~\cite{sl_cole04}; Carrera
    et   al.~\cite{sl_carrera07}; Starkenburg et al.~\cite{sl_starkenburg10}).

 \subsubsection{Medium--resolution combined with spectral synthesis}

    An alternative method to infer the metallicity of individual stars relies
    on medium--resolution stellar spectra ($R\sim 6500$ at 8500\AA) in
    combination with spectral synthesis (Kirby, Guhathakurta \& Sneden
    \cite{sl_kirby08}).  A plethora of individual absorption lines within a
    broad spectral range from 6400\,\AA\ to 9000\,\AA\ is used in a Bayesian
    approach, where the observed spectra are compared to a grid of synthetic
    stellar spectra with a range of T$_{\rm eff}$, log\,$g$, and
    composition. The adopted metallicity for each individual star is the
    metallicity of the best--fit template spectrum to the observed one. This
    technique is different from the Ca\,T method, in the sense that the entire
    available spectroscopic features in the observed spectrum are used, not
    only to derive the metallicity, but also to simultaneously pinpoint all
    stellar parameters (effective temperature, surface gravity, and an
    empirical microturbulence). We note that the spectral resolution of this
    method is equivalent to the one of the Ca\,T method. A complete
    description of the medium--resolution method is given in Kirby et
    al.~(\cite{sl_kirby10}; and references therein), while possible
    systematics affecting the derived metallicities as well as a comparison
    with high--resolution spectroscopic metallicities are discussed in Kirby
    et al.~(\cite{sl_kirby09, sl_kirby10}).

\subsection{Spectroscopic sample}

    We use two sources for the spectroscopic metallicities, both adopted from 
    available literature data. 

    The first spectroscopic sample consists of Ca\,T--based spectroscopic
    metallicities for all five dSphs. For Carina, the Ca\,T data are adopted
    from Koch et al.~(\cite{sl_koch06}), for Leo\,II from Koch et
    al.~(\cite{sl_koch07}), for Fornax from Battaglia et
    al.~(\cite{sl_battaglia06}), for Sextans from Battaglia et
    al.~(\cite{sl_battaglia11}; and private communication), and for Sculptor
    from Battaglia et al.~(\cite{sl_battaglia08}). We refer to these
    publications for the description of the spectroscopic observations and
    analysis. In all cases, the strength of the Ca\,T lines is used as a
    metallicity indicator for the individual red giant stars from either low
    or medium resolution spectroscopy. The spectroscopic metallicities for
    Carina, Leo\,II, Fornax and Sculptor are inferred through the calibration
    in the sense of R97, while for Sextans the calibration defined in
    Starkenburg et al.~(\cite{sl_starkenburg10}) is used. The main difference
    between the calibration in Starkenburg et al.~(\cite{sl_starkenburg10})
    and in the earlier calibrations is that in the new calibrations the
    relation between metallicity and line strength is not linear, while it is
    in the earlier calibrations. In addition, the Starkenburg et
    al.~(\cite{sl_starkenburg10}) calibration is valid for metallicities from
    $-$4\,dex to $-$0.5\,dex, while the previous ones are calibrated from
    $-$2\,dex to $-$0.2\,dex (see Cole et al.~\cite{sl_cole04}). In the
    following, all the Ca\,T--based spectroscopic metallicities mentioned are
    given on the CG97 metallicity scale, as originally provided in the
    literature.

    The second spectroscopic sample refers to spectroscopic metallicities
    derived using medium--resolution spectroscopy (MRS) combined with spectral
    synthesis and adopted from Kirby et al.~(\cite{sl_kirby10}). Their dSph
    sample includes MRS metallicities for Sculptor, Sextans, Fornax, and
    Leo\,II. We refer to Kirby et al.~(\cite{sl_kirby10,sl_kirby09}) and
    Kirby, Guhathakurta \& Sneden (\cite{sl_kirby08}) for the description of
    the spectroscopic observations and analysis of the method. We note that
    the MRS metallicities form a metallicity scale of their own.

\subsection{Photometric sample}

    The photometry of Carina, Fornax and Sculptor is adopted from Walker et
    al.~(\cite{sl_walker09}; and private communication), of Leo\,II from
    Bellazzini et al.~(\cite{sl_bellazzini05}), and of Sextans from Lee et
    al.~(\cite{sl_lee03}). We refer to these works for further details on the
    photometric observations and analysis. In all the studied dSphs the final
    photometric datasets are placed on a common \textit{V, I} photometry scale
    in the Johnsons--Cousins photometric system (Carina: Walker et
    al.~\cite{sl_walker07}; Fornax: Walker et al.~\cite{sl_walker06}; Leo\,II:
    Bellazzini et al.~\cite{sl_bellazzini05}; Sextans: Lee et
    al.~\cite{sl_lee03}; Sculptor: Walker et al.~\cite{sl_walker06}, Coleman,
    Da Costa, \& Bland-Hawthorn \cite{sl_coleman05}). We note that for Carina,
    Fornax, and Sculptor, there is WFI (Wide Field Imager camera at MPG/ESO
    2.2\,m telescope) photometry available for the spectroscopic targets
    (Carina: Koch et al.~\cite{sl_koch06}; Fornax and Sculptor: Battaglia et
    al.~\cite{sl_battaglia06, sl_battaglia08}, and private communication), but
    we choose not to use these datasets since they are poorly calibrated.
%______________________________________________________________

\section{Results}

\subsection{Color--magnitude diagrams}

    We show the color--magnitude diagrams (CMDs; small grey dots) of
%
%%%%%% FIGURE 1 - ALL %%%%%%%%%%%%% Two column figure (place early!) %%%%%%  
 \begin{figure*}                                               
    \centering
       \includegraphics[width=6cm,clip]{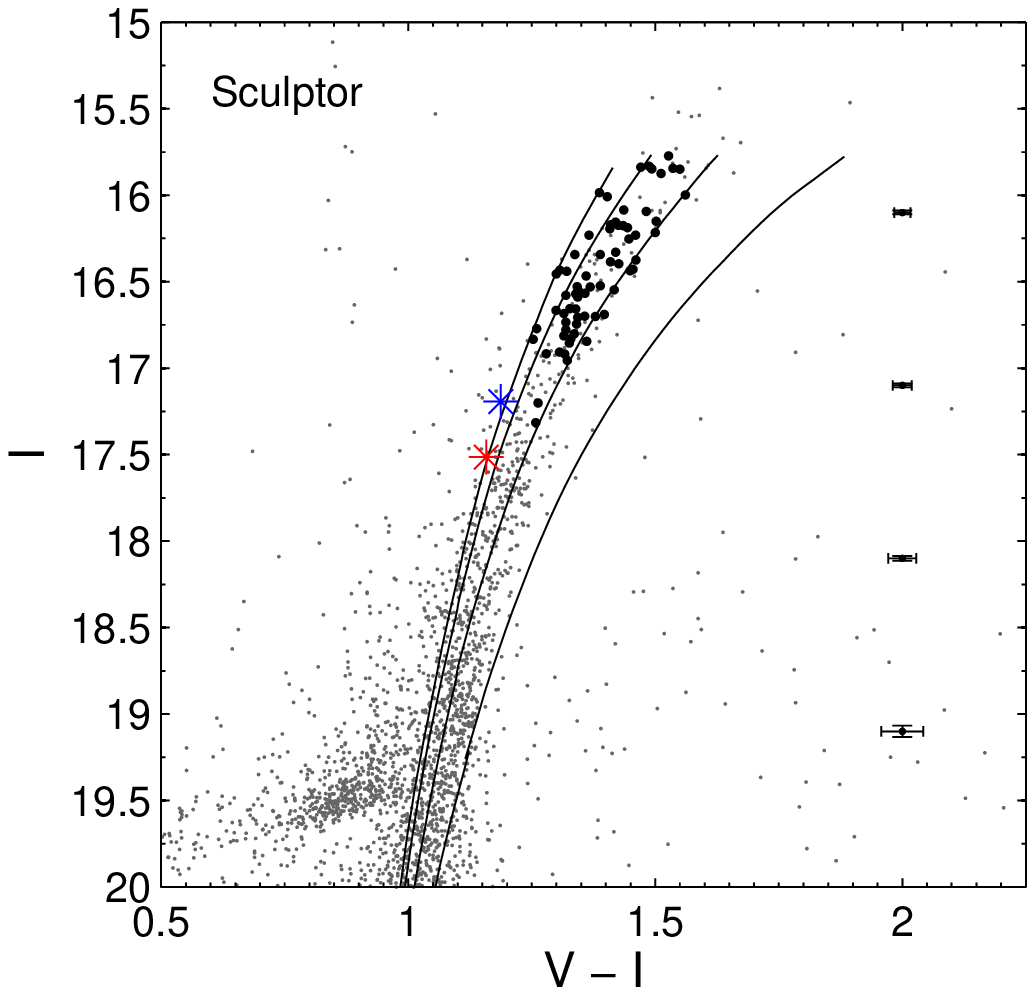}
       \includegraphics[width=6cm,clip]{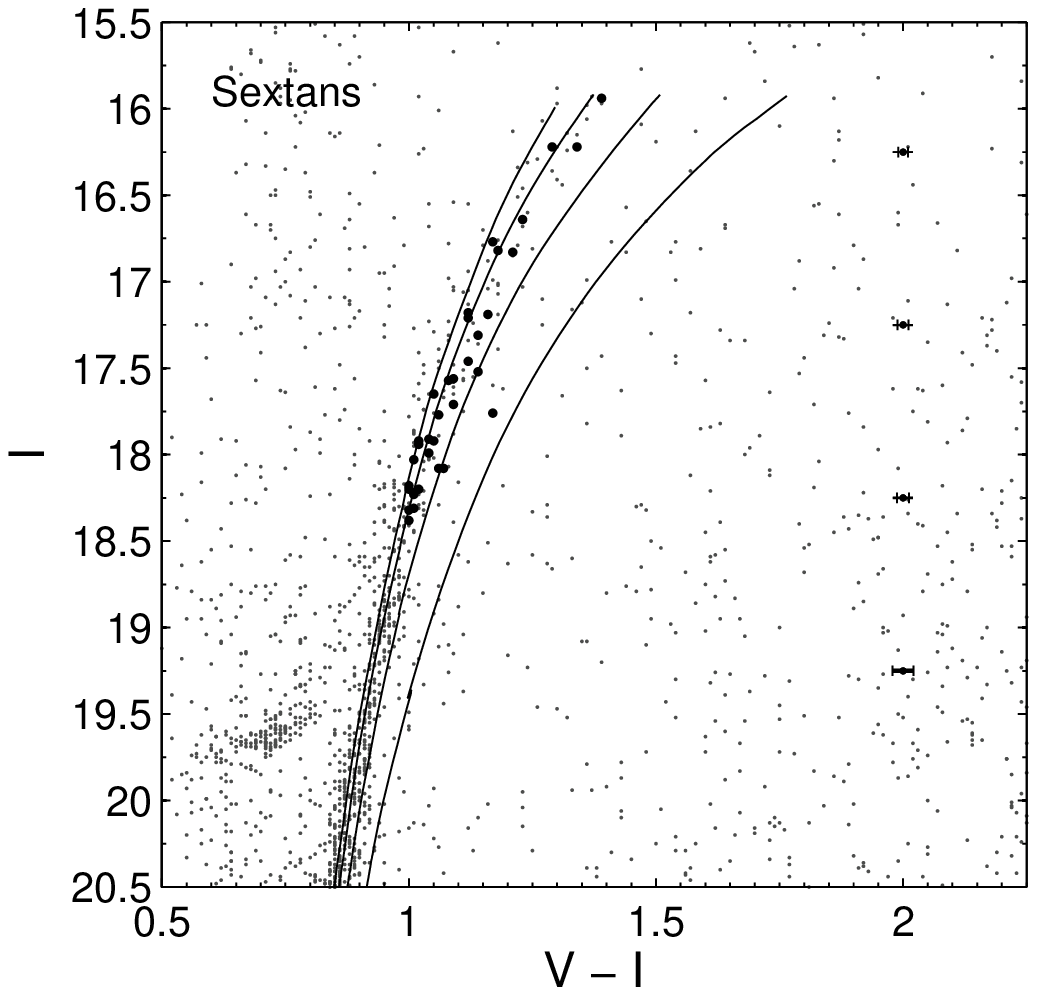}
       \includegraphics[width=6cm,clip]{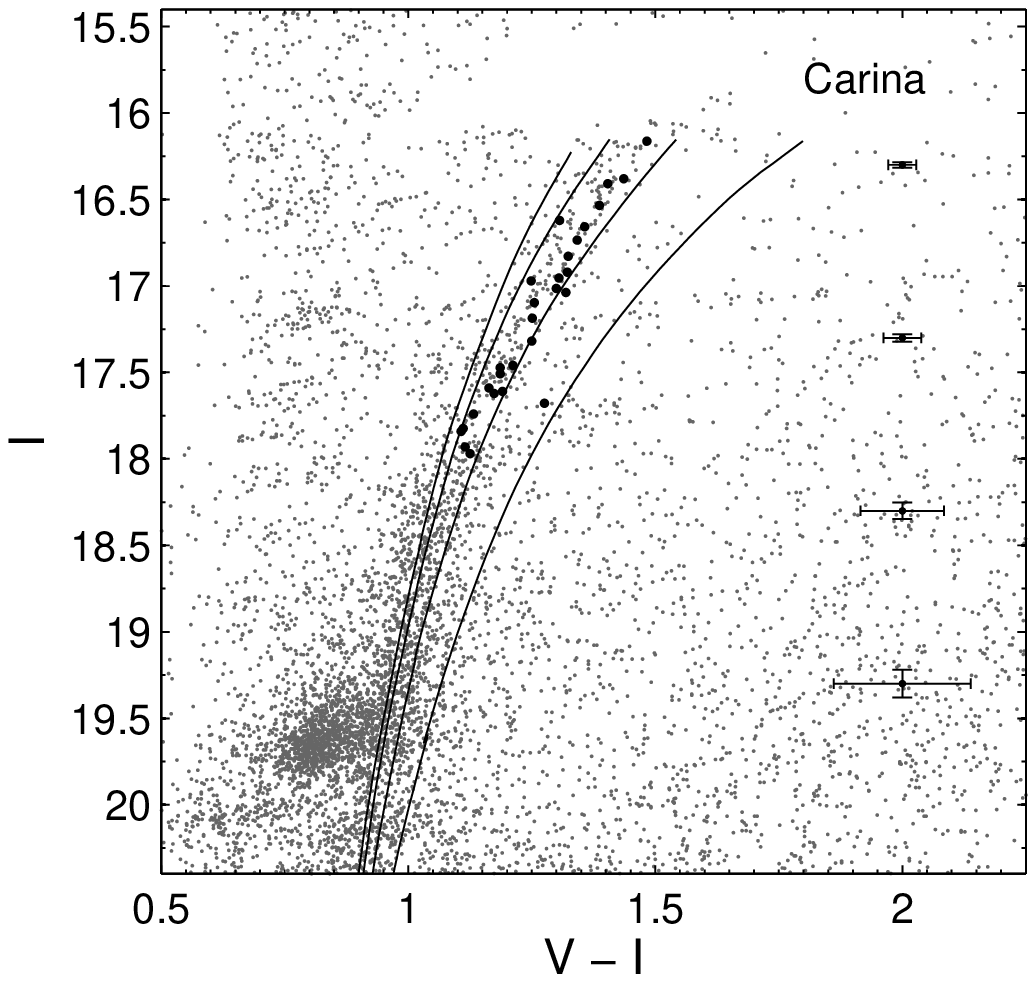}
       \includegraphics[width=6cm,clip]{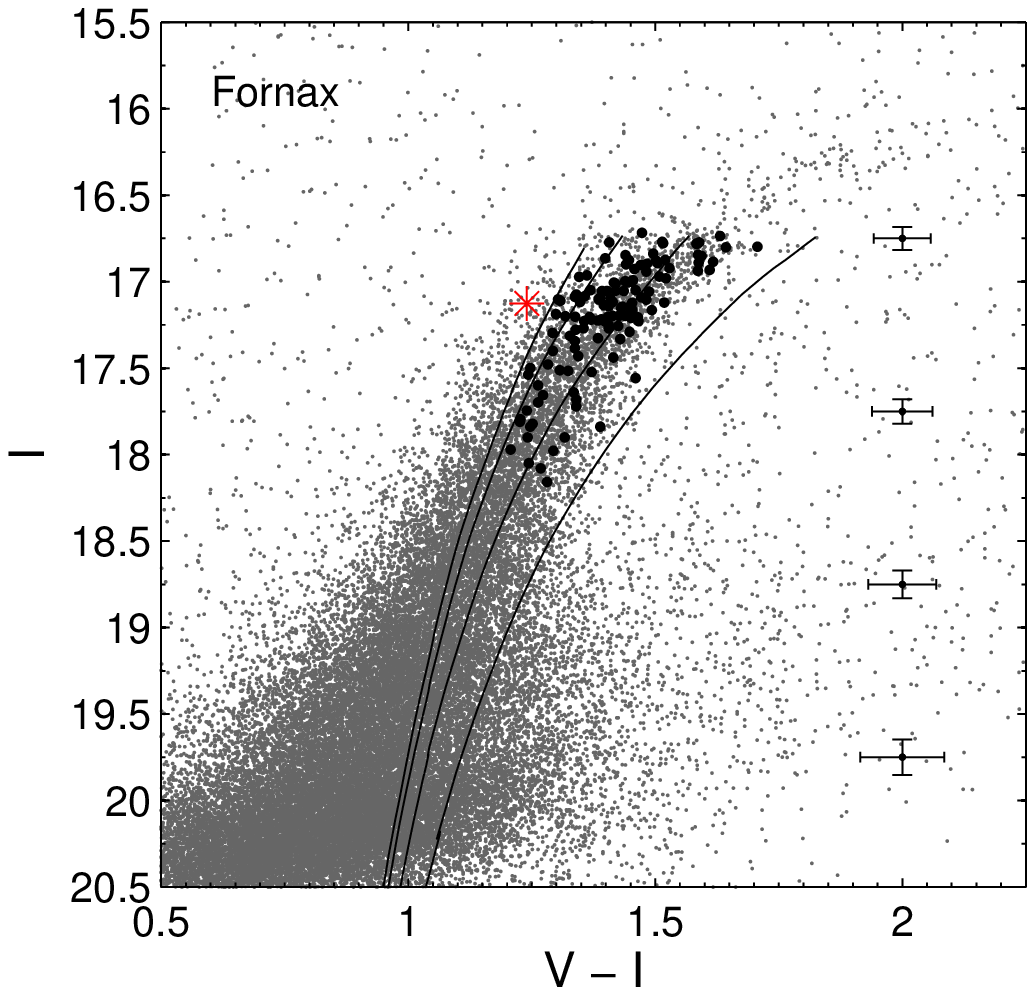}
       \includegraphics[width=6cm,clip]{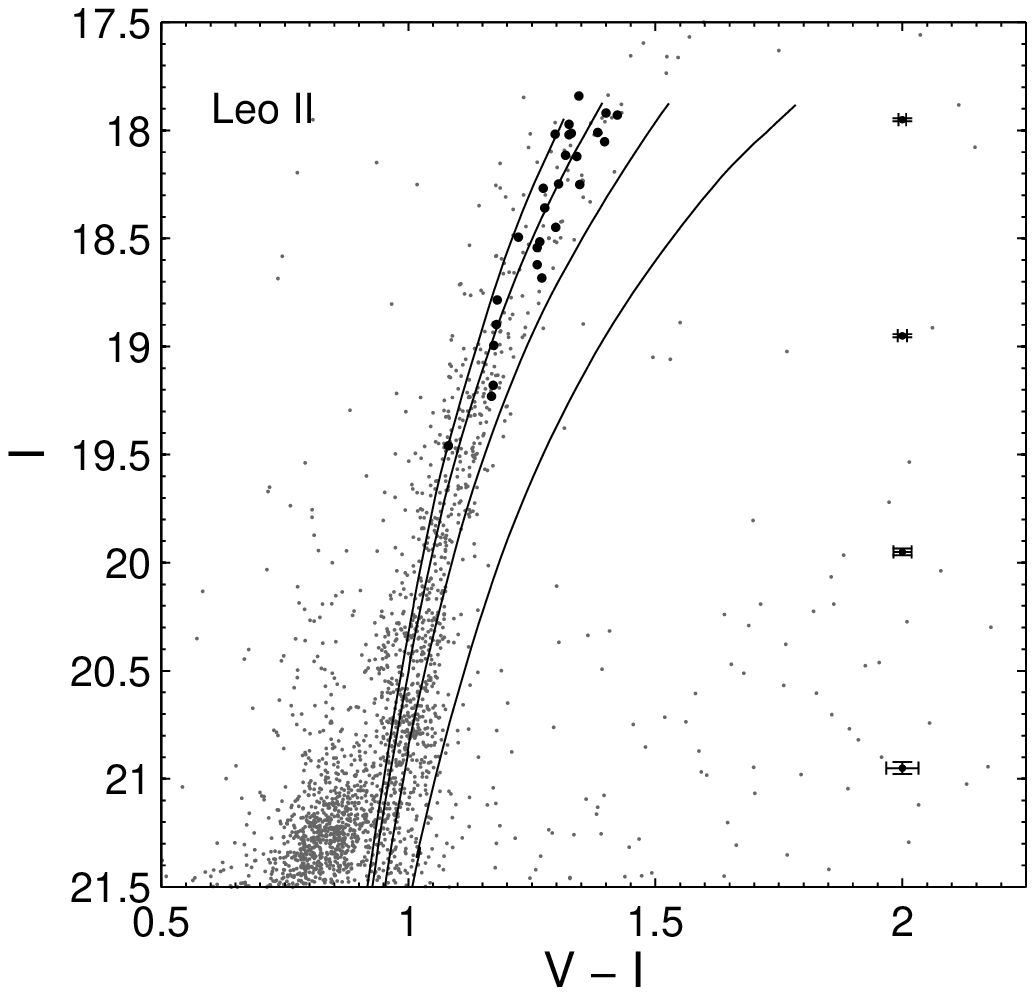}
    \caption{Color--magnitude diagrams of the five Galactic dSphs, shown as
      small grey dots. Dartmouth isochrones are overplotted as solid lines,
      for a fixed age of 12.5\,Gyr, a range in metallicities from $-$2.5 to
      $-$1\,dex, and a step size of 0.5\,dex. The thick black dots correspond
      to the stars in common to the photometric and Ca\,T--based spectroscopic
      sample. The error bars correspond to the photometric errors. In the CMD
      of Sculptor, the red and blue asterisks indicate the location of the
      extremely metal--poor stars with a high--resolution iron abundance of
      $-$3.96\,dex and $-$3.48\,dex, respectively, identified by Tafelmeyer et
      al.~(\cite{sl_tafelmeyer10}; red: Scl\,07--50; blue: Scl\,07--49), while
      the photometric metallicities obtained for the same stars are
      $-$2.55\,dex and $-$2.71\,dex, respectively. In the CMD of Fornax, again
      the red asterisk indicates the location of the extremely metal--poor
      star with a high--resolution iron abundance of $-$3.66\,dex (Tafelmeyer
      et al.~\cite{sl_tafelmeyer10}; Frx\,05--42), while the photometric
      metallicity assigned to the same star is $-$2.98\,dex.}
\label{sl_cmds}% 
\end{figure*}
%%%%%%%%%%%%%%%%%%%%%%%%%%%%%%%%%%%%%%%%%%%%%%%%%%%%%%%%%%%%%%%%%%%%%%%%%%%%%
%
    the five Galactic dSphs Sculptor, Sextans, Carina, Fornax, and Leo\,II in
    Fig.~\ref{sl_cmds}, along with Dartmouth isochrones (Dotter et
    al.~\cite{sl_dotter07,sl_dotter08}) overplotted with a fixed age of
    12.5\,Gyr and metallicities ranging from $-$2.5\,dex to $-$1\,dex with a
    step size of 0.5\,dex. The thick black dots represent the stars in common
    between the photometric and spectroscopic samples, corresponding to 68
    stars for Sculptor, 28 for Carina, 36 for Sextans, 132 for Fornax, and 27
    for Leo\,II, here quoting the stars in common with the Ca\,T sample. We
    note that these numbers are further reduced after applying the
    spectroscopic metallicity cuts in addition to the photometric metallicity
    cuts, as explained later on in the analysis.

\subsection{Photometric metallicities}

    For the five studied Galactic dSphs, we derive their photometric
    metallicities using linear interpolation between Dartmouth isochrones with
    a fixed age of 12.5\,Gyr, with a range in metallicities from $-$2.5\,dex
    to $-$0.3\,dex and with a step size of 0.05\,dex (e.g., Lianou et
    al.~\cite{sl_lianou10}). In practice, for each star under consideration,
    we interpolate between the two closest isochrones that bracket its color
    in order to find its metallicity. We use Dartmouth isochrones, since they
    give the best simultaneous fit to the full stellar distribution within a
    CMD as demonstrated by, e.g., Glatt et al.~(\cite{sl_glatt08a},
    \cite{sl_glatt08b}). We correct the magnitudes and colors of the
    theoretical isochrones for foreground Galactic extinction in the
    \textit{V}--band and \textit{I}--band and for the distance moduli, listed
    in columns (2), (3) and (4) of Table~\ref{table1b}, respectively. We
    analyse all bona--fide RGB stars that lie within 3\,mag below the
    \textit{I}--band magnitude of the TRGB, listed in Table~\ref{table1b},
    regardless of whether they were observed spectroscopically, in order to
    derive the {\em mean} photometric metallicity properties and compare them
    with the {\em mean} spectroscopic properties. We impose a metallicity cut
    on the derived photometric metallicities so as to only include stars that
    fall within the theoretical isochrones' metallicity range of $-$2.5\,dex
    to $-$0.3\,dex used in the interpolation method, thus excluding any
    extrapolated values. Additionally, we require the photometric metallicity
    uncertainties to be less than 0.2\,dex, which is the upper limit of
    accuracy typically achieved in low--resolution spectroscopic studies. The
    random uncertainties of the photometric metallicities are estimated by
    accounting for the photometric errors. For that purpose, each star is
    varied by its photometric uncertainties (both in color and magnitude) and
    re--fit using the same isochrone interpolation code. The 1\,$\sigma$
    scatter of the output random realisations is then adopted as the
    metallicity error for each star. 

%
%%%%%% FIGURE 2 - ALL %%%%%%%%%%%%% Two column figure (place early!) %%%%%%  
 \begin{figure*}                                               
    \centering
       \includegraphics[width=4.55cm,clip]{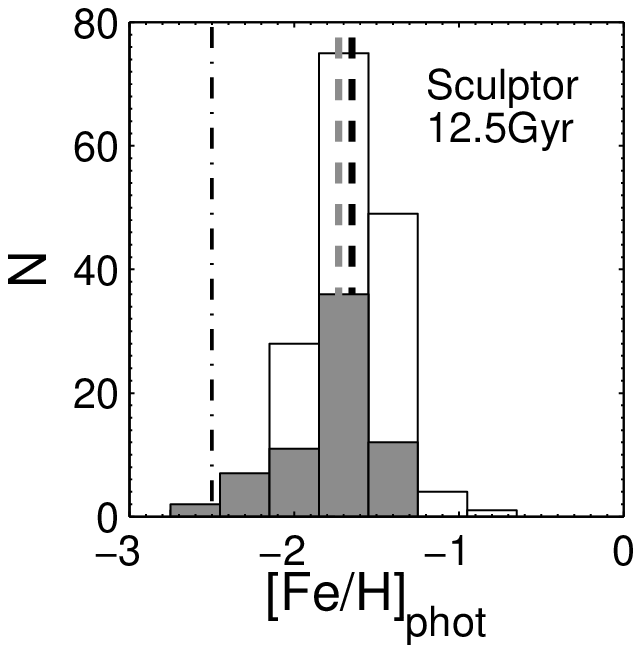}
       \includegraphics[width=4.55cm,clip]{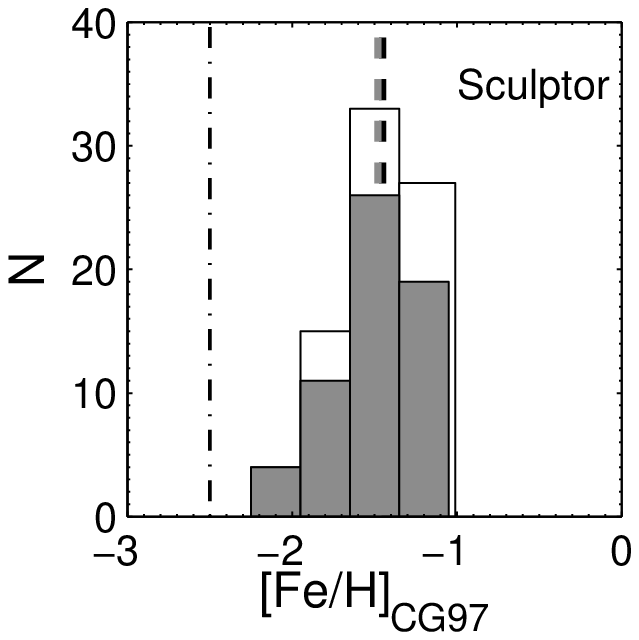}
       \includegraphics[width=4.55cm,clip]{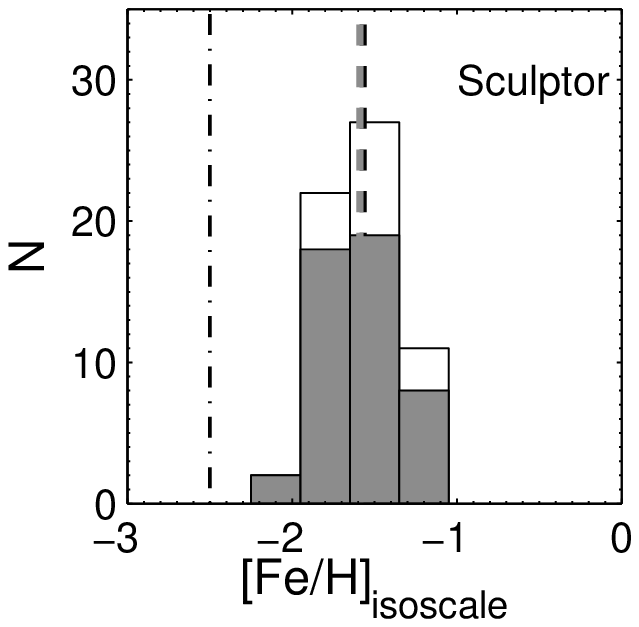}
       \includegraphics[width=4.55cm,clip]{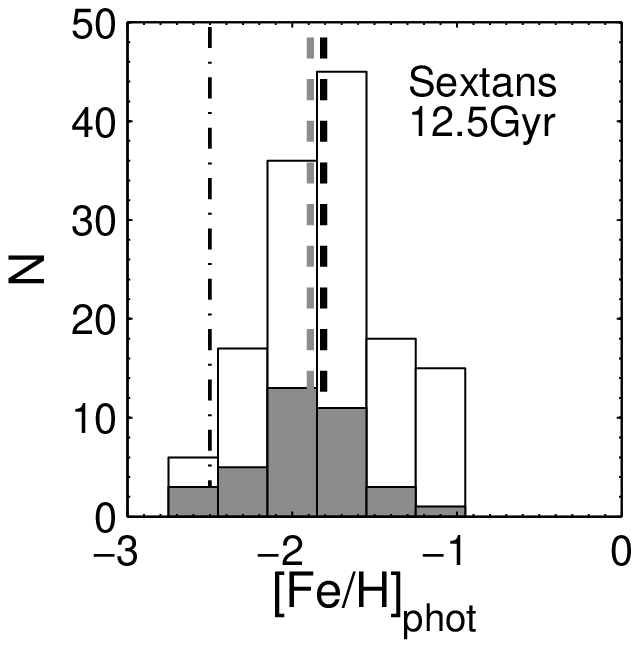}
       \includegraphics[width=4.55cm,clip]{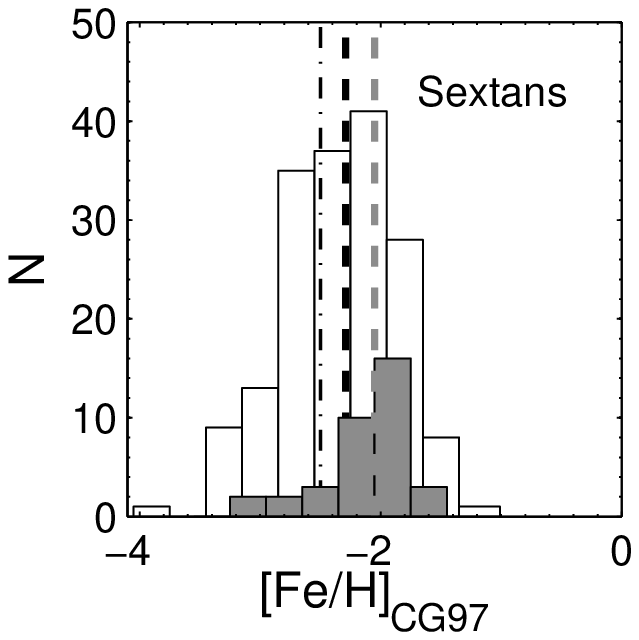}
       \includegraphics[width=4.55cm,clip]{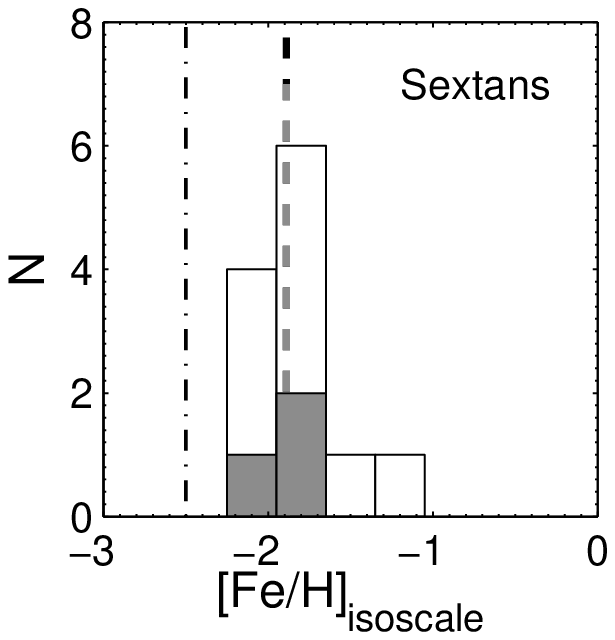}
       \includegraphics[width=4.55cm,clip]{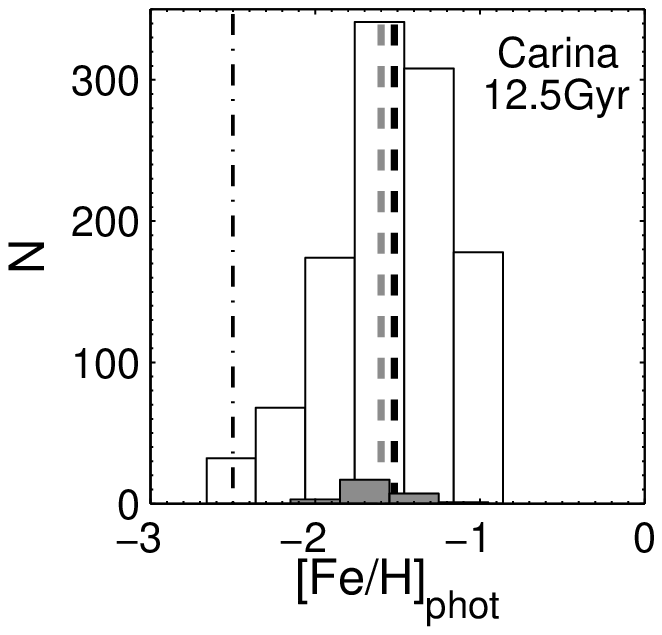}
       \includegraphics[width=4.55cm,clip]{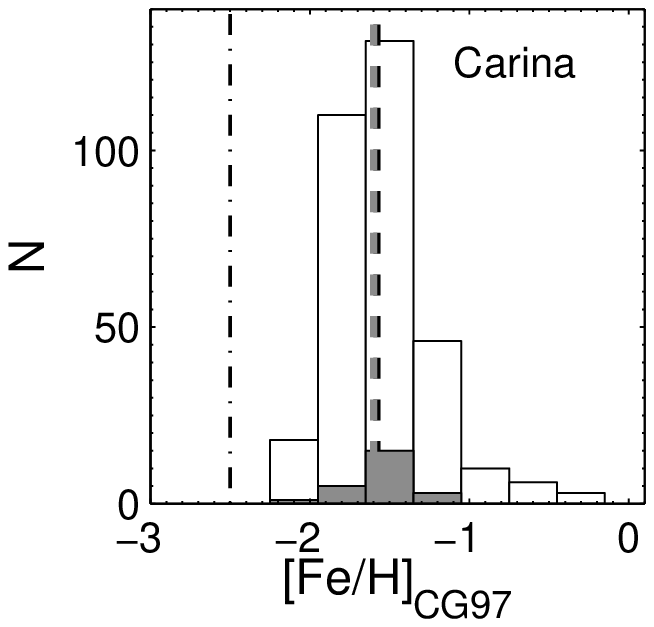}
       \includegraphics[width=4.55cm,clip]{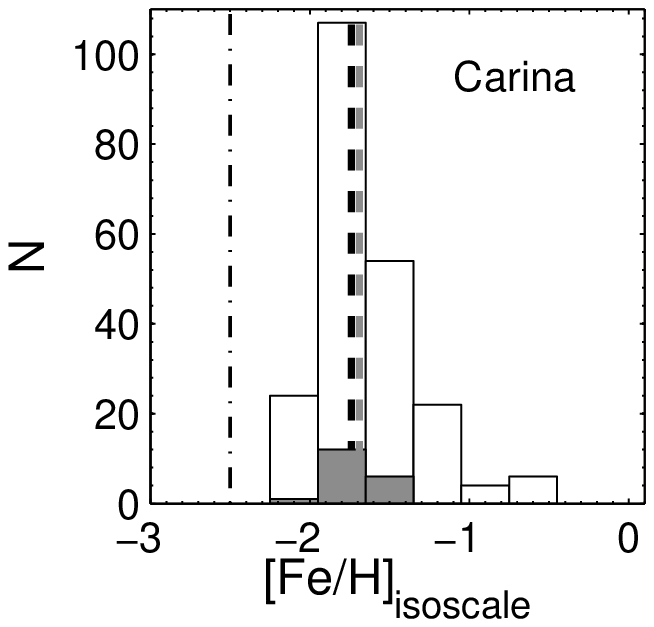}
       \includegraphics[width=4.55cm,clip]{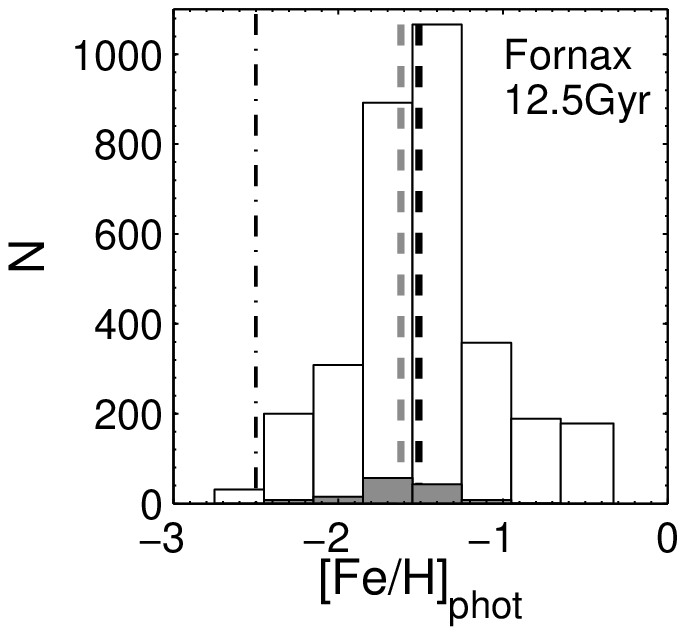}
       \includegraphics[width=4.55cm,clip]{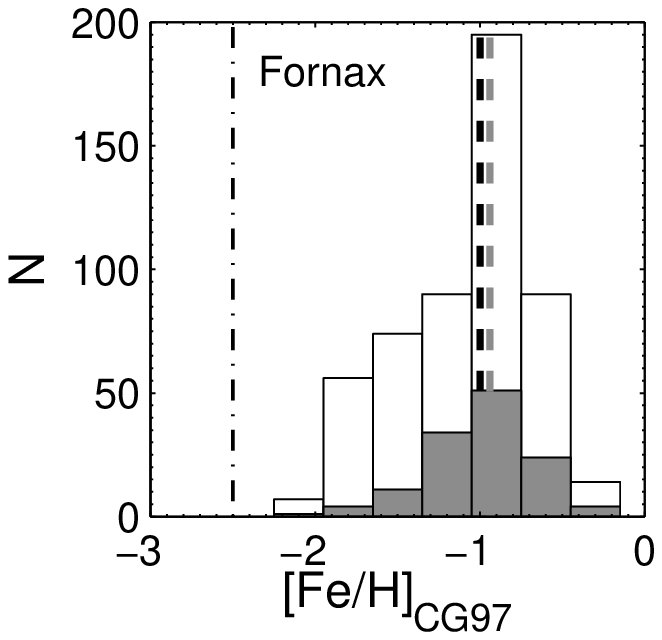}
       \includegraphics[width=4.55cm,clip]{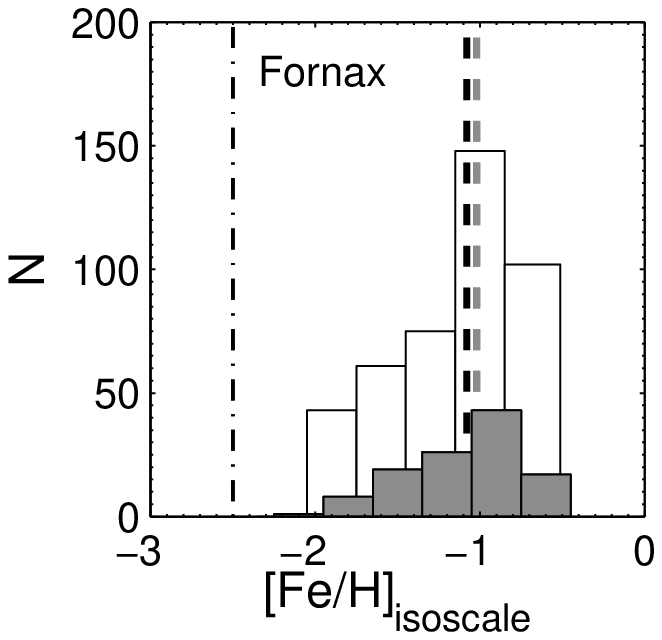}
       \includegraphics[width=4.55cm,clip]{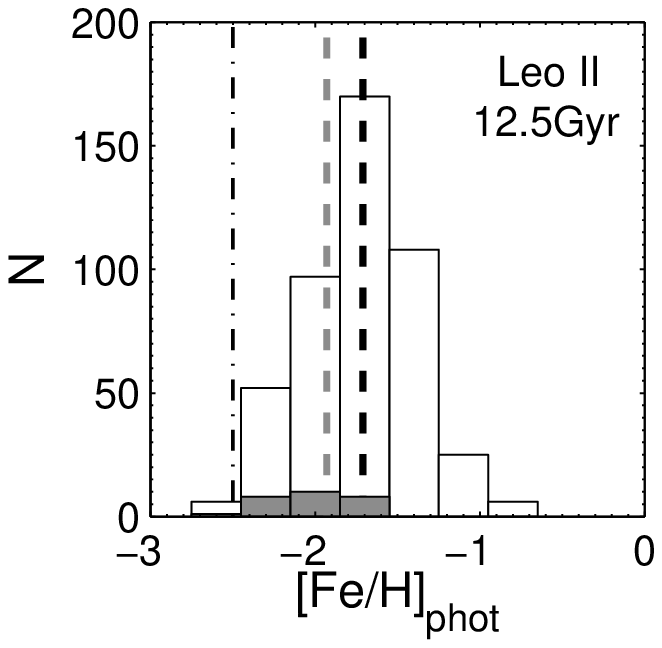}
       \includegraphics[width=4.55cm,clip]{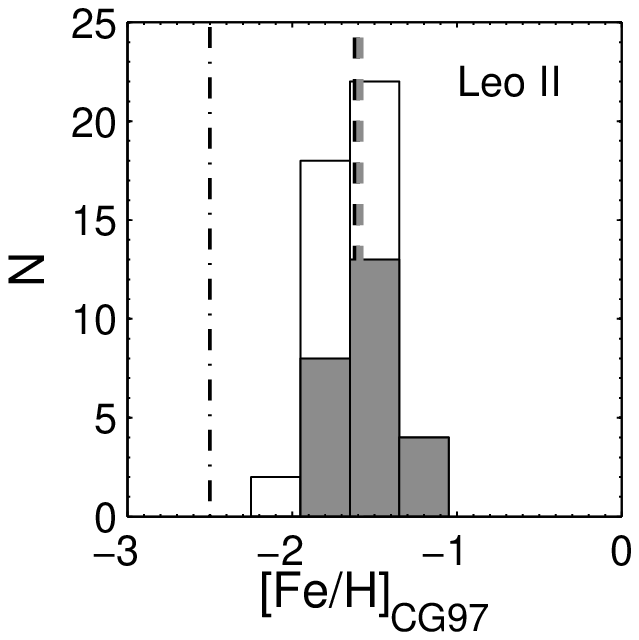}
       \includegraphics[width=4.55cm,clip]{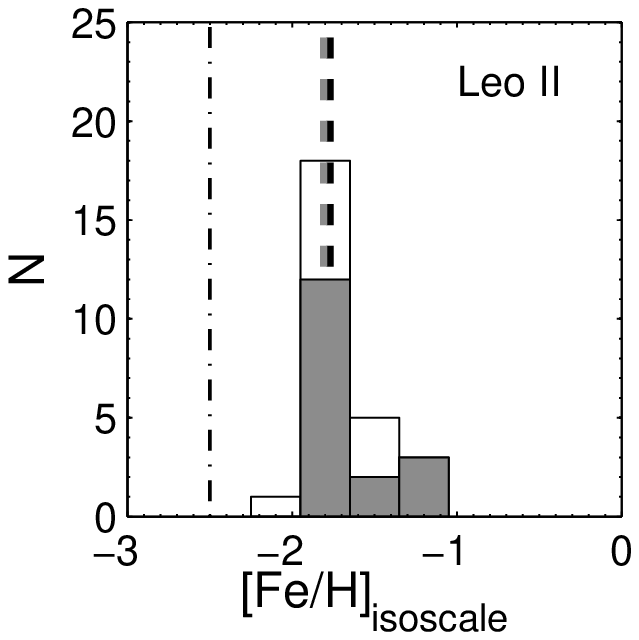}
    \caption{Left panels: The white histograms show the photometric MDFs of
      all the stars within 3\,mag below the TRGB, for Sculptor, Sextans,
      Carina, Fornax, and Leo\,II, while the shaded histograms show the same
      but only for the common stars. 
      Middle panels: The white histograms show the Ca\,T--based MDFs on the
      CG97 metallicity scale for the full available spectroscopic sample, as
      described on the text, while the shaded histograms show the same but
      only for the stars in common.  
      Right panels: The open histograms show the Ca\,T--based MDFs on the
      isoscale (discussed in Section 3.3), while the shaded histograms show
      the same but only for the common stars. In all panels, the vertical
      black dashed line corresponds to the median metallicity of the full
      sample, while the grey dashed line corresponds to the one of the common 
      stars. The vertical dotted--dashed line in all panels corresponds to the 
      isoschrone's metal--poor limit of roughly $-$2.5\,dex. Note that for
      Sextans on the isoscale MDF (right panel), the total number of the
      common stars is three (shaded histogram), while the one for the full
      sample is twelve (white histogram). }  
    \label{sl_4mdf}% 
\end{figure*}
%%%%%%%%%%%%%%%%%%%%%%%%%%%%%%%%%%%%%%%%%%%%%%%%%%%%%%%%%%%%%%%%%%%%%%%%%%%%%
%
    We show the derived photometric metallicity distribution functions (MDFs)
    with the white histograms in Fig.~\ref{sl_4mdf} for Sculptor, Sextans,
    Carina, Fornax, and Leo\,II.

\subsection{Dartmouth isochrones metallicity scale}

    In order to perform a direct comparison between the photometric and the
    spectroscopic metallicities, it is important to clarify to which
    metallicity scale the photometric metallicities conform. The photometric
    metallicities are tied to the isochrone models that are used in the
    interpolation method. The Dartmouth isochrones used here are not
    explicitly tied to any of the spectroscopic, standard abundance scales
    (i.e., ZW84; CG97; KI03; CBG09). Their metallicities are rather based on
    the mass fractions of the heavy elements and hydrogen in the models along
    with the adopted solar abundances. In that sense, the photometric
    metallicities based on the Dartmouth isochrones form a metallicity scale
    on their own. However, the Dartmouth models tend to lie close to the ZW84
    and to the KI03 metallicity scales (A.~Dotter, private communication; see
    also Dotter et al.~\cite{sl_dotter10}).

%%%%%% TABLE 2 %%%%%%%%%%%%%%%%%%%%%%%%%%%%%%%%%%%%%%%%%%%%%%%%%%%%
\begin{table}
\begin{minipage}[t]{\columnwidth}
\caption[]{Galactic GC metallicities in different metallicity scales.}
\label{tablea} 
\centering
\renewcommand{\footnoterule}{}
\begin{tabular}{l c c c }

\hline\hline
 [Fe/H]       &47\,Tuc             &NGC\,3201             &NGC\,6397          \\
 (1)          &(2)                 &(3)                   &(4)                \\
\hline

 ZW84         &$-0.71\pm0.05$     &$-1.53\pm0.03$         &$-1.94\pm0.02$     \\

 CG97         &$-0.78\pm0.02$     &$-1.24\pm0.03$         &$-1.76\pm0.03$     \\

 KI03         &$-0.70\pm0.09$     &$-1.56\pm0.10$         &$-2.02\pm0.07$     \\

 CBG09        &$-0.743\pm0.026$   &$-1.495\pm0.073$       &$-1.993\pm0.060$   \\

 KM08,11      &$-0.76\pm0.01$     &\dots                  & $-2.10\pm0.02$    \\

 Dartmouth    &$-0.70$            &$-1.50$                &$-2.10$             \\

\hline

\end{tabular}
\footnotetext[0]{Notes.-- ZW84 stands for the Zinn \& West (\cite{sl_zinn84})
  metallicity scale; CG97 stands for the Carretta \& Gratton
  (\cite{sl_carretta97}) metallicity scale; KI03 stands for the Kraft \& Ivans
  (\cite{sl_kraft03}) metallicity scale; CBG09 stands for the Carretta et
  al.~(\cite{sl_carretta09}) metallicity scale; KM08\,11 stands for the Koch
  \& McWilliam (\cite{sl_koch08_47tuc}; 2011, in prep.) differential
  metallicity scale; Dartmouth stands for the metallicities derived using
  isochrone fitting (Dotter et al.~\cite{sl_dotter10}).}
\end{minipage}
\end{table}
%%%%%%%%%%%%%%%%%%%%%%%%%%%%%%%%%%%%%%%%%%%%%%%%%%%%%%%%%%%%%%%%%%%%%%%%%%%%%%%%%%%%
%
    In Table~\ref{tablea} we show as an example the mean metallicities of
    three Galactic GCs with metallicities on the ZW84 (adopted from R97; their
    Table 2, column 5), CG97 (adopted from R97; their Table 2, column 6),
    KI03, CBG09, and Koch \& McWilliam (\cite{sl_koch08_47tuc}; 2011, in
    prep.; hereafter KM08,\,11) metallicity scales, as well as the metallicity
    derived through isochrone fitting using Dartmouth isochrones (Dotter et
    al.~\cite{sl_dotter10}). Indeed, the metallicities derived using isochrone
    fitting agree better with the metallicities on the ZW84, KI03 and CBG09 
%
%%%%%% FIGURE 3 - ALL %%%%%%%%%%%%% Two column figure (place early!) %%%%%%  
 \begin{figure}
    \centering                                                               
       \includegraphics[width=5.5cm,clip]{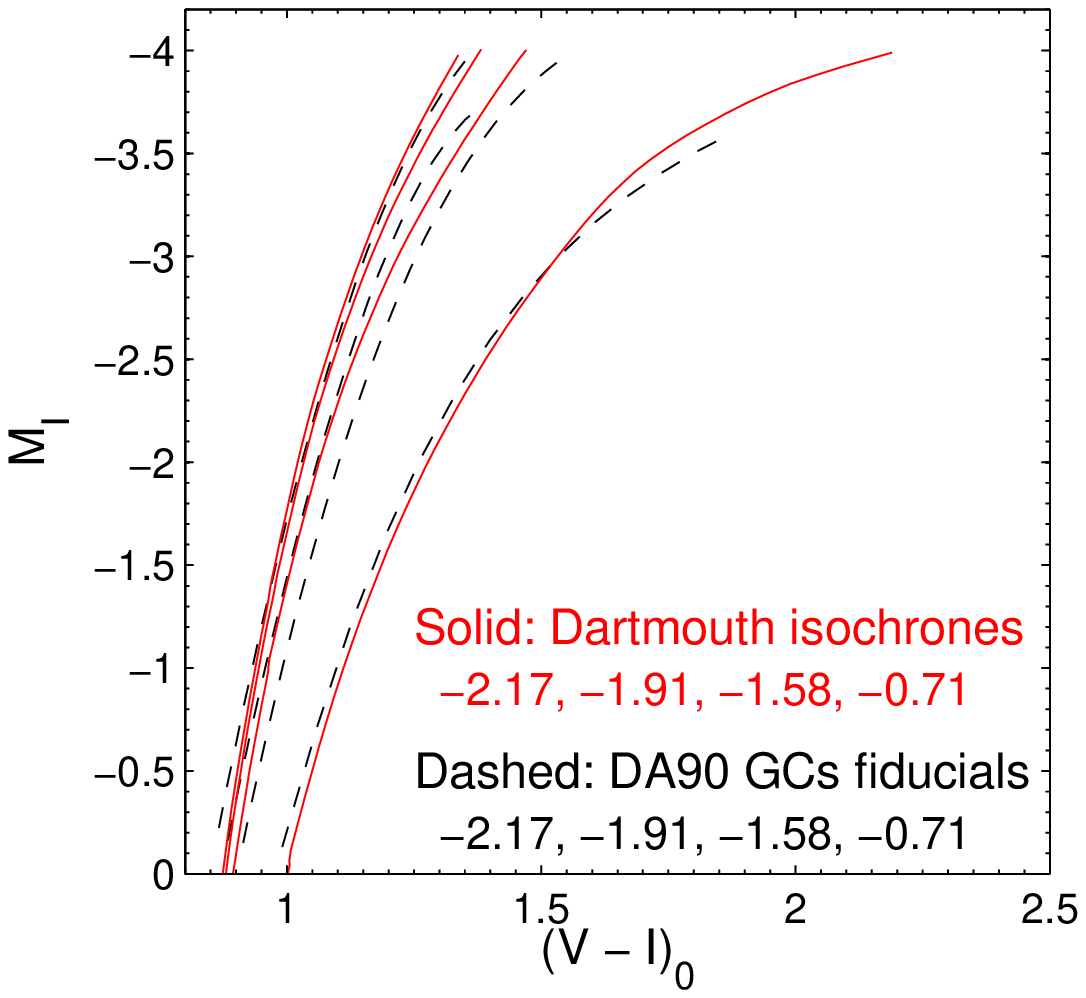}                  
       \includegraphics[width=5.5cm,clip]{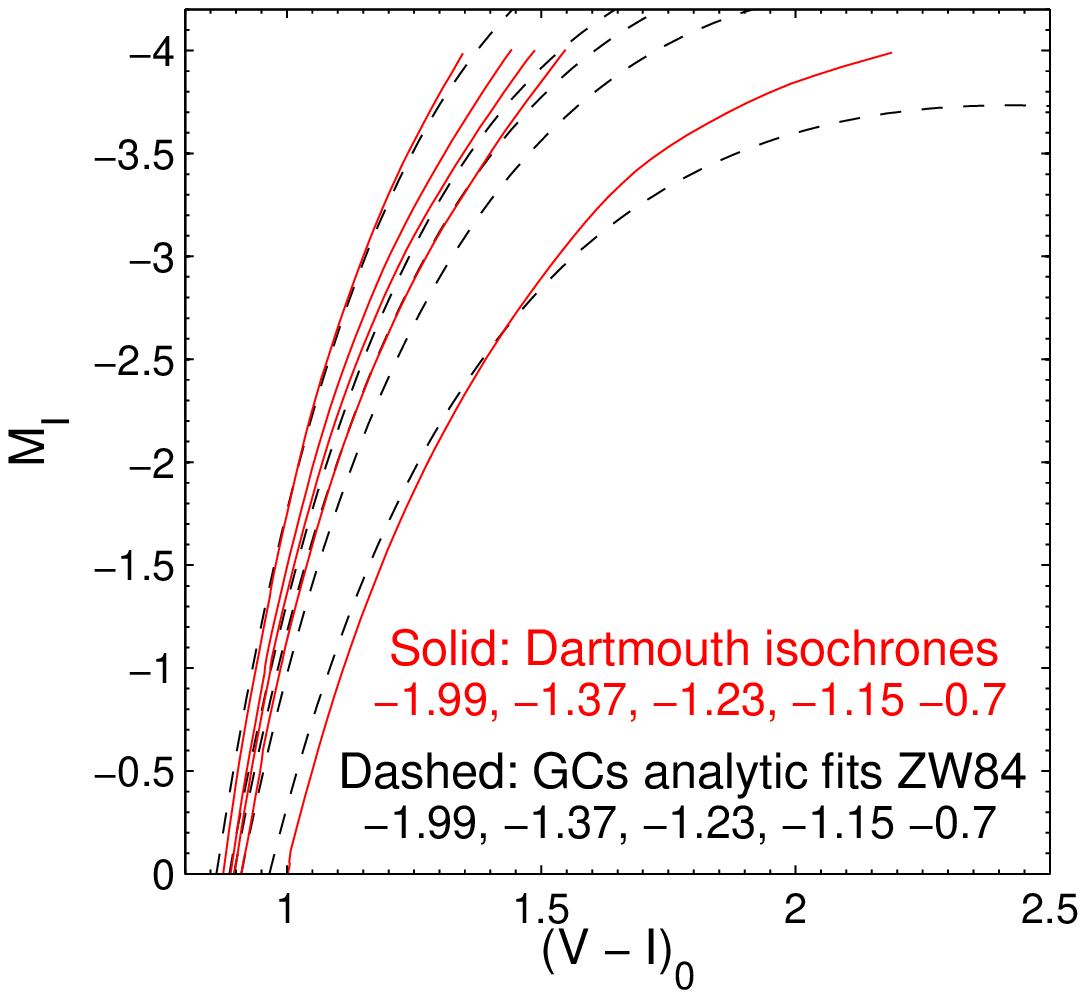}                  
       \includegraphics[width=5.5cm,clip]{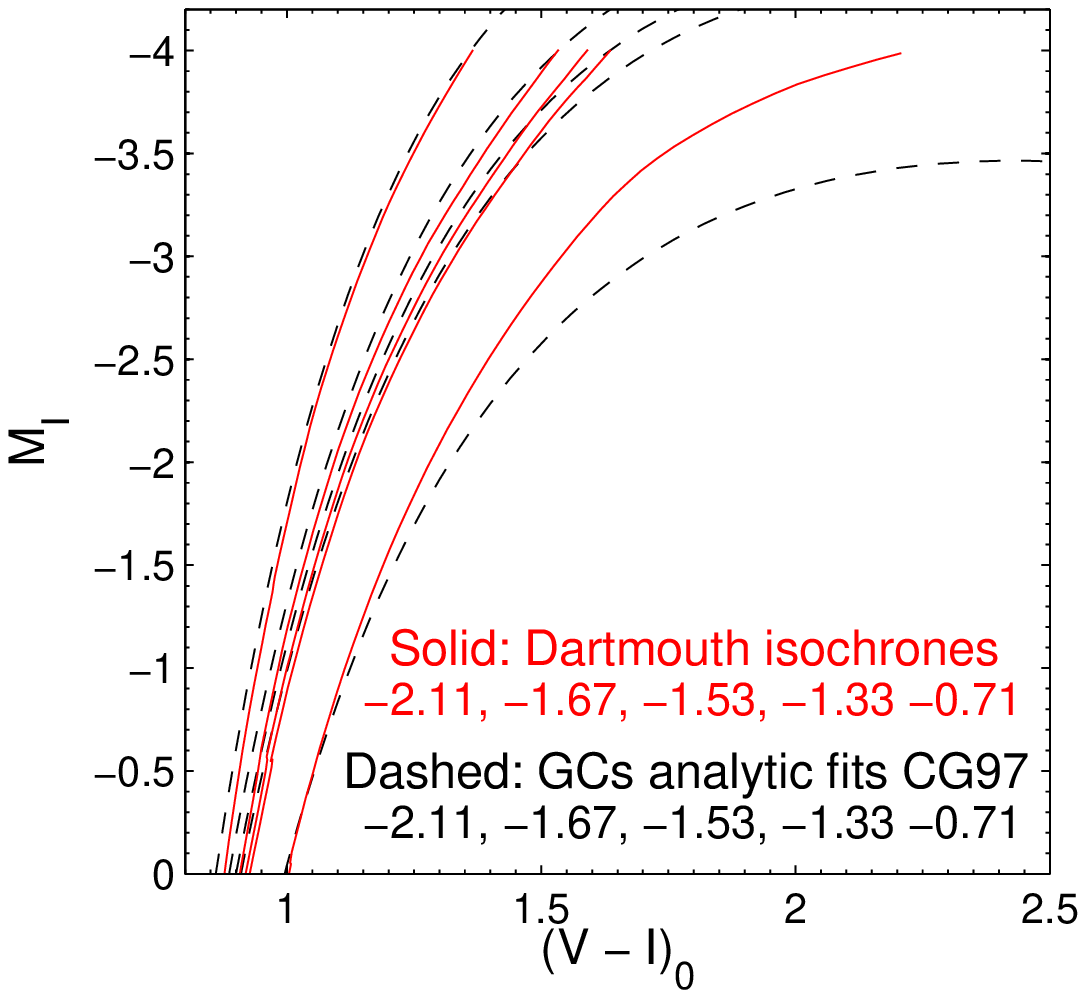}                  
    \caption{Upper panel: Galactic GC fiducials from DA90 shown as black
      dashed lines, along with Dartmouth isochrones shown as red solid
      lines. From left to right, the Galactic GC fiducials correspond to
      M\,15, NGC\,6397, M\,2 and 47\,Tuc, where the  GCs have a metallicity of
      $-$2.17, $-$1.91, $-$1.58 and $-$0.71 dex, respectively, on the ZW84
      scale. The isochrones correspond to an age of 12.5\,Gyr and have the
      same metallicities as the Galactic GCs, from left to right. 
      Middle panel: Analytic functions of Galactic GC fiducial loci from
      Saviane et al.~(\cite{sl_saviane00a}) on the ZW84 scale (black dashed
      lines) along with Dartmouth isochrones of 12.5\,Gyr (red solid
      lines). The metallicities correspond to $-$2.11, $-$1.67, $-$1.53,
      $-$1.33, and $-$0.71 dex from left to right, respectively.   
      Lower panel: The same as in the middle panel but for the CG97
      metallicity scale. The metallicities correspond to $-$1.99, $-$1.37,
      $-$1.23, $-$1.15, and $-$0.7 dex from left to right, respectively.}  
    \label{sl_fid}%                                                  
\end{figure}                                                                 
%%%%%%%%%%%%%%%%%%%%%%%%%%%%%%%%%%%%%%%%%%%%%%%%%%%%%%%%%%%%%%%%%%%%%%%%%%%%%
%
    scales and the differential reference scale based on high--resolution
    spectroscopy of KM08,\,11 than with the CG97 metallicty
    scale. Furthermore, in Fig.~\ref{sl_fid}, upper panel, we plot the
    Galactic GC fiducials of M\,15, NGC\,6397, M\,2, and Tuc\,47, adopted from
    DA90, along with Dartmouth isochrones for a fixed age of 12.5\,Gyr. The
    metallicities of the Galactic GC fiducials are $-$2.17, $-$1.91, $-$1.58,
    and $-$0.71 dex, respectively, on the ZW84 scale (adopted from DA90), and
    the same metallicities are chosen for the Dartmouth isochrones. In the
    middle and lower panels we plot the analytic fits to the fiducial loci of
    GC RGBs adopted from Saviane et al.~(\cite{sl_saviane00a}), on the ZW84
    and CG97 metallicity scales, respectively. In this comparison, the
    Dartmouth isochrones provide an excellent approximation to the DA90
    fiducials on the ZW84 scale at the metal--poor and metal--rich end of the
    fiducials with slightly worse agreement at metallicities in between. The
    Dartmouth isochrones reproduce the GC RGB slopes very well. With the
    exception of the metal--poor end, the analytic fits to GC RGBs by Saviane
    et al.~(\cite{sl_saviane00a}) on the ZW84 metallicity scale show offsets
    compared to the Dartmouth isochrones (Fig.~\ref{sl_fid}, middle
    panel). When comparing the Dartmouth isochrones to Saviane's et
    al.~analytic fits on the CG97 scale, the slopes are quite similar, but the
    fiducials of Saviane et al.~are systematically too metal--poor. At the
    metal--rich end ($\sim-$0.7\,dex), the isochrones and the fiducials show
    substantial discrepancies.

%
%%%%%% FIGURE 4 - ALL %%%%%%%%%%%%% Two column figure (place early!) %%%%%%  
 \begin{figure}
    \centering
       \includegraphics[width=6cm,clip]{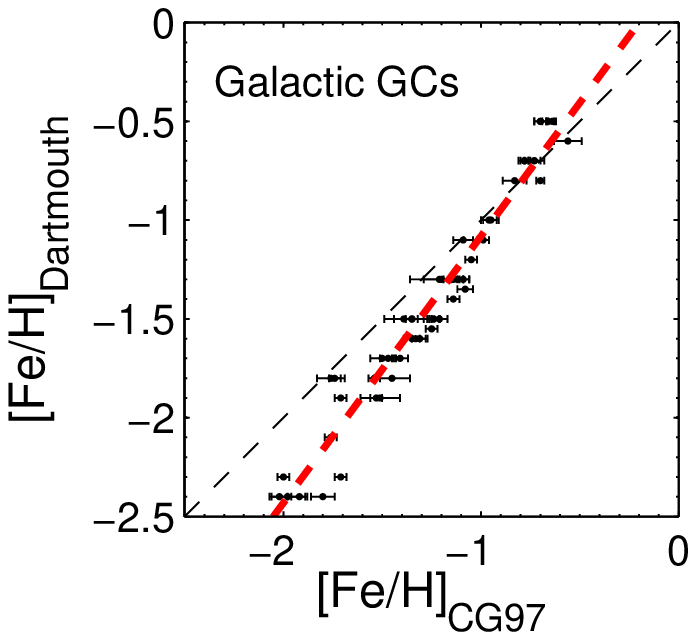}
    \caption{Transformation from the CG97 metallicity scale to the Dartmouth
      isochrone metallicity scale, simply called ``isoscale''. The red thick
      dashed line corresponds to the error--weighted linear least squares fit
      to the data. The black thin dashed line corresponds to unity.} 
    \label{sl_isoscale}% 
\end{figure}
%%%%%%%%%%%%%%%%%%%%%%%%%%%%%%%%%%%%%%%%%%%%%%%%%%%%%%%%%%%%%%%%%%%%%%%%%%%%%
%  
    Therefore, considering that the Dartmouth isochrones form a metallicity
    scale of their own, we transform the CG97 spectroscopic metallicities of
    all five dSphs to the metallicity scale defined by the Dartmouth
    isochrones. We call the metallicity scale defined by the Dartmouth
    isochrones simply ``isoscale''. In order to  derive this transformation,
    we use the metallicities of those Galactic GCs in Dotter et
    al.~(\cite{sl_dotter10}; their Table~2), derived using isochrone fitting
    that also have metallicities on  the CG97 metallicity scale as derived in
    R97 (their Table~2). In total we include 50 Galactic GCs. The resulting
    transformation of the metallicities from the CG97 to the isoscale reads as
    follows: 
    \begin{equation} 
        [{\rm Fe/H}]_{\rm isoscale} = 1.35_{(\pm0.04)} [{\rm Fe/H}]_{\rm CG97} + 0.27_{(\pm0.06)},
    \end{equation}
    and holds within the metallicity range of
    $-2.02\leq$[Fe/H]$_{CG97}\leq-0.5$\,(dex), as given by the availability of
    GCs with [Fe/H] on the CG97 metallicity scale that also have [Fe/H] based
    on the Dartmouth isochrone fitting. The transformation is plotted in
    Fig.~\ref{sl_isoscale}. The reversed metallicity transformation from the
    isoscale to the CG97 metallicity scale reads as follows: 
    \begin{equation} 
        [{\rm Fe/H}]_{\rm CG97} = 0.70_{(\pm0.02)} [{\rm Fe/H}]_{\rm isoscale} - 0.26_{(\pm0.04)},
    \end{equation}
    and holds within the metallicity range of
    $-2.4\leq$[Fe/H]$_{isoscale}\leq-0.5$\,(dex), which is defined by the
    available metallicities of the GCs based on the isochrone fitting. 

    Since we do not have enough Galactic GCs that have metallicities in both
    the MRS metallicity scale and the Dotter et al.~(\cite{sl_dotter10})
    sample, we perform the comparison between the photometric metallicities
    and the MRS metallicities without transforming the latter ones to the
    scale of the former. Therefore, in the later sections, the comparison of
    the MRS and photometric metallicities will not be performed on the
    isoscale.

\subsection{Spectroscopic metallicities}

    The white histograms in the middle and right panels of Fig.~\ref{sl_4mdf}
    show the Ca\,T--based spectroscopic MDFs on the CG97 metallicity scale and
    on the isoscale, for Sculptor, Sextans, Carina, Fornax, and Leo\,II. These
    histograms were derived using the whole available spectroscopic sample
    with the requirement that the spectroscopic metallicities (of all dSphs
    apart from Sextans) are within the range of
    $-2\leq$\,[Fe/H]$_{CG97}\leq-0.2$\,(dex) for the CG97 metallicity scale
    (Cole et al.~\cite{sl_cole04}), and
    $-2.02\leq$\,[Fe/H]$_{isoscale}\leq-0.5$\,(dex) for the isoscale. Within
    this metallicity range the linear calibration between [Fe/H] and
    W$^{\prime}$ (or $\Sigma$W) is valid (Cole et al.~\cite{sl_cole04}).

    For Sextans on the CG97 metallicity scale, we adopt the metallicity range
    of $-4\leq$\,[Fe/H]$_{CG97}\leq-0.5$\,(dex) (Battaglia et
    al.~\cite{sl_battaglia11}), where the revised calibration of the Ca\,T of
    Starkenburg et al.~(\cite{sl_starkenburg10}) has been applied to derive
    [Fe/H] based on the sum of the Ca\,T lines, $\Sigma$W, on the CG97
    metallicity scale.

%
%%%%%% FIGURE 5 - ALL %%%%%%%%%%%%% Two column figure (place early!) %%%%%%  
 \begin{figure}                                               
    \centering
       \includegraphics[width=4.3cm,clip]{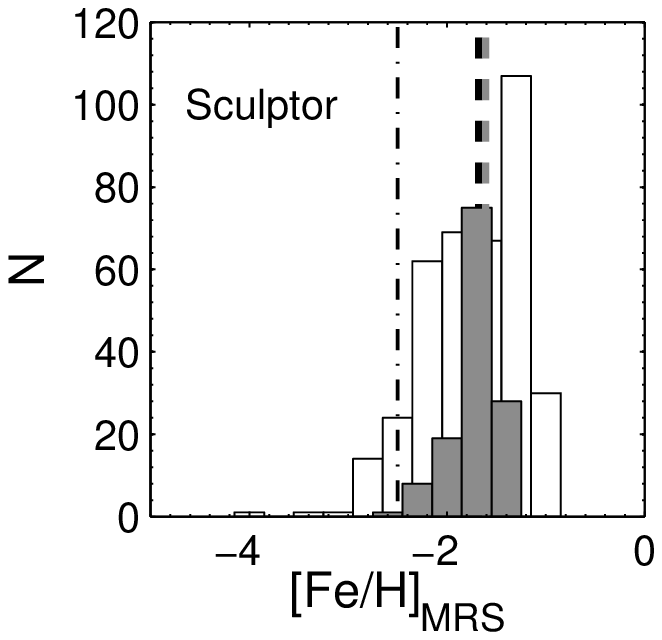}
       \includegraphics[width=4.3cm,clip]{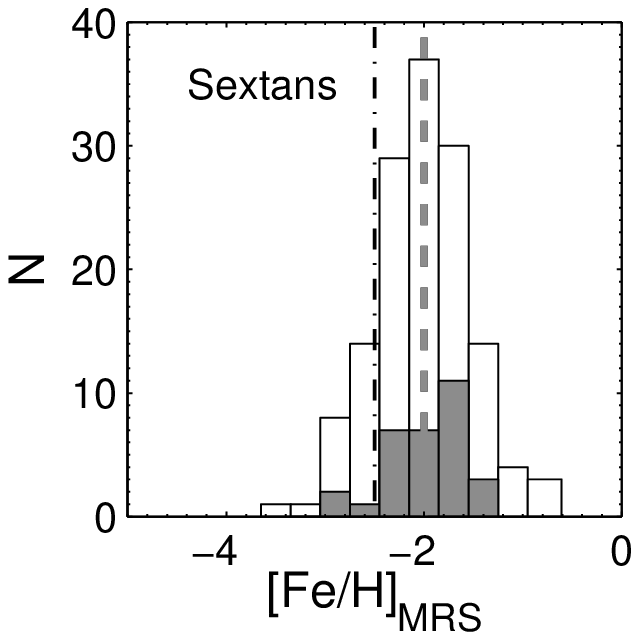}
       \includegraphics[width=4.3cm,clip]{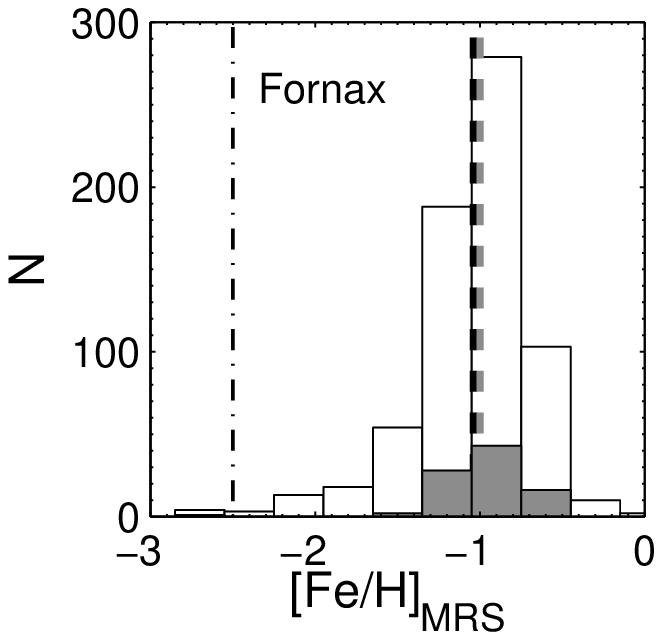}
       \includegraphics[width=4.3cm,clip]{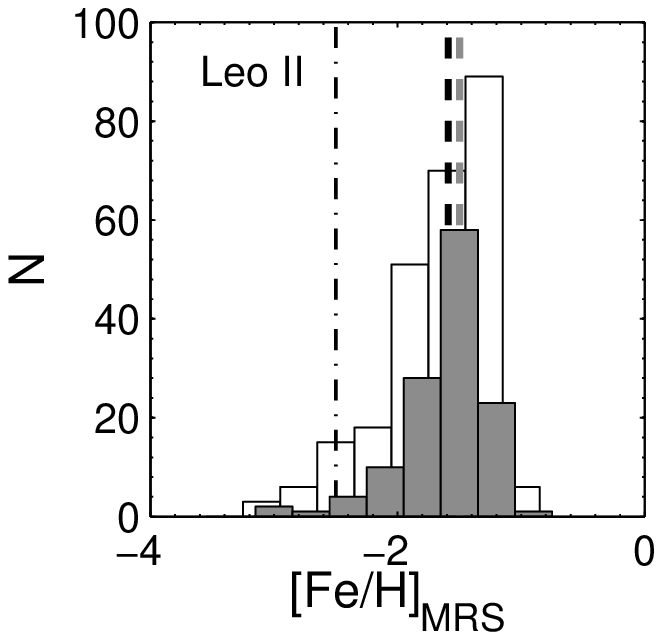}
    \caption{The white histograms show the MRS MDFs for the full available MRS
      sample for Sculptor, Sextans, Fornax, and Carina. The shaded histograms
      show the MRS MDFs for the stars in common to both the MRS and
      photometric samples. In all panels, the vertical black dashed line
      corresponds to the median metallicity value of the full sample, while
      the grey dashed line corresponds to the median metallicity value of the
      common stars. The vertical dotted--dashed line corresponds to the lower
      limit of the most metal--poor isochrone used of roughly $-$2.5\,dex.} 
    \label{sl_mrsmdf}% 
\end{figure}
%%%%%%%%%%%%%%%%%%%%%%%%%%%%%%%%%%%%%%%%%%%%%%%%%%%%%%%%%%%%%%%%%%%%%%%%%%%%%
%
    In Fig.~\ref{sl_mrsmdf} we show the MRS--based MDFs of the whole available
    spectroscopic sample (white histograms) for Sculptor, Sextans, Fornax,
    and Leo\,II. 
 
\subsection{Photometric and spectroscopic metallicities of the common stars }

    The photometric and spectroscopic MDFs for those stars with both
    photometric and Ca\,T--based spectroscopic measurements are shown as the
    shaded histograms in Fig.~\ref{sl_4mdf} for Sculptor, Sextans, Carina,
    Fornax, and Leo\,II. In order to construct the photometric MDFs of the
    common stars, only those stars within the photometric metallicity range of
    $-2.5<$\,[Fe/H]$_{phot}\leq-0.3$\,dex and with an error of less than
    0.2\,dex are retained. The distribution of the common stars falls within
    3$r_c$, 1.5$r_c$, 2$r_c$, 5$r_c$, 1.5$r_c$ for Sculptor, Sextans, Carina,
    Fornax, and Leo\,II, respectively, where r$_c$ denotes the core radius of
    each dSph adopted from Irwin \& Hatzidimitriou (\cite{sl_irwin95}). For
    the spectroscopic MDFs of the common stars in addition to the photometric
    metallicity cuts we impose the Ca\,T spectroscopic metallicity cuts as
    described in Section 3.4. The number of the common stars that have
    metallicities both on the CG97 metallicity scale and on the isoscale are
    listed in Table~\ref{table3b}, except for Sextans. As shown in
    Fig.~\ref{sl_4mdf}, Sextans only has three stars for which we have both
    reliable photometry and Ca\,T spectroscopy. Thus we exclude Sextans from
    any further analysis regarding common stars.

    The differences of the Ca\,T spectroscopic (on the isoscale) minus
    the photometric metallicities versus the \textit{V}--band magnitudes 
%
%%%%% FIGURE 6 - ALL %%%%%%%%%%%%% One column figure (place early!) %%%%%%
 \begin{figure}
    \centering
       \includegraphics[width=4cm,clip]{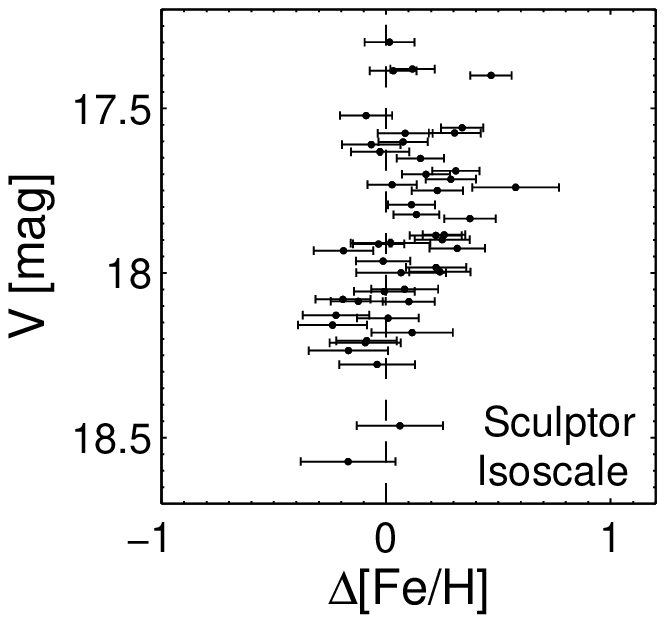}  
       \includegraphics[width=4cm,clip]{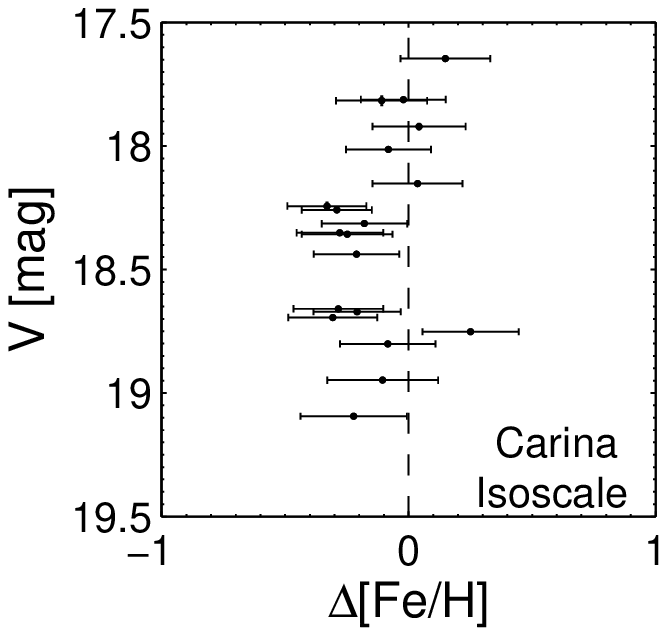}
       \includegraphics[width=4cm,clip]{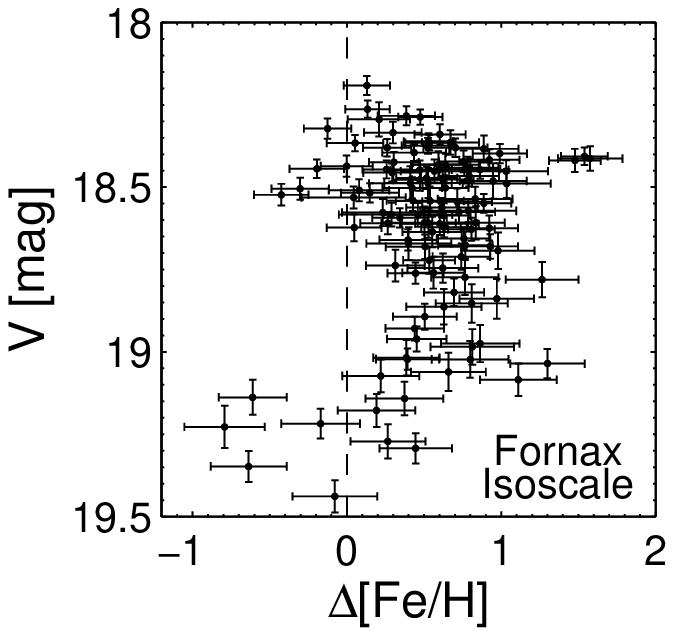}
       \includegraphics[width=4cm,clip]{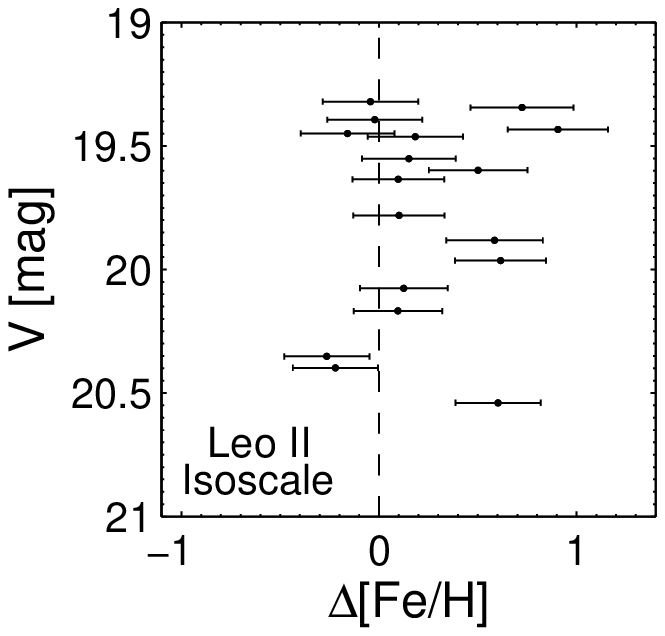}
       \caption{\textit{V}--band magnitude versus the difference in
         spectroscopic minus the photometric metallicities on the isoscale for
         Sculptor, Carina and Fornax, and Leo\,II.}
       \label{sl_fehvmag}%
\end{figure}
%%%%%%%%%%%%%%%%%%%%%%%%%%%%%%%%%%%%%%%%%%%%%%%%%%%%%%%%%%%%%%%%%%%%%%%%%%
%
    are shown in Fig.~\ref{sl_fehvmag} for Sculptor, Carina, Fornax, and
    Leo\,II. There is a slight trend of the metallicity differences to become
    negative as the \textit{V}--band magnitude becomes fainter, and to become
    positive as the \textit{V}--band magnitude becomes brighter. This trend is
    not significant, as indicated by the Pearson correlation coefficients of
    $-$0.44, $-$0.31, $-$0.2, $-$0.15 for Sculptor, Carina, Fornax, and
    Leo\,II, respectively. We note that the spacing between the isochrones
    decreases towards the metal--poor end so that an unambiguous assignment of
    metallicities becomes difficult. This results in an inability of the
    photometric metallicities to reproduce the Ca\,T spectroscopic metal--poor
    tail of the MDF (cf.~Koch et al.~\cite{sl_koch08b}). On the other hand,
    the Ca\,T method has its largest sensitivity at the metal--poor end.

    The shaded histograms in Fig.~\ref{sl_mrsmdf} show the MRS--based MDFs of
    the common stars. Again, only stars with photometric metallicities within
    the range of $-$2.5\,dex to $-$0.3\,dex are included. The number of the
    common stars for each dSph is listed in Table~\ref{table3b}.

\section{Discussion}

    In a stellar system with a complex SFH where both old and
    intermediate--age stellar populations are present, its RGB contains stars
    belonging to the full age range of approximately 1.5\,Gyr and older ages
    (Salaris, Cassisi \& Weiss \cite{sl_salaris02}), depending on the details
    of the stellar system's SFH. Thus the assumption of a single old age for
    the stellar populations and therefore for the isochrones used in the
    interpolation holds only in the case of a negligible intermediate--age
    population. In dSphs, the initial star formation may have lasted as long
    as 3\,Gyr or even longer (Marcolini et al.~\cite{sl_marcolini08}; Ikuta \&
    Arimoto~\cite{sl_ikuta02}), thus leading to large metallicity dispersions
    (Grebel, Gallagher \& Harbeck \cite{sl_grebel03}). In the case of dSphs
    dominated by old populations with ages larger than 10\,Gyr, this extended
    star formation does not substantially affect their photometric
    metallicities and can lead to photometric metallicity differences of
    individual stars of only approximately 0.1\,dex, as demonstrated in Lianou
    et al.~(\cite{sl_lianou10}) using isochrone grids of two different ages
    (12.5\,Gyr and 10.5\,Gyr) and a range in metallicities in M\,81 group
    dSphs. Here we explore the effects of the presence of intermediate--age
    populations on deriving photometric metallicities under the assumption of
    a single old age. For that purpose, we compare the photometric
    metallicities with spectroscopic metallicities derived through the Ca\,T
    method and through the MRS method, on a star--by--star basis.

    It is clear that in the presence of intermediate--age populations we do
    not expect a priori that there will be an agreement between the
    photometric and spectroscopic metallicities. The existence of mixed--age
    populations is expected to lead to an overestimate of the photometric
    metallicities towards the metal--poor part. At fixed metallicities, an
    intermediate--age population would lie bluewards in color on the RGB as
    compared to an old population. Thus, intermediate--age populations would
    be assigned more metal--poor metallicities than their true value, if they
    were erroneously 
%
%%%%% FIGURE 7 - ALL %%%%%%%%%%%%% One column figure (place early!) %%%%%%
 \begin{figure}
    \centering
       \includegraphics[width=6cm,clip]{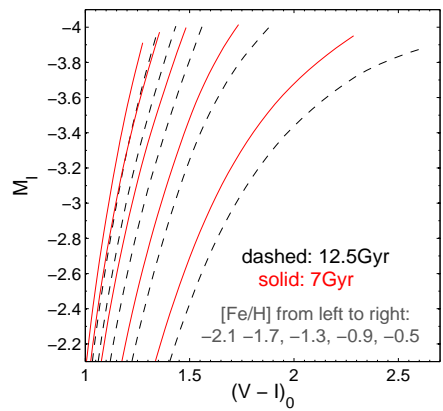}  
       \caption{Dartmouth isochrones for an age of 12.5\,Gyr (black dashed)
         and 7\,Gyr (red solid). The intermediate--age isochrone lies
         bluewards from the older isochrone at a fixed metallicity. Note that
         for a constant age and different values of [Fe/H] the slope of the
         RGB changes, while for a varying age and a constant [Fe/H] the slope
         of the RGB changes very little.} 
       \label{oldbias}%
\end{figure}
%%%%%%%%%%%%%%%%%%%%%%%%%%%%%%%%%%%%%%%%%%%%%%%%%%%%%%%%%%%%%%%%%%%%%%%%%%
%
    assumed to be old. This photometric ``metal--poor bias'' is demonstrated
    in Fig.~\ref{oldbias}, where isochrones of two fixed ages of 12.5\,Gyr and
    7\,Gyr are overplotted with metallicities ranging from $-$2.1 to
    $-$0.5\,dex.

    \subsection{Mean metallicity properties} 

%
%%%%% TABLE 3 %%%%%%%%%%%%%%%%%%%%%%%%%%%%%%%%%%%%%%%%%%%%%%%%%%%%
\begin{table*}
\caption[]{{\em Mean} metallicity properties corresponding to the white
  histograms of Fig.~\ref{sl_4mdf} and Fig.~\ref{sl_mrsmdf}.}
\label{table2b}
\centering
\renewcommand{\footnoterule}{}
\begin{tabular}{l c c c c c c c c c c c}

\hline

Galaxy     &\multicolumn{2}{c}{[Fe/H]$_{phot}$} &  &\multicolumn{2}{c}{[Fe/H]$_{CG97}$} &    &\multicolumn{2}{c}{[Fe/H]$_{isoscale}$} &  &\multicolumn{2}{c}{[Fe/H]$_{MRS}$}  \\ 
                                                                                                                                                                                          
\cline{2-3}
\cline{5-6}
\cline{8-9}                                                                                                                                                                            
\cline{11-12}

          &mean$\pm\sigma$  &median     &         &mean$\pm\sigma$   &median      &        &mean$\pm\sigma$  &median        &         &mean$\pm\sigma$  &median             \\     
          &(dex)            &(dex)      &         &(dex)             &(dex)       &        &(dex)            &(dex)         &         &(dex)            &(dex)              \\     
\hline                                                                                                                                                                                    
                                                                                                                                                                                          
Sculptor  &$-1.67\pm0.25$  &$-1.65$     &        &$-1.47\pm0.26$    &$-1.45$      &        &$-1.57\pm0.23$   &$-1.57$       &         &$-1.74\pm0.48$   &$-1.68$            \\     
                                                                                                                                                                                           
Sextans   &$-1.79\pm0.37$  &$-1.81$     &        &$-2.33\pm0.46$    &$-2.29$      &        &$-1.81\pm0.25$   &$-1.89$       &         &$-2.00\pm0.48$   &$-2.00$            \\     
                                                                                                                                                                                          
Carina    &$-1.54\pm0.35$  &$-1.52$     &        &$-1.55\pm0.29$    &$-1.58$      &        &$-1.65\pm0.30$   &$-1.74$       &         &...              &...                \\     
                                                                                                                                                                                         
Fornax    &$-1.49\pm0.43$  &$-1.51$     &        &$-1.08\pm0.40$    &$-1.00$      &        &$-1.15\pm0.39$   &$-1.08$       &         &$-1.04\pm0.36$   &$-1.00$            \\     
                                                                                                                                                                                          
Leo\,II   &$-1.72\pm0.35$  &$-1.71$     &        &$-1.63\pm0.23$    &$-1.61$      &        &$-1.73\pm0.22$   &$-1.77$       &         &$-1.69\pm0.42$   &$-1.59$            \\      

\hline   
                      
\end{tabular} 
\end{table*}
%%%%%%%%%%%%%%%%%%%%%%%%%%%%%%%%%%%%%%%%%%%%%%%%%%%%%%%%%%%%%%%%%%%%%%%%%%%%%%%%%%%%
%
    We list the {\em mean} photometric and {\em mean} spectroscopic
    metallicity properties for the five studied dSphs in Table~\ref{table2b},
    where both the mean and median metallicity values are listed, as well as
    the standard deviation. For the Ca\,T spectroscopic metallicities we list
    the {\em mean} metallicity properties both on the CG97 metallicity scale
    and on the isoscale. The {\em mean} properties were derived from the full
    available data sets, corresponding to the white histograms shown in
    Fig.~\ref{sl_4mdf} and Fig.~\ref{sl_mrsmdf}. Therefore they do not
    correspond to the common stars.
   
  \subsubsection{Sculptor and Sextans}

    In the case of Sculptor and Sextans, the difference between their median
    photometric and median Ca\,T spectroscopic metallicities on the isoscale
    is 0.08\,dex. The typical photometric metallicity uncertainties have a
    median of 0.13\,dex and 0.06\,dex for Sculptor and Sextans, respectively,
    while the typical spectroscopic uncertainties have a median of 0.11\,dex
    and 0.15\,dex for Sculptor and Sextans, respectively. Sculptor and Sextans
    are dominated by old populations. The fraction of the old populations is
    more than 86\% in Sculptor and 100\% in Sextans (Orban et
    al.~\cite{sl_orban08}). It is therefore reassuring that the photometric
    and Ca\,T spectroscopic metallicities on the isoscale are in such a good
    agreement.

    There is good agreement also between the median photometric metallicity of
    Sculptor and its median Ca\,T spectroscopic metallicity on the CG97
    metallicity scale, with a difference of 0.2\,dex, while typical
    spectroscopic metallicity uncertainties on the CG97 metallicity scale have
    a median of 0.05\,dex. This is not the case for Sextans, where the
    difference of the median spectroscopic and the photometric metallicity
    amounts to 0.48\,dex, based on 173 stars (Battaglia et
    al.~\cite{sl_battaglia11}) (white histogram in the middle panel of
    Fig.~\ref{sl_4mdf}). This mismatch between the two median metallicity
    values for Sextans can be explained if one considers that the individual
    Ca\,T spectroscopic metallicities on the CG97 metallicity scale include
    metallicity values as metal--poor as $-$4\,dex (Battaglia et
    al.~\cite{sl_battaglia11}), while the individual photometric metallicities
    were restricted to $-$2.5\,dex, which is the most metal--poor value of
    metallicity provided for the Dartmouth isochrones. Therefore, the
    different selection criteria in terms of metallicity ranges used for the
    metallicities on the CG97 metallicity scale and the photometric
    metallicities may account for this large difference, which further
    suggests that such a comparison may not be appropriate for a galaxy with
    as metal--poor stars as in Sextans.

    Finally, the agreement between the medians of the photometric and MRS
    metallicities is quite good in the case of Sculptor where their difference
    amounts to only 0.03\,dex, whereas for Sextans their difference amounts to
    0.19\,dex. Typical MRS metallicity uncertainties have a median of
    0.12\,dex and 0.19\,dex for Sculptor and Sextans, respectively.
 
  \subsubsection{Carina, Fornax, and Leo\,II}

    Carina, Fornax, and Leo\,II have complex star formation and chemical
    enrichment histories that produced substantial intermediate--age
    populations, each in different amounts. The difference of their median
    photometric metallicity from their median spectroscopic metallicity is
    indeed non--zero, with the tendency of the median photometric
    metallicities to be more metal--poor than the spectroscopic ones. 

    We can qualitatively estimate the expected metal--poor bias by comparing
    the median metallicity derived assuming a purely old population with the
    median metallicity derived assuming a mixture of the stellar
    populations. In order to find the median metallicity of a mixture of
    stellar populations, we use the fraction of the total stellar mass formed
    within the last 10\,Gyr and 1\,Gyr (\textit{f}$_{10G}$, \textit{f}$_{1G}$,
    respectively; Orban et al.~\cite{sl_orban08}, their Table~1; reproduced in
    Table~\ref{table3b}) in conjuction with the mean mass--weighted age
    ($\tau$; Orban et al.~\cite{sl_orban08}, their Table~1; reproduced in
    Table~\ref{table3b}) for Carina, Fornax, and Leo\,II. For that purpose, we
    run the interpolation code with a constant age of the isochrones equal  to
    $\tau$. Then, we randomly assign
    \textit{f}$_{inter}=($\textit{f}$_{10G}-$\textit{f}$_{1G}$)\% of the stars
    within our RGB sample metallicities as derived using isochrones with
    constant ages equal to $\tau$. The remaining
    100\,$-$\,\textit{f}$_{inter}$ of the stars are assigned their original
    metallicities, assuming that they are of a constant, old age of
    12.5\,Gyr. In all cases, the range in metallicities is varied from
    $-$2.5\,dex to $-$0.3\,dex. The median metallicity of the mixture of
    stellar populations is $-$1.34\,dex for Carina, $-$1.35\,dex for Fornax,
    and $-$1.56 for
%
%%%%% TABLE 4 %%%%%%%%%%%%%%%%%%%%%%%%%%%%%%%%%%%%%%%%%%%%%%%%%%%%
\begin{table}
\caption[]{{\em Mean} metallicity properties corresponding to the mixture
  of stellar fractions as described in Section 4.1.2.} 
\label{tablemix}
\centering
\renewcommand{\footnoterule}{}
\begin{tabular}{l c c c}

\hline

Galaxy          &   &\multicolumn{2}{c}{[Fe/H]$_{mixture}$}  \\ 
                                                                                            
\cline{3-4}

                &       &mean$\pm\sigma$  &median  \\     
                &       &(dex)            &(dex)   \\     
\hline                                                    
                                                                                                                    
Carina          &       &$-1.38\pm0.32$   &$-1.34$ \\     
                                                          
Fornax          &       &$-1.36\pm0.37$   &$-1.35$ \\     
                                                          
Leo\,II         &       &$-1.57\pm0.42$   &$-1.56$ \\      

\hline   
                      
\end{tabular} 
\end{table}
%%%%%%%%%%%%%%%%%%%%%%%%%%%%%%%%%%%%%%%%%%%%%%%%%%%%%%%%%%%%%%%%%%%%%%%%%%%%%%%%%%%%
%
    Leo\,II. These values are listed in Table~\ref{tablemix}, along with the
    mean metallicities and the standard deviations. The difference of the
    median metallicities when assuming a purely old stellar population and
    when assuming a mixture of stellar populations results in 0.17\,dex,
    0.16\,dex, and 0.15\,dex, respectively, for the aforementioned dSphs
    making them more metal--rich in the case of mixed--age populations. The
    differences of the median photometric from the median spectroscopic (on
    the isoscale) metallicities in the case of Carina and Leo\,II are
    consistent with such mixtures of the stellar populations, while in the
    case of Fornax the difference is larger than that computed using its
    respective admixture. This suggests that either a higher fraction
    \textit{f}$_{inter}$ and\,/\,or a younger age $\tau$ is required in order
    to produce such a difference. The age $\tau$ of Fornax is 7.4\,Gyr (Orban
    et al.~\cite{sl_orban08}). If we assume 100\% of the stars within our RGB
    sample to have formed with this age we derive a median photometric
    metallicity of $-$1.29\,dex.  Subtracting the latter value from the median
    photometric metallicity assuming a purely old population does not lead to
    the same difference as that between the photometric and spectroscopic (on
    the isoscale) medians. Therefore, an age $\tau$ much younger than 7.4\,Gyr
    is required in order to justify the difference between the median
    photometric and spectroscopic (on the isoscale) metallicities. It turns
    out that all the stars in our RGB sample would need to have an age of
    $\sim$4\,Gyr in order to force agreement between the photometric and
    spectroscopic metallicities. Moreover, we compute that (84, 72, 55)\% of
    the stars within our RGB sample would have an age of approximately (3.5,
    3, 2)\,Gyr, respectively. Therefore, stars within our RGB sample with a
    fraction \textit{f}$_{inter}$ from 55\% to 100\% and with ages from 2\,Gyr
    to 4\,Gyr, respectively, are required in order to produce the observed
    difference between the median photometric metallicity and the median
    spectroscopic metallicity on the isoscale for Fornax. These age ranges are
    consistent with the findings of Coleman \& de Jong (\cite{sl_coleman08}),
    i.e., that Fornax experienced a strong burst of star formation during the
    last 3--4\,Gyr. We also note that Fornax has the largest age spread ever
    found in any Galactic dSph, extending to ages as young as 100--200\,Myr
    (Grebel \& Stetson \cite{sl_grebel99b}).

    \subsection{General error sources for photometric metallicities}

    For those stars with both spectroscopic and photometric measurements, we
    show the photometric versus the Ca\,T metallicities
%
%%%%% FIGURE 8 - ALL %%%%%%%%%%%%% One column figure (place early!) %%%%%%
 \begin{figure*}
    \centering
       \includegraphics[width=5cm,clip]{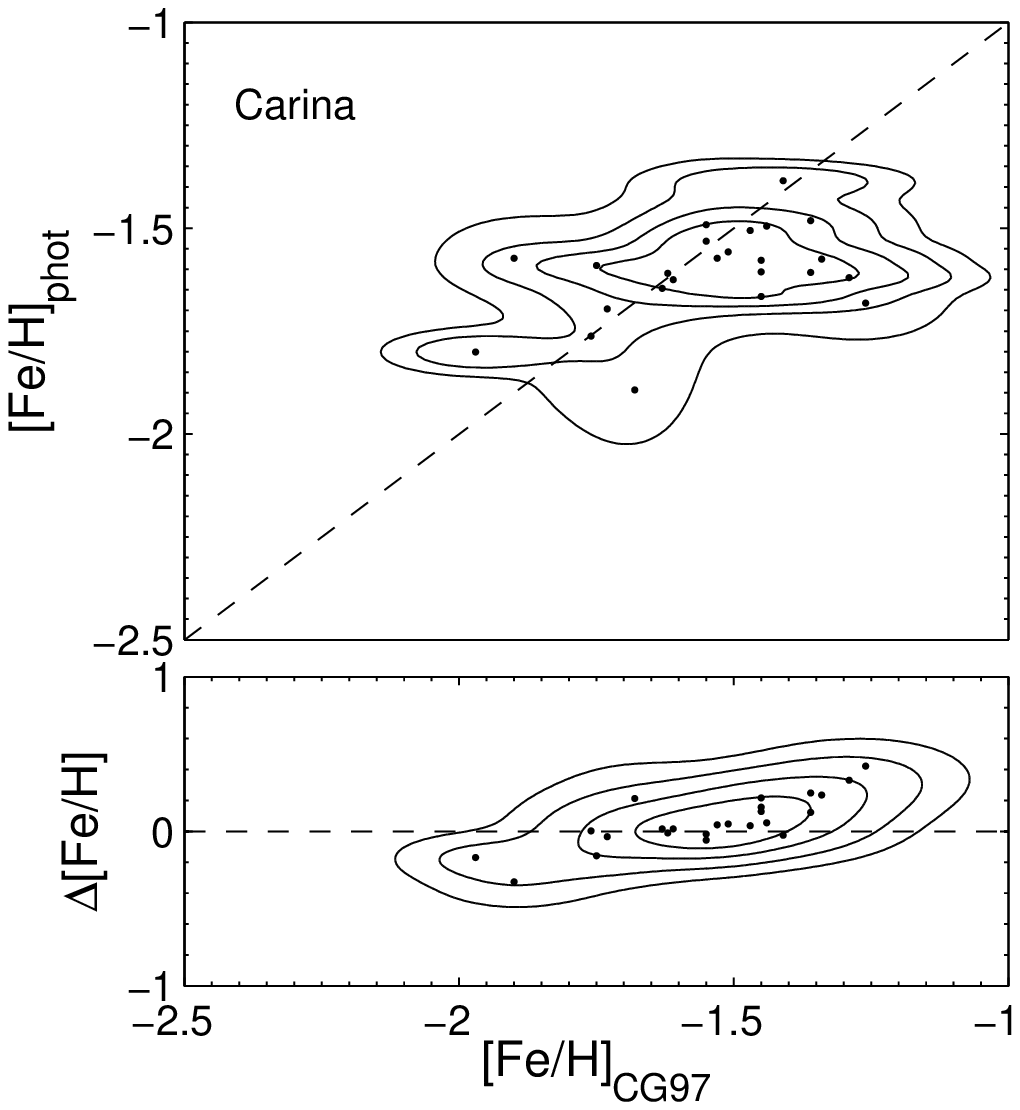}
       \includegraphics[width=5cm,clip]{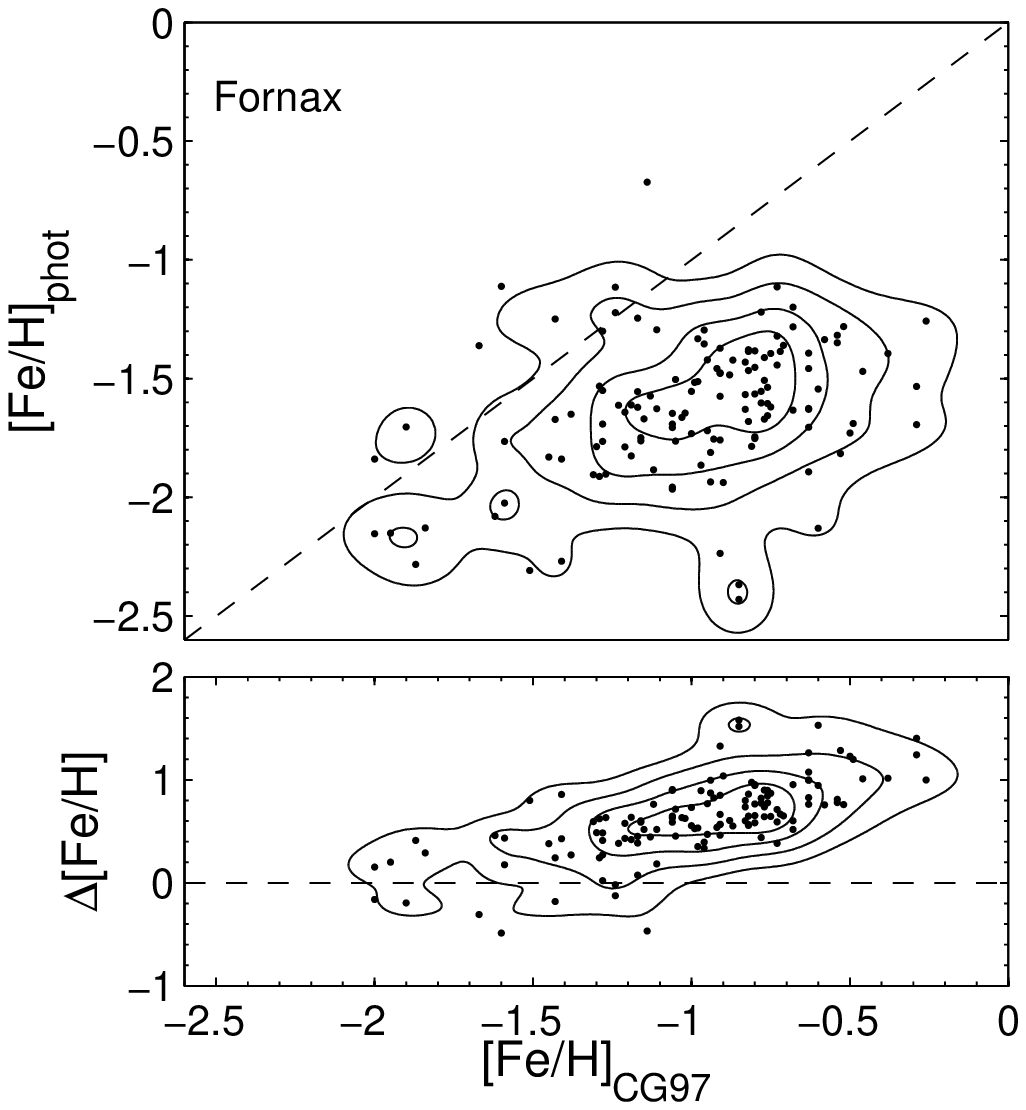}
       \includegraphics[width=5cm,clip]{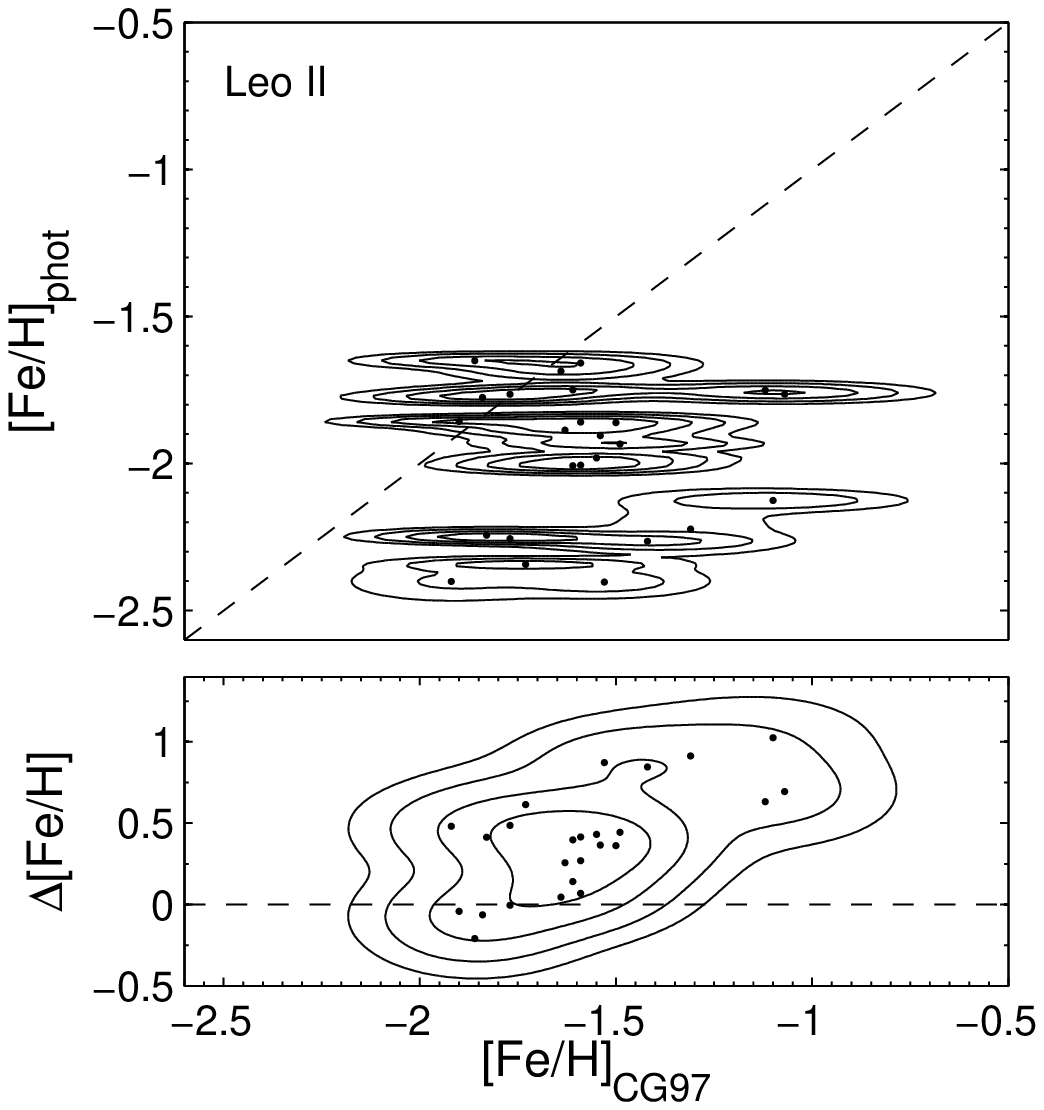}
       \includegraphics[width=5cm,clip]{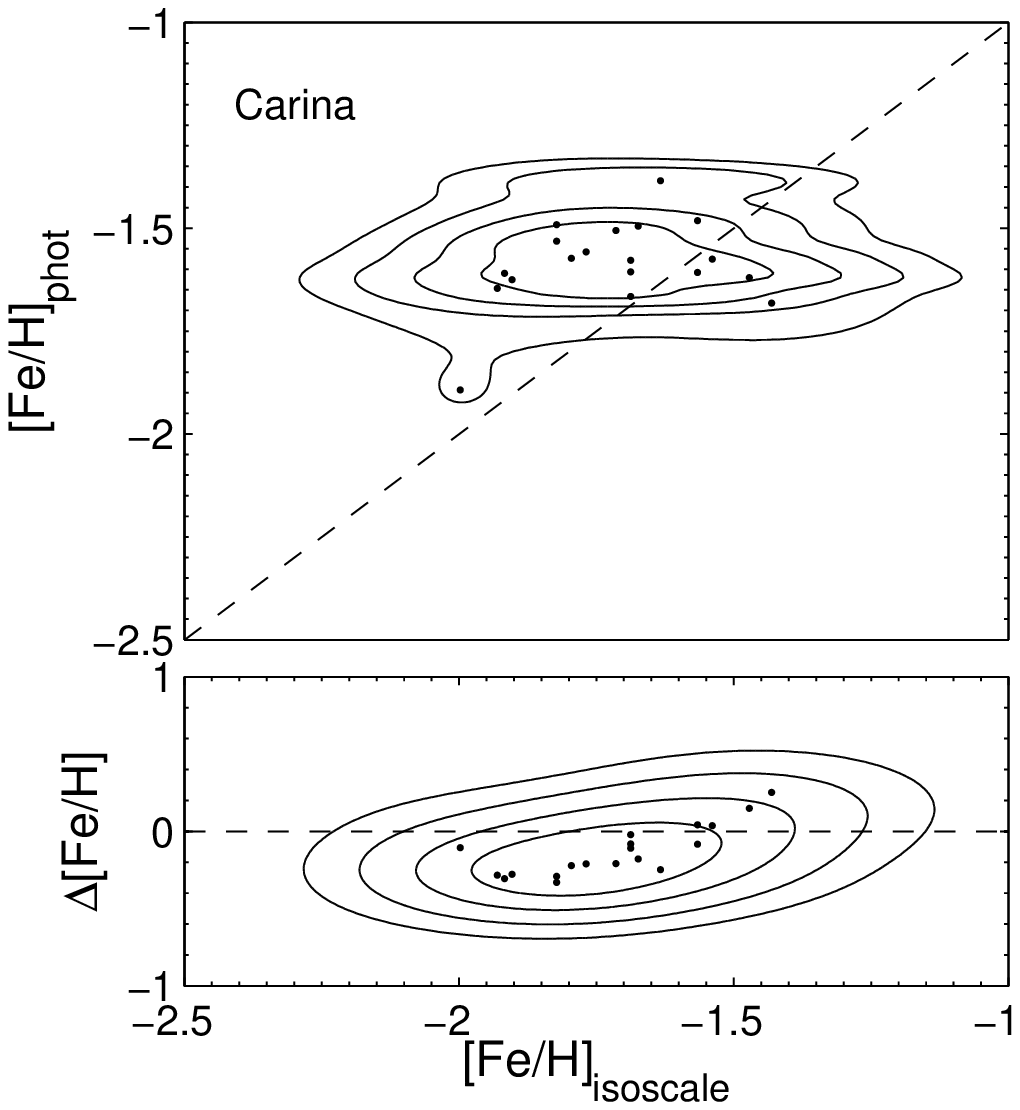}
       \includegraphics[width=5cm,clip]{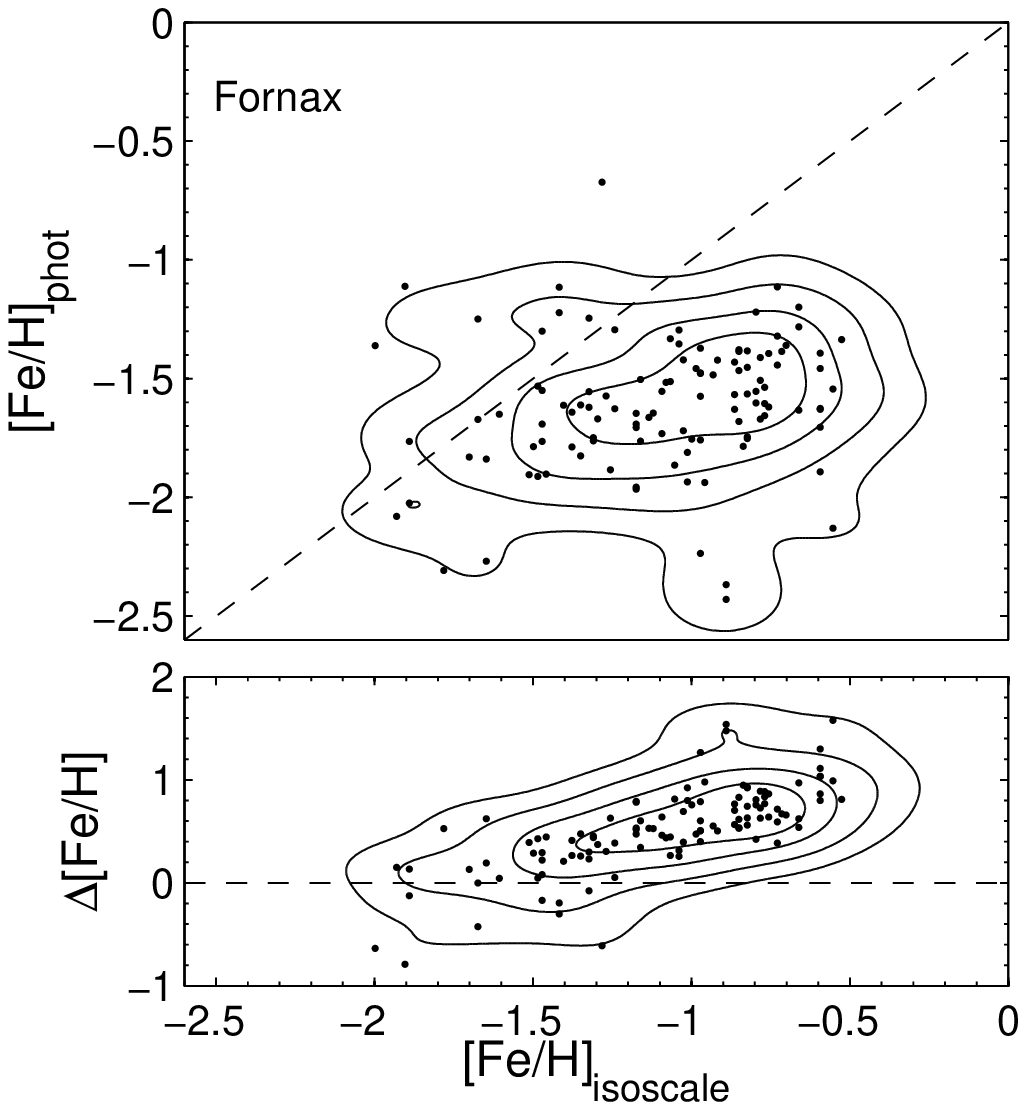}
       \includegraphics[width=5cm,clip]{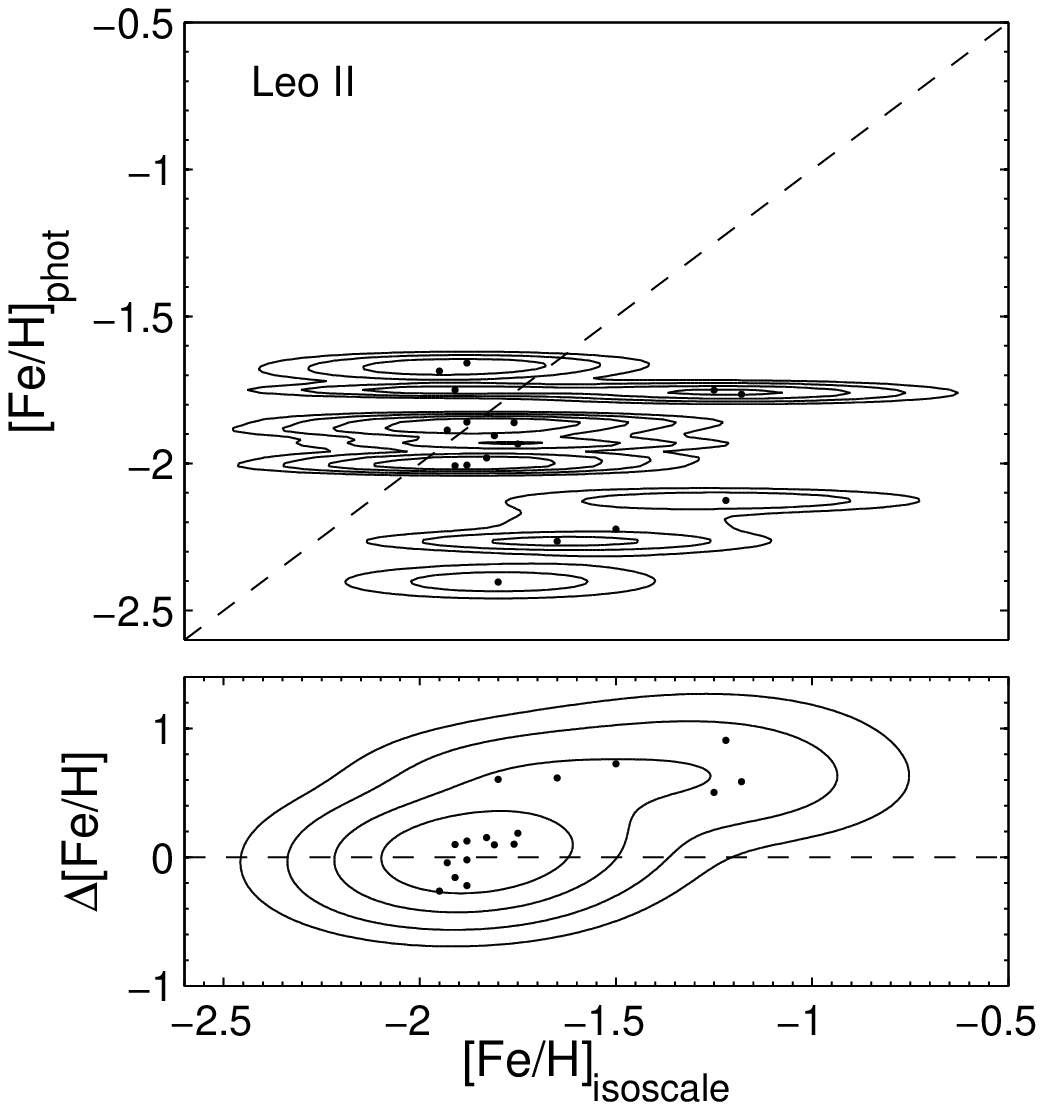}
       \includegraphics[width=5cm,clip]{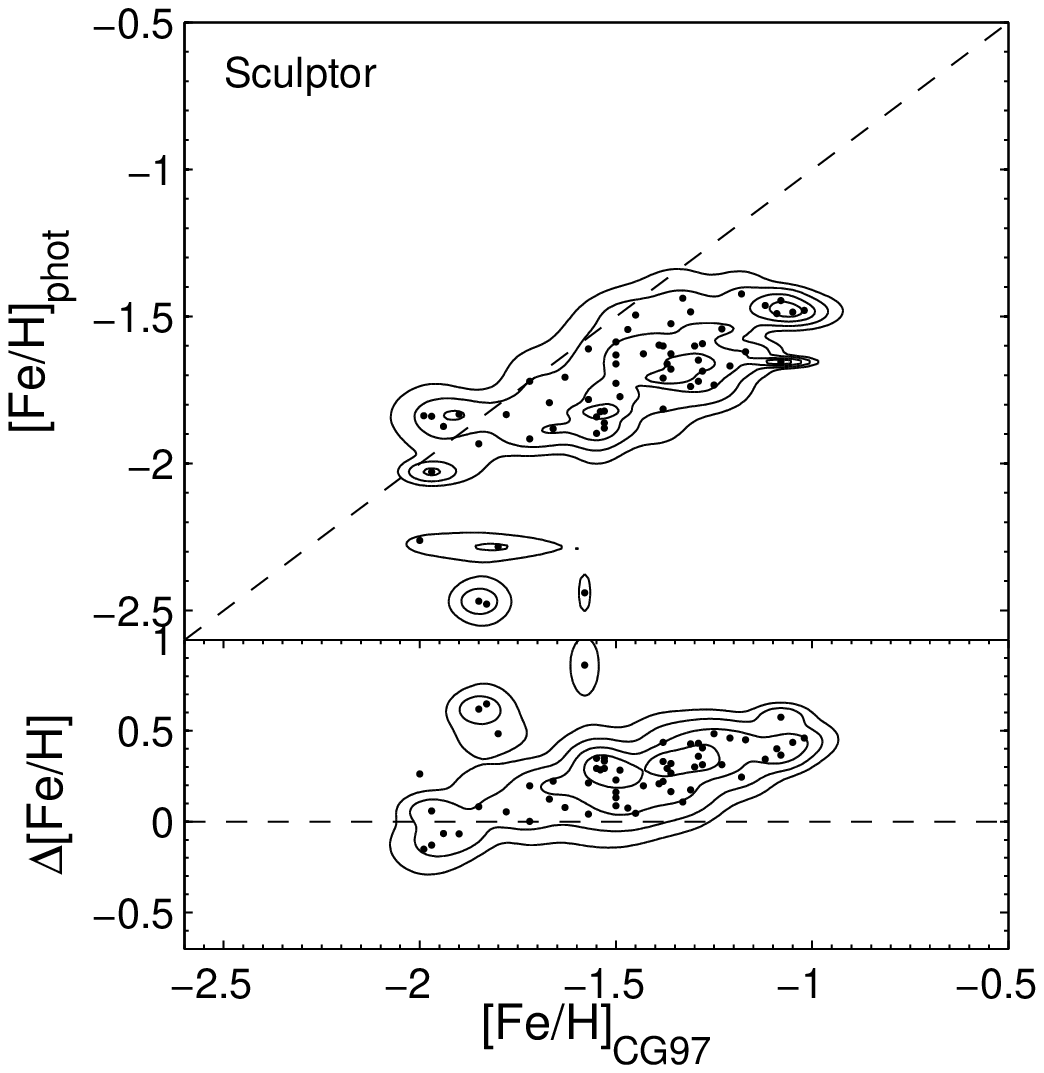}
       \includegraphics[width=5cm,clip]{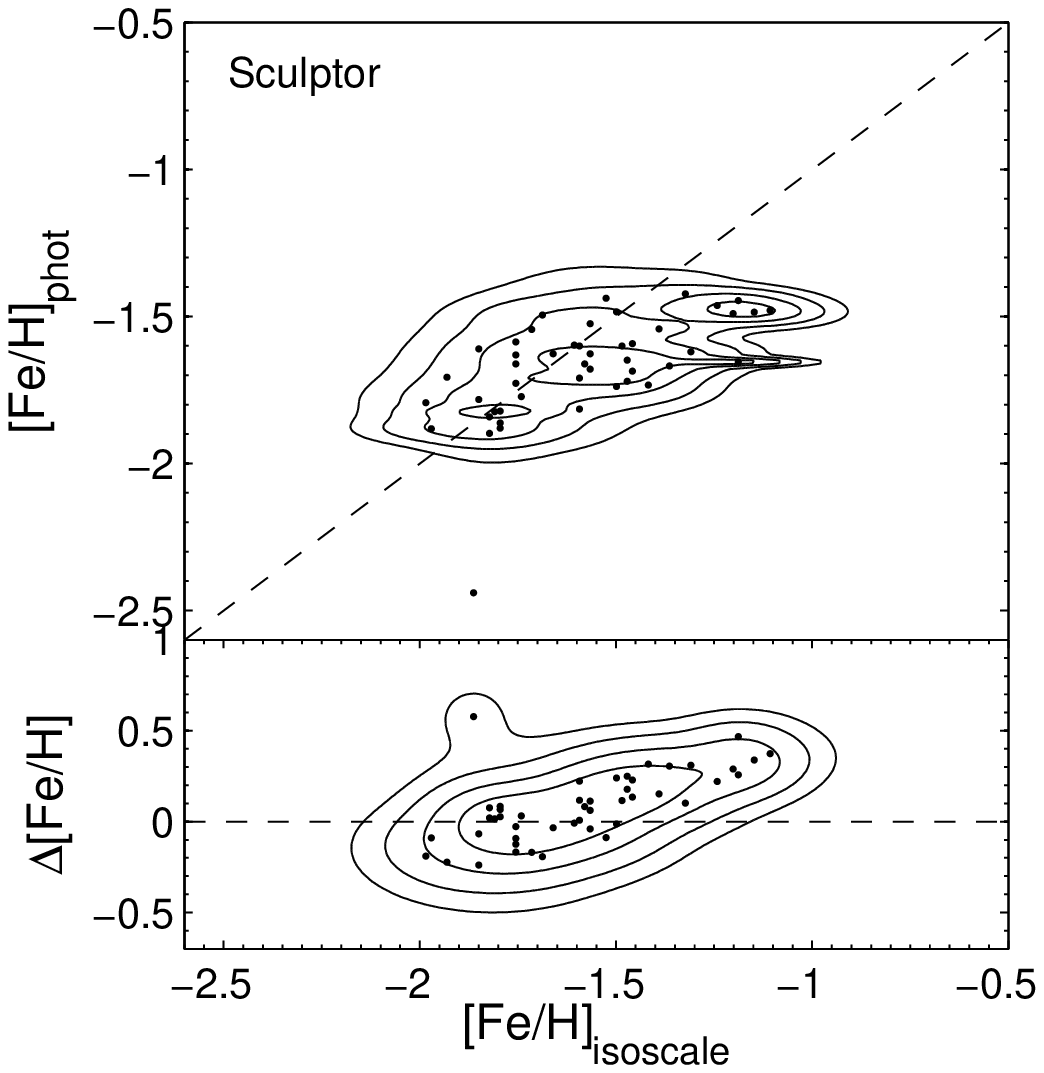}
       \caption{The first two rows show, from left to right, the photometric
         versus the Ca\,T spectroscopic metallicities (upper panels), as well
         as the residuals of the comparison (lower panels), on the CG97
         metallicity scale and on the isoscale for Carina, Fornax, and
         Leo\,II. The last row shows the same for Sculptor. The
         error--weighted contours range from 0.5 to 2.5\,$\sigma$ in steps of
         0.5. The dashed line indicates unity.}
       \label{sl_diffeh}%
\end{figure*}
%%%%%%%%%%%%%%%%%%%%%%%%%%%%%%%%%%%%%%%%%%%%%%%%%%%%%%%%%%%%%%%%%%%%%%%%%%
%
    in the upper panels of Fig.~\ref{sl_diffeh}, separately on the CG97
    metallicity scale and on the isoscale for Carina, Fornax, Leo\,II, and
%
%%%%%% FIGURE 9 - ALL %%%%%%%%%%%%% Two column figure (place early!) %%%%%%  
 \begin{figure*}                                               
    \centering
       \includegraphics[width=4.5cm,clip]{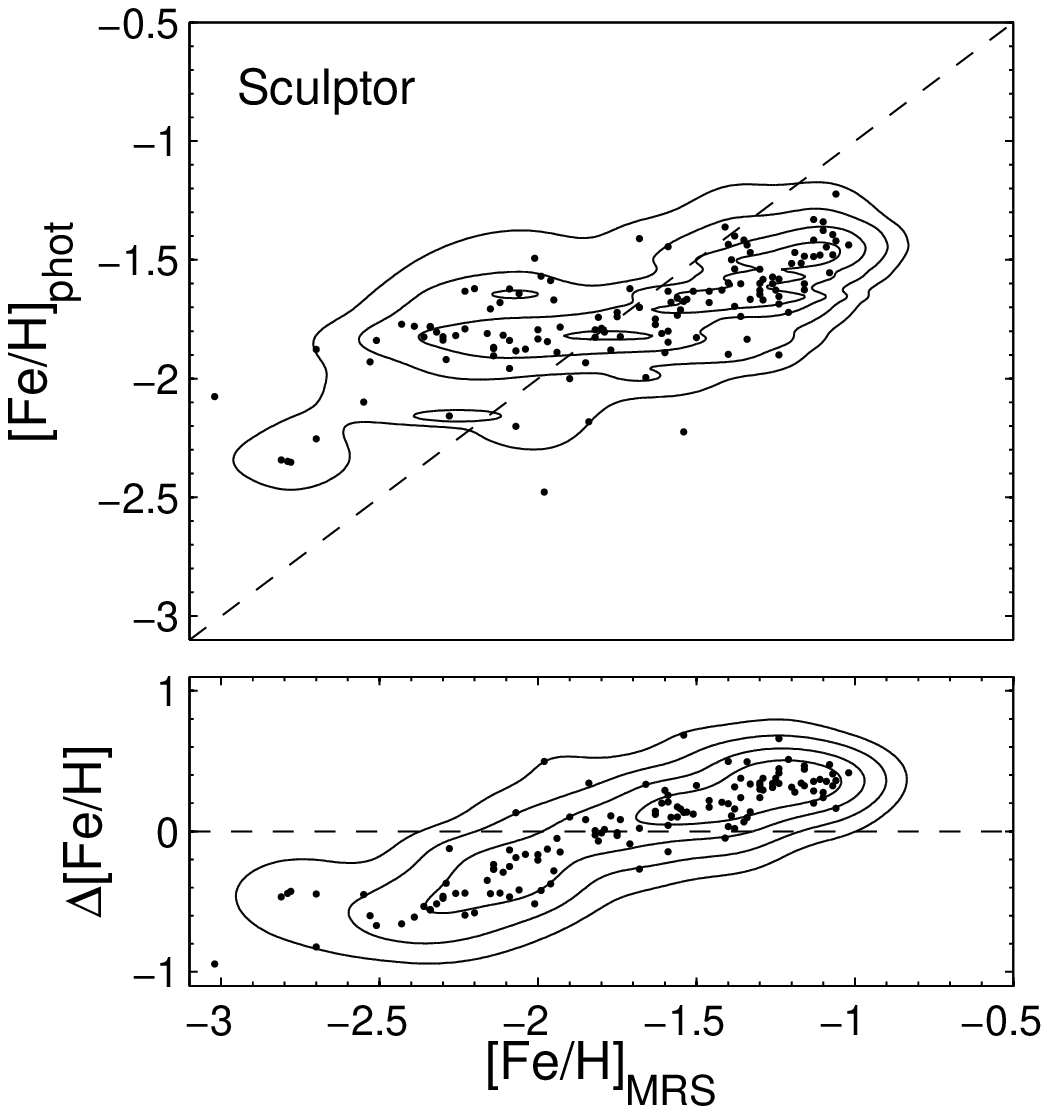}
       \includegraphics[width=4.5cm,clip]{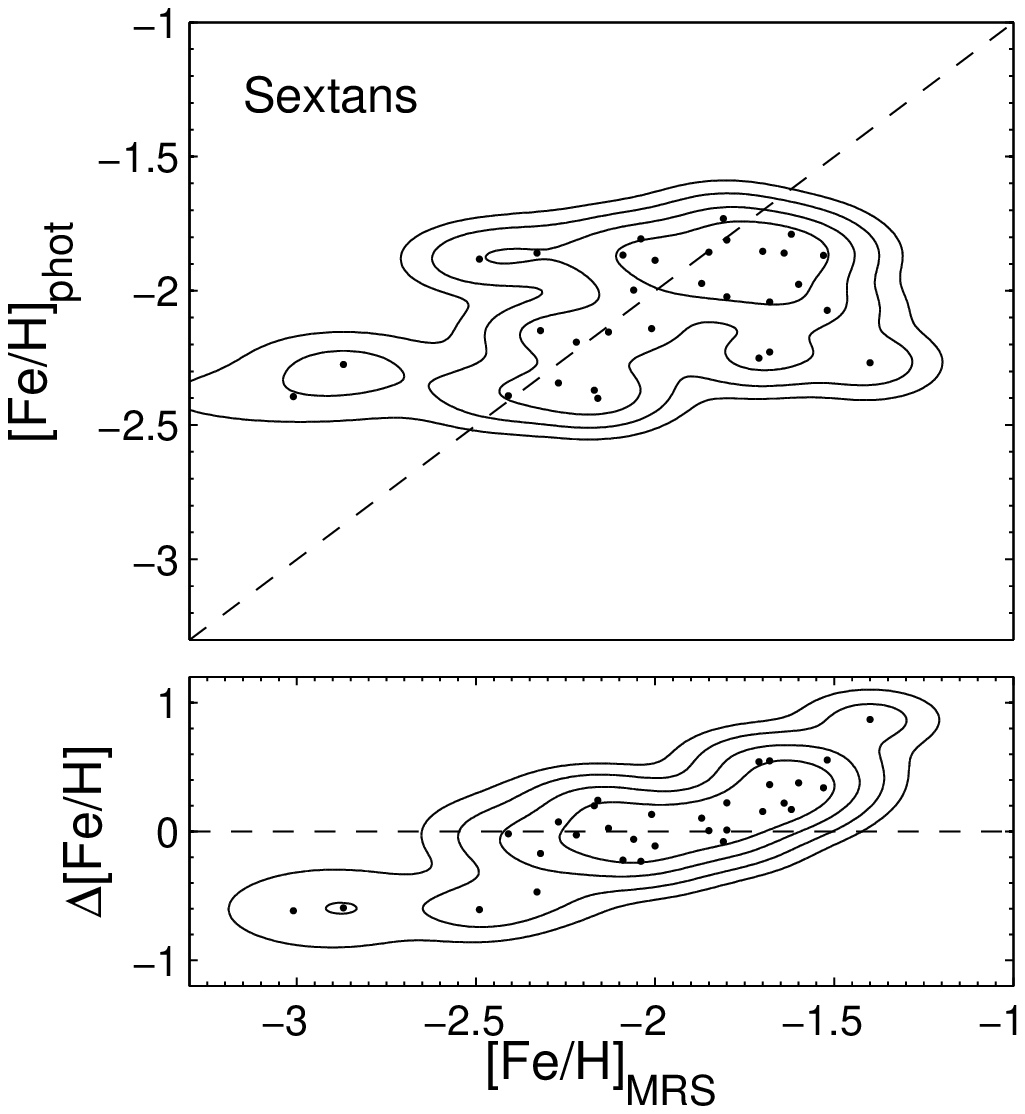}
       \includegraphics[width=4.5cm,clip]{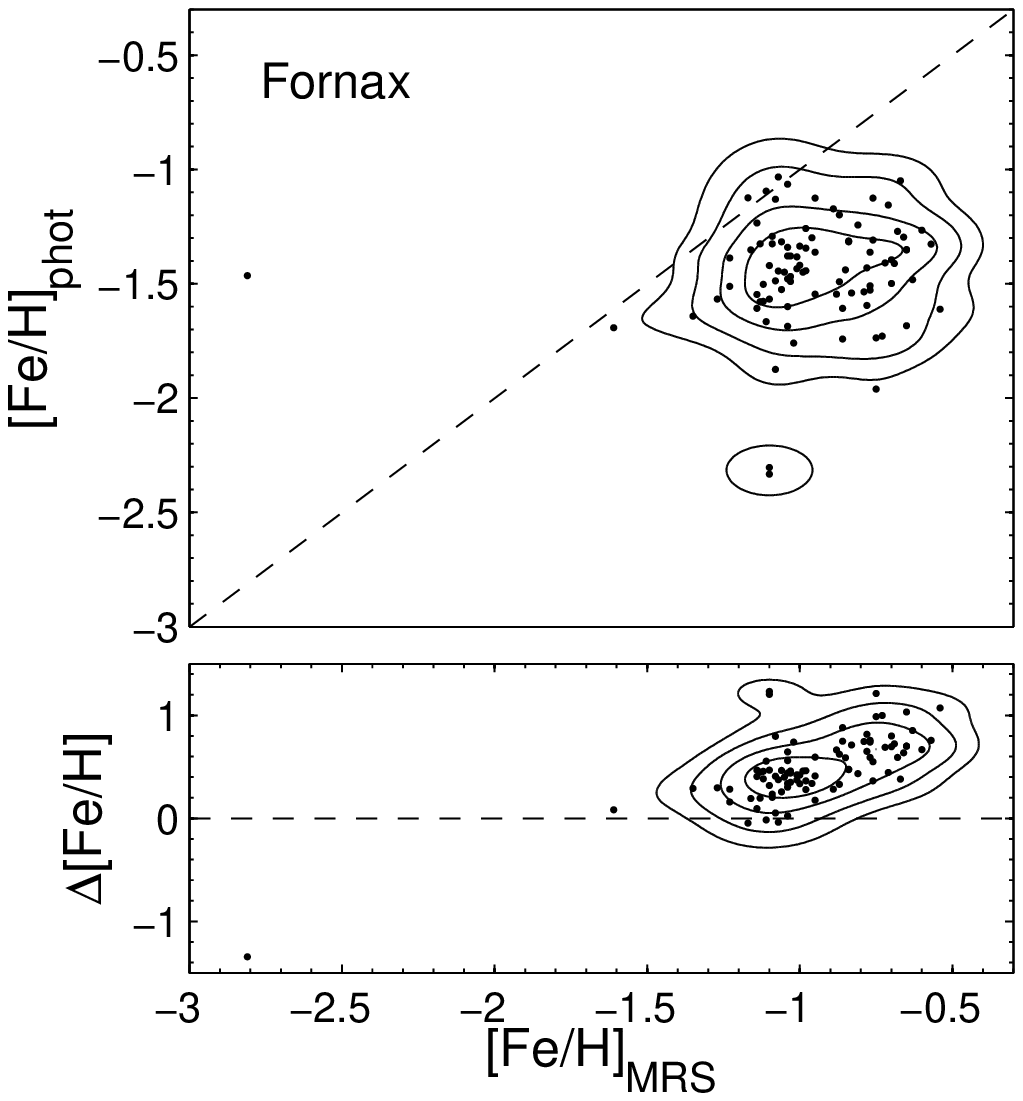}
       \includegraphics[width=4.5cm,clip]{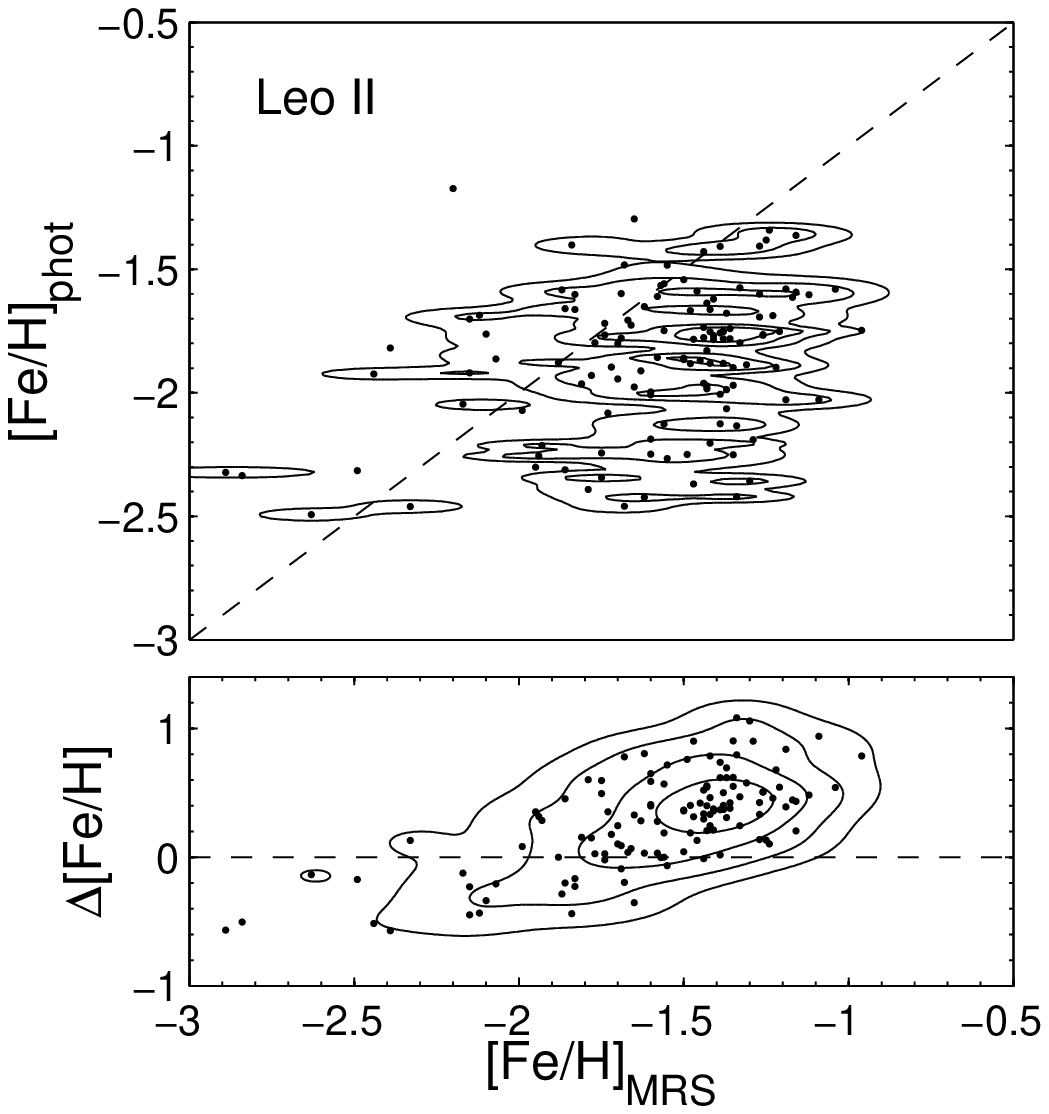}
    \caption{ Photometric metallicities versus MRS metallicities (upper
      panels), as well as their residuals (lower panels), in the sense of MRS
      metallicities minus photometric metallicities versus MRS metallicities,
      for Sculptor, Sextans, Fornax and Leo\,II. The error--weighted contours
      range from 0.5 to 2.5\,$\sigma$ in steps of 0.5. The dashed line
      indicates unity.}
\label{sl_kirby1}% 
\end{figure*}
%%%%%%%%%%%%%%%%%%%%%%%%%%%%%%%%%%%%%%%%%%%%%%%%%%%%%%%%%%%%%%%%%%%%%%%%%%%%%
%
    Sculptor. For those stars with both MRS and photometric measurements, the
    photometric versus the MRS metallicities is shown in Fig.~\ref{sl_kirby1}
    for Sculptor, Sextans, Fornax, and Leo\,II. In all cases, the lower panels
    each show the residuals of the comparison. $\Delta$\,[Fe/H] is always the
    difference between the spectroscopic metallicities minus the photometric
    metallicities.
    
    A positive difference $\Delta$[Fe/H] means that the spectroscopic
    metallicities are more metal--rich than the photometric metallicities. The
    photometric systematic uncertainty that can contribute to a positive
    difference $\Delta$[Fe/H] is the photometric metal--poor bias due to the
    presence of an intermediate--age population. This metal--poor bias has
    been estimated to amount to up to 0.4\,dex for a star--by--star comparison
    when deriving photometric metallicities assuming a single old age of
    12.5\,Gyr for the underlying population as compared to assuming a single
    age of 8.5\,Gyr (Lianou et al.~\cite{sl_lianou10}), with the maximum
    difference occurring at the metal--poor end.

    Possible sources of uncertainties that can contribute to the photometric
    metallicities are the distance modulus and reddening. In the case of
    Sculptor, a variation of the distance modulus by $\pm$0.14\,mag, which
    corresponds to a relative error of less than 1\%, leads to a relative
    difference of photometric metallicities of 10\%, where an increase of the
    distance modulus by 0.14\,mag leads to more metal--poor metallicities. A
    variation of the reddening by $\pm$0.02\,mag, which corresponds to a
    relative error of less than 16\%, leads to a relative difference of the
    photometric metallicities of 12\%, where an increase in the reddening by
    0.02\,mag leads again to more metal--poor metallicities.

    Another source of uncertainty originates from the assumption of a
    scaled--solar composition of the [$\alpha$/Fe] chosen for the Dartmouth
    isochrones. Again, using Sculptor as a test case, we choose an
    [$\alpha$/Fe] equal to $+$0.2\,dex, constant through the whole range of
    the photometric metallicities.  This choice leads to more metal--poor
    individual photometric metallicities, approximately by 6\% for the
    relative differences, while the  median photometric metallicity gets more
    metal--poor by 0.1\,dex (see also Kalirai et
    al.~\cite{sl_kalirai06}). Individual $\alpha$--element ratios for Sculptor
    indicate that [$\alpha$/Fe] has an average value of approximately zero
    (Venn et al.~2004; their Figure~2 and Figure~7), across the range of
    metallicities we consider here for the isoscale of
    $-2.02\leq$[Fe/H]$_{CG97}\leq-0.5$\,(dex). For the dSphs that have more
    complex SFHs, the choice of an [$\alpha$/Fe] of zero is further justified
    by the range of ages of the stars present, where their [$\alpha$/Fe] tends
    to approach solar values, as discussed in Koch et
    al.~(\cite{sl_koch08b}). Furthermore, for Carina and for the metallicity
    ranges we consider here, the [$\alpha$/Fe] ratio has an average value of
    approximately zero (Koch et al.~\cite{sl_koch08a}; their Figure~2, left
    bottom panel), and the same holds for Fornax (Letarte et
    al.~\cite{sl_letarte10}; Venn et al.~\cite{sl_venn04}; their Figure~2). 

    The motivation of keeping the $\alpha$--element enhancement constant
    through the whole metallicity range, in contrast to what high--resolution
    $\alpha$--element abundances indicate (e.g., Koch et
    al.~\cite{sl_koch08a}, Cohen \& Huang \cite{sl_cohen09,sl_cohen10}; and
    references therein), stems from the fact that the exact position of the
    ``knee'', formed by a plateau of constant [$\alpha$/Fe] as a function of
    [Fe/H] and by the declining values of [$\alpha$/Fe] towards the
    metal--rich end due to the SN\,Ia contribution, depends on the star
    formation and chemical evolution history of each dSph (e.g., Marcolini et
    al.~\cite{sl_marcolini08}). In more distant galaxies it is not possible to
    obtain high--resolution measurements or to measure individual stellar
    [$\alpha$/Fe] ratios. Therefore one cannot infer the location of the knee,
    whose position could be used to fit [$\alpha$/Fe] as a function of the
    [Fe/H] range (Cohen \& Huang \cite{sl_cohen09,sl_cohen10}) and then use
    such a function as an input for the [$\alpha$/Fe] in the isochrones. Thus,
    one needs to choose a constant [$\alpha$/Fe] value for the whole
    metallicity range used in the isochrones. We note that the present day
    high--resolution spectroscopic measurements of metallicities of the dSphs
    studied here permit the determination of the location of the knee in
    conjuction with detailed chemical evolution models in the cases of
    Sculptor and Carina (Geisler et al.~\cite{sl_geisler07}; Koch et
    al.~\cite{sl_koch08a}; Lanfranchi, Matteucci \& Cescutti
    \cite{sl_lanfranchi06}), while in the remaining dSphs only limits on the
    position of the knee can  be placed (e.g., Koch \cite{sl_koch09}; Tolstoy,
    Hill \& Tosi \cite{sl_tolstoy09}; Lanfranchi \& Matteucci
    \cite{sl_lanfranchi10}; and references therein).

  \subsection{Sculptor}

  \subsubsection{Ca\,T versus photometric metallicities}

    In the case of an old--age dominated population such as in Sculptor, one
    would expect the photometric metallicities to match the spectroscopic
    metallicities once everything has been placed on the same metallicity
    scale. This is not what is observed in Fig.~\ref{sl_diffeh} for Sculptor,
    where there is an excess of stars with positive $\Delta$[Fe/H] that
    increases towards the Ca\,T spectroscopic metal--rich end. On the
    isoscale, the median value of the difference $\Delta$[Fe/H] is equal to
    0.08\,dex with a full range of 0.82\,dex, while typical metallicity
    uncertainties have a median of 0.05\,dex (photometric) and 0.11\,dex
    (Ca\,T). The minimum $\Delta$[Fe/H] is equal to $-$0.24\,dex and the
    maximum is equal to 0.58\,dex. The slope of the difference in
    metallicities $\Delta$[Fe/H] versus the spectroscopic metallicity is
    listed in Table~\ref{table3b}. On the CG97 metallicity scale, the median
    value of $\Delta$[Fe/H] is equal to 0.28\,dex with a full range of
    1.01\,dex.
   
    There is a very good agreement between the metallicities of the two
    methods within the metallicity range from $-$2\,dex to $-$1.5\,dex, as
    shown in Fig.~\ref{sl_diffeh}. The median difference of the spectroscopic
    metallicity minus the photometric metallicity in this metallicity range is
    0.01\,dex, while the full range of the difference in metallicities remains
    the same, equal to 0.82\,dex. 

  \subsubsection{High--resolution versus photometric metallicities}

    The spectroscopic metallicities in our samples may be as metal--poor as
    $-$4\,dex. Tafelmeyer et al.~(\cite{sl_tafelmeyer10}) find two extremely
    metal--poor stars in Sculptor that are shown by red and blue asterisks in
    the CMD of Sculptor (Fig.~\ref{sl_cmds}; red: Scl\,07--50; blue:
    Scl\,07--49). These extremely metal--poor stars have a high--resolution
    Fe\,I abundance of $-$3.96\,dex and $-$3.48\,dex, respectively, while they
    are assigned a photometric metallicity of $-$2.55\,dex and  $-$2.71\,dex
    in our study, respectively. These photometric metallicities are rejected
    since they are extrapolated values outside the color and metallicity range
    used for the isochrone interpolation. Although it may be useful to have
    Dartmouth isochrones with more metal--poor values than $-$2.5\,dex, the
    spacing of the isochrones in the metal--poor part gets increasingly
    narrow, so even small photometric uncertainties lead to very large
    metallicity uncertainties.

%
%%%%% FIGURE 10 - ALL %%%%%%%%%%%%% One column figure (place early!) %%%%%%
 \begin{figure}
    \centering
       \includegraphics[width=5cm,clip]{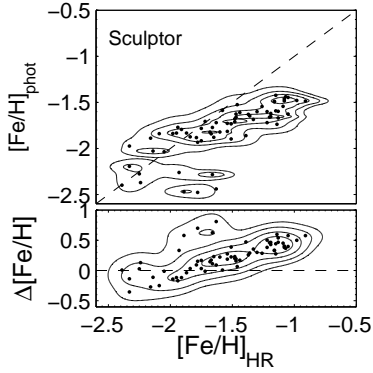}
       \caption{Photometric metallicity, as well as the residuals, as a
         function of the high--resolution spectroscopic metallicity. The
         error--weighted contours range from 0.5 to 2.5\,$\sigma$ in steps of
         0.5. The dashed line indicates unity.} 
       \label{sl_scuhr}%
\end{figure}
%%%%%%%%%%%%%%%%%%%%%%%%%%%%%%%%%%%%%%%%%%%%%%%%%%%%%%%%%%%%%%%%%%%%%%%%%%
%
    In Fig.~\ref{sl_scuhr}, we show the high--resolution spectroscopic
    metallicities from Battaglia et al.~(\cite{sl_battaglia08}) versus the
    photometric metallicities, as well as their residuals. Given that Sculptor
    is an old--age dominated system, we do not expect to have any significant
    age effects affecting the metallicities. We note that in
    Fig.~\ref{sl_scuhr} (upper panel), there exist some stars towards the
    photometric metal--poor end that have more metal--rich high--resolution
    metallicities, which could be indicative of a metal--poor bias or may be
    due to the decreased metallicity sensitivity of the photometric method for
    bluer RGB colors. Again, there is the tendency of $\Delta$[Fe/H] to
    increase towards the high--resolution spectroscopic metal--rich end. The
    median of the differences is 0.23\,dex, while the relative differences
    $\Delta$[Fe/H]$/$[Fe/H]$_{HR}$ in this case are approximately
    16\%. Typical high--resolution metallicity uncertainties have a median of
    0.1\,dex. The full range of the differences is equal to 1.17\,dex, with
    minimum diffefence of $-$0.36\,dex and maximum difference of 0.81\,dex. 

  \subsubsection{MRS versus photometric metallicities}

    The same trend of an excess of stars with positive $\Delta$[Fe/H] is
    observed in Fig.~\ref{sl_kirby1}, where we compare the MRS metallicities
    with the photometric metallicities. In addition, in some cases there is
    also an excess of stars with negative $\Delta$[Fe/H], indicative of stars
    with spectroscopic metallicities more metal--poor than the photometric
    ones. In any case, the discrepancy between the MRS metallicities and the
    photometric metallicities is higher than in the case of the Ca\,T--based
    spectroscopic metallicities. The median of the differences between the
    photometric and MRS metallicities is 0.1\,dex, while the full range of the
    differences is 1.62\,dex, which is double the full range in the case of
    the Ca\,T  on the isoscale versus photometric metallicities. Typical MRS
    metallicity uncertainties have a median of 0.11\,dex. The minimum
    $\Delta$[Fe/H] is equal to $-$0.94\,dex and the maximum is equal to
    0.68\,dex. The slope of the difference in metallicities $\Delta$[Fe/H]
    versus the MRS metallicity is listed in Table~\ref{table3b}. 

  \subsubsection{Ca\,T versus MRS metallicities}

    The Ca\,T and MRS metallicities of a given star differ from each other
%
%%%%%% FIGURE 11 - ALL %%%%%%%%%%%%% Two column figure (place early!) %%%%%%  
 \begin{figure}                                               
    \centering
       \includegraphics[width=4cm,clip]{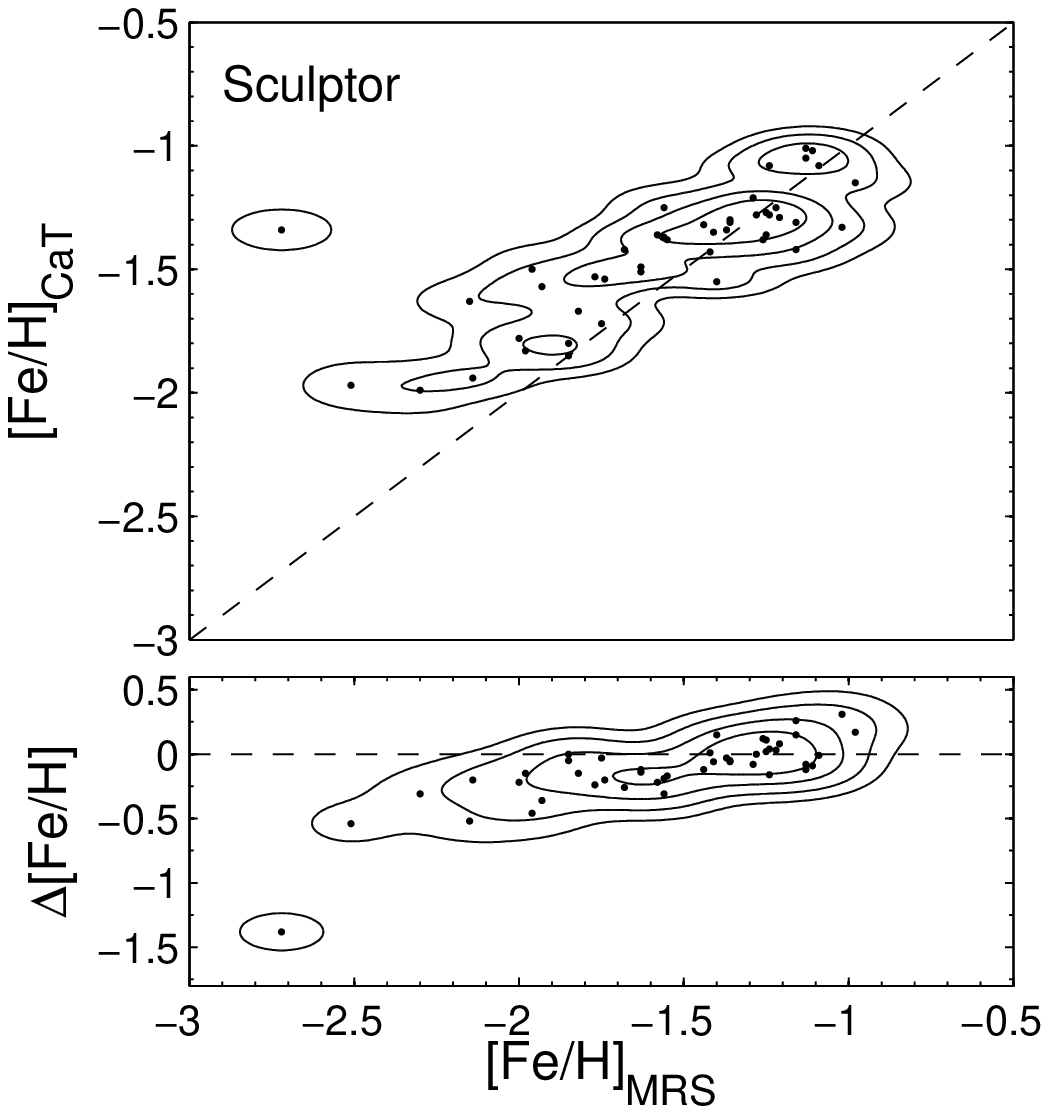}
       \includegraphics[width=4cm,clip]{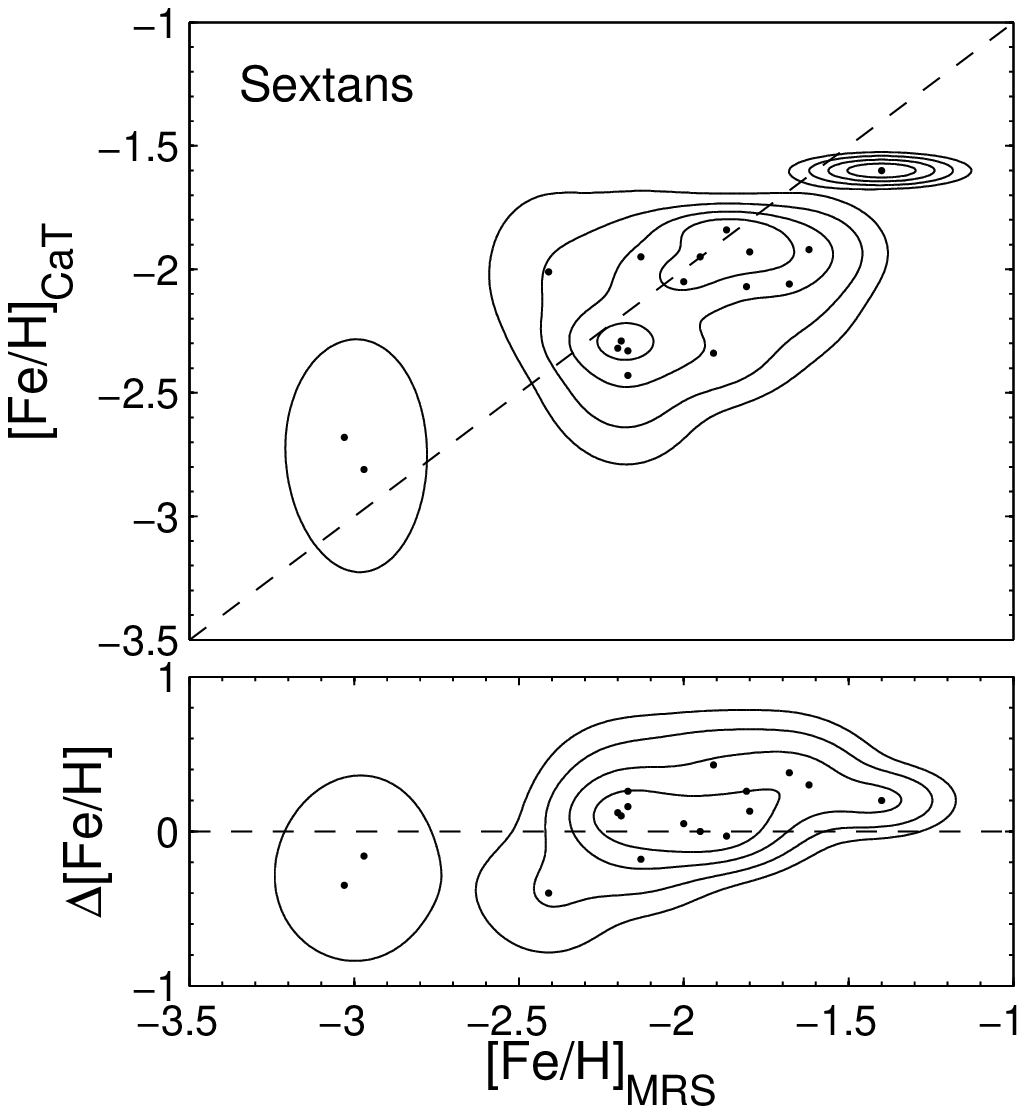}
       \includegraphics[width=4cm,clip]{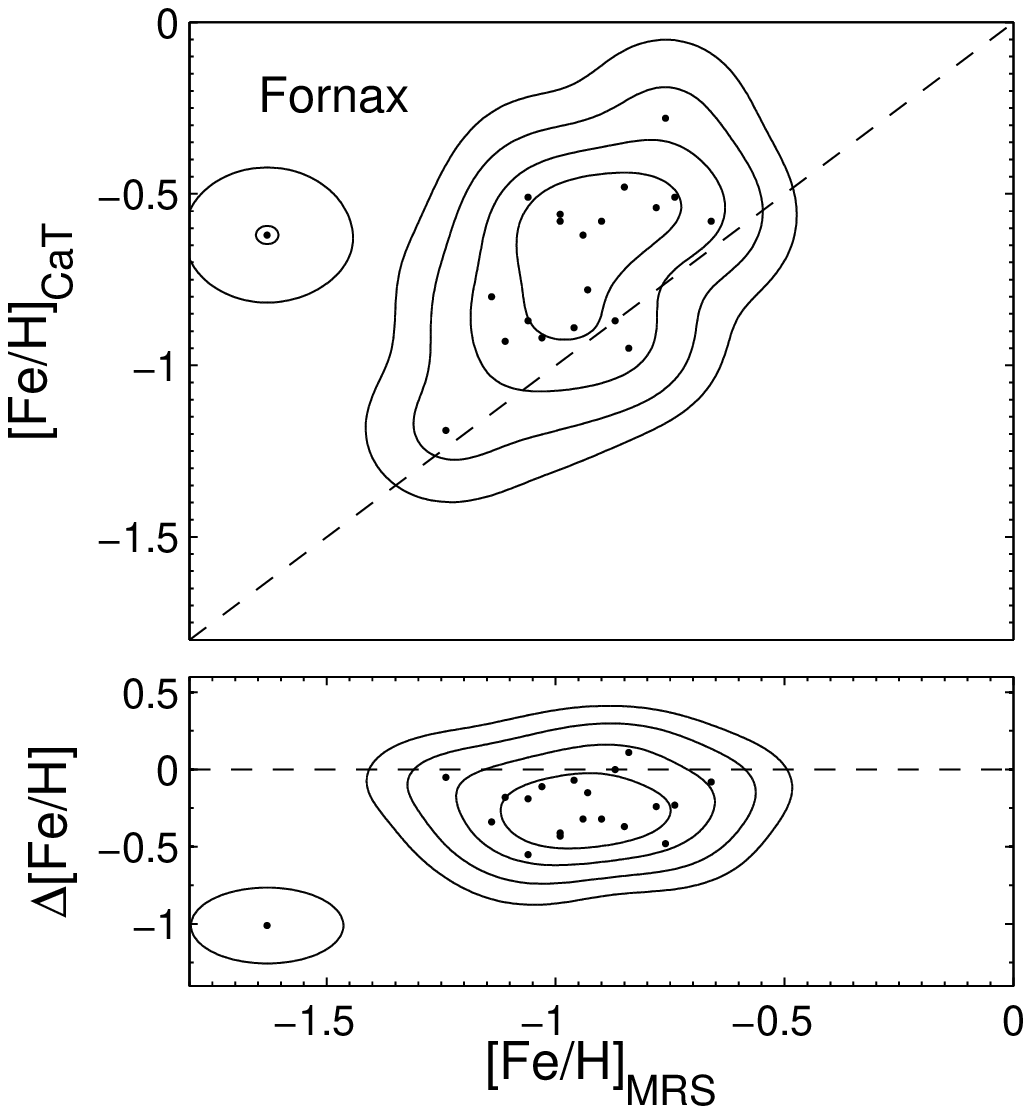}
       \includegraphics[width=4cm,clip]{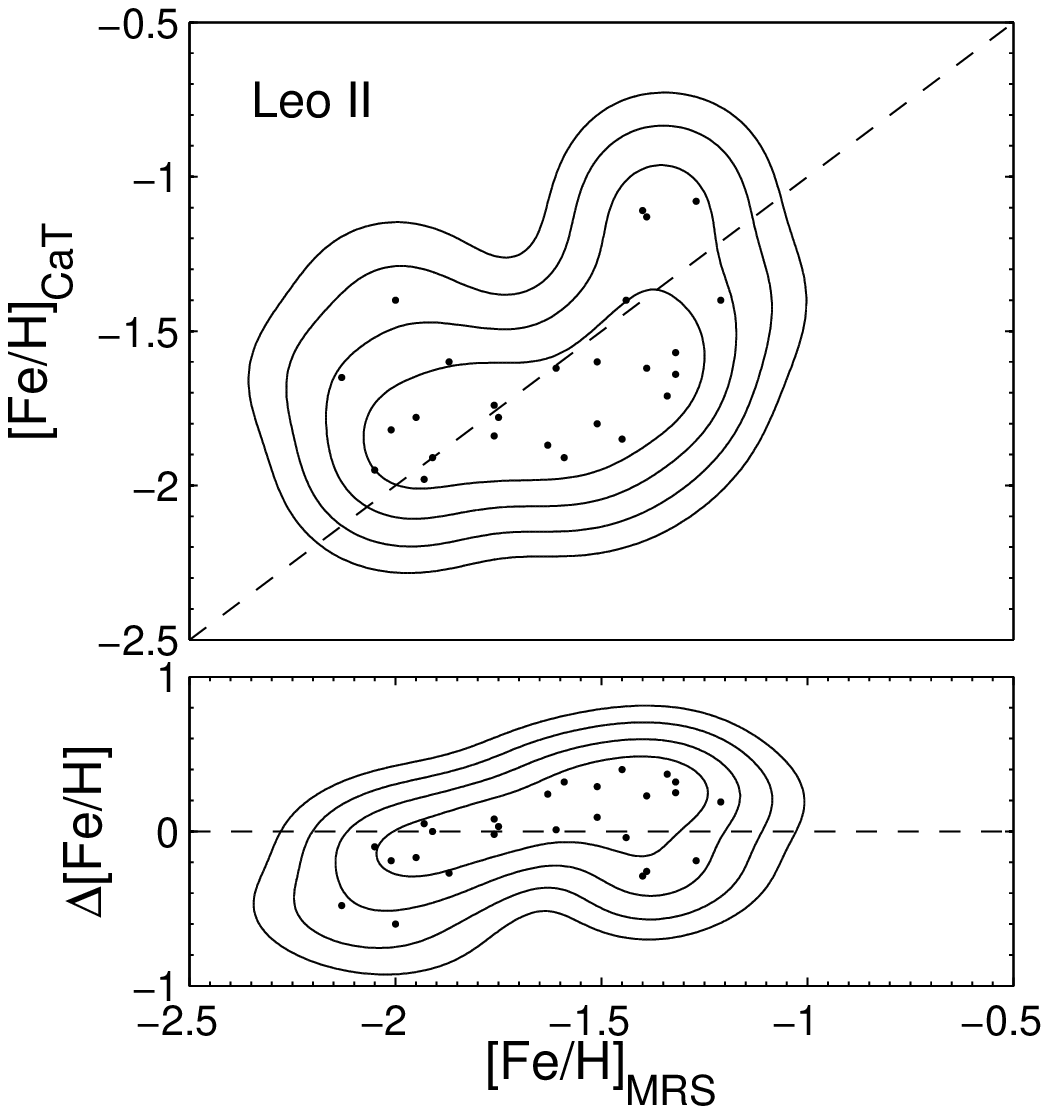}
    \caption{ From left to right, the upper panels show the Ca\,T
      spectroscopic metallicities versus the MRS metallicities for Sculptor,
      Sextans, Fornax, and Leo\,II, while the lower panels show their
      residuals. The error--weighted contours range from 0.5 to 2.5\,$\sigma$
      in steps of 0.5. The dashed line indicates unity.}
    \label{sl_kirby2}% 
\end{figure}
%%%%%%%%%%%%%%%%%%%%%%%%%%%%%%%%%%%%%%%%%%%%%%%%%%%%%%%%%%%%%%%%%%%%%%%%%%%%%
%
    (Fig.~\ref{sl_kirby2}), and from high--resolution measurements (Battaglia
    et al.~\cite{sl_battaglia08} for Ca\,T; Kirby et al.~\cite{sl_kirby09} for
    MRS). Shetrone et al.~(\cite{sl_shetrone09}) discuss that the discrepancy
    between the Ca\,T--based and MRS metallicities for Leo\,II may be due to
    the different metallicity scales, among other factors. It would be thus
    interesting to compare the Ca\,T and MRS metallicities when both are
    placed to a common metallicity scale, such as the isoscale. Such a
    comparison needs to be postponed until more Galactic GC metallicities on
    the MRS scale are observed that are in common with the Dotter et
    al.~(\cite{sl_dotter10}) GC sample.
    
    In the case of Sculptor, the Ca\,T versus the MRS metallicities are shown
    in the upper left panel of Fig.~\ref{sl_kirby2}. The median of the
    differences between the Ca\,T and MRS metallicities is 0.09\,dex, while
    the full range of the differences is 1.69\,dex. We note that the latter
    full range of the difference in metallicities is twice the full range as
    compared to the Ca\,T versus photometric metallicities, for Sculptor, but
    of the same order as compared to the MRS versus the photometric
    metallicities case. Typical metallicity uncertainties have a median of
    0.06\,dex (Ca\,T) and 0.11\,dex (MRS). The minimum $\Delta$[Fe/H] is equal
    to $-$1.38\,dex and the maximum is equal to 0.31\,dex. If we focus on the
    Ca\,T metallicity range from $-$2\,dex to $-$1.5\,dex, then the median of
    the metallicity differences is equal to 0.2\,dex, with a full range of the
    differences of 0.69\,dex, while the minimum value of the difference is
    $-$0.54\,dex and the maximum value of the difference is 0.15\,dex. In this
    metallicity range, the median of the metallicity differences is double the
    one in the case of the Ca\,T on the isoscale versus the photometric
    metallicities. In the metallicity range where the Ca\,T is the most
    sensitive, there seems to be similarly highly discrepant to the MRS
    metallicities.

  \subsubsection{Age effects in Sculptor}

    Sculptor is an old--age dominated system, with more than 86\% of its stars
    having ages larger than 10\,Gyr (Orban et al.~\cite{sl_orban08}).  The
    presence of old, low--luminosity AGB stars may contribute to the
    $\Delta$[Fe/H] becoming positive due to the photometric metal--poor bias,
    but the overall mean metallicity properties are unaffected (e.g.~Lianou et
    al.~\cite{sl_lianou10}). Hurley--Keller et al.~(\cite{sl_hurley-keller99})
    discuss that Sculptor may contain two old stellar populations, based on
    its horizontal branch morphology, while Harbeck et
    al.~(\cite{sl_harbeck01}) find a population gradient and Tolstoy et
    al.~(\cite{sl_tolstoy04}) identify two kinematically distinct ancient
    components. De Boer et al.~(\cite{sl_deboer11}) suggest that star
    formation in Sculptor ceased 7\,Gyr ago. Furthermore, Menzies et
    al.~(\cite{sl_menzies11}) identify two AGB variables, one of which
    suggests that stars as recent as 1--2\,Gyr ago may have formed, consistent
    with the age distribution modelling that Revaz et al.~(\cite{sl_revaz09})
    derive.

    Motivated by the recent findings of de Boer et al.~(\cite{sl_deboer11}),
    we use an isochrone of 7\,Gyr to derive the photometric metallicities of
    Sculptor. The new median metallicity becomes more metal--rich by
    0.28\,dex, which would place a metal--rich limit on the median metallicity
    of Sculptor assuming that 100\% of the stars have an age of 7\,Gyr. We
    further use the fractions of the total stellar mass of Sculptor given in
    Orban et al.~(\cite{sl_orban08}) to derive the intermediate--age stellar
    fraction, which is equal to \textit{f}$_{10G}-$\textit{f}$_{1G}$. Then, we
    randomly assign a corresponding fraction of the stars in our RGB sample to
    this intermediate--age population with metallicities derived using the
    7\,Gyr isoschrones, while the remaining fraction of stars is assigned
    metallicities based on the 12.5\,Gyr isochrones. The derived median
    metallicity of this mixture becomes more--metal rich by 0.02\,dex. If we
    ask instead what the fraction of the intermediate--age stars (with an age
    of 7\,Gyr) is that would produce a metallicity difference of the order of
    that between the Ca\,T (placed on the isoscale) and photometric median
    metallicities of 0.08\,dex, we find a fraction of 24\%. That is, 24\%
    intermediate--age stars with ages of 7\,Gyr are needed in order to result
    in a difference of the median photometric metallicity of the mixed--age
    population and the median photometric metallicity assuming a purely old
    system of 0.08\,dex.

  \subsection{Sextans}

    Sextans is the only dSph in our sample that consists of purely old stellar
    populations (Orban et al.~\cite{sl_orban08}; Lee et
    al.~\cite{sl_lee09}). As described earlier, we do not perform a comparison
    of individual common stars between the Ca\,T and photometric metallicies
    since we do not have enough stars in common. Here, we only compare the MRS
    versus the photometric metallicities, as well as the Ca\,T--based versus
    the MRS metallicities. 

  \subsubsection{MRS versus photometric metallicities}

    The second panel from the left in Fig.~\ref{sl_kirby1} shows the MRS
    metallicity versus the photometric metallicity, as well as the
    residuals. Since Sextans is a purely old system, one would again expect a
    very good agreement between the MRS and photometric metallicities. As
    shown in Fig.~\ref{sl_kirby1}, this is not the case. The median of the
    differences between the photometric and MRS metallicities is 0.07\,dex,
    while the full range of the differences is 1.48\,dex. The relative
    differences are 11\%. The slope of the difference in  metallicities
    $\Delta$[Fe/H] versus the MRS metallicity is listed in
    Table~\ref{table3b}. Typical spectroscopic uncertainties have a median of
    0.12\,dex.  

  \subsubsection{Ca\,T versus MRS metallicities}

    In the upper right panel of Fig.~\ref{sl_kirby2} we show the Ca\,T versus
    the MRS metallicities for Sextans. In this case, again a high scatter is
    observed between the two spectroscopic methods. The median of the
    differences between the Ca\,T and MRS metallicities is 0.12\,dex, while
    the full range of the differences is 0.83\,dex. Typical metallicity
    uncertainties  have a median of 0.15\,dex (Ca\,T) and 0.12\,dex (MRS). The
    spectroscopic measurements are assumed to be independent of age.

 \subsection{Carina, Fornax and Leo\,II}

    Carina, Leo\,II and Fornax have a significant fraction of
    intermediate--age stars that lead to an age--metallicity degeneracy along
    the RGB, as shown in Fig.~10 of Koch et al.~(\cite{sl_koch06}) for Carina,
    in Fig.~8 of Koch et al.~(\cite{sl_koch07}) for Leo\,II, and in Fig.~22 of
    Battaglia et al.~(\cite{sl_battaglia06}) for Fornax. These complex SFHs
    will affect the photometric metallicities in the sense of the photometric
    metal--poor bias discussed earlier.
 
   \subsubsection{Ca\,T versus photometric metallicities}

    In all cases, just as with Sculptor, Fig.~\ref{sl_diffeh} shows a trend of
    increasingly positive $\Delta$\,[Fe/H] with increasing [Fe/H]. The same
    trend is observed in the study of Gullieuszik et
    al.~(\cite{sl_gullieuszik07}; their Figure 13) for Fornax, where they
    compare their photometric metallicities, derived from near--IR colors,
    with the Ca\,T spectroscopic measurements of Battaglia et
    al.~(\cite{sl_battaglia06}) and Pont et al.~(\cite{sl_pont04}). In our
    study, the positive differences of the metallicities are attributed to the
    presence of intermediate--age stars, which have bluer colors than old
    stars at a given metallicity, as demonstrated in Fig.~\ref{oldbias}. The
    negative differences can be attributed to the poorer resolution of the
    isochrones towards the metal--poor end. The median $\Delta$[Fe/H] is
    0.18\,dex, 0.13\,dex, and 0.52\,dex for Carina, Leo\,II and Fornax,
    respectively, while the full range of $\Delta$[Fe/H] is approximately
    0.58\,dex,  1.17\,dex, and 2.37\,dex respectively. The values quoted refer
    to the Ca\,T--based spectroscopic metallicities placed on the
    isoscale. Typical spectroscopic uncertainties have a median of 0.17\,dex,
    0.24\,dex, and 0.16\,dex, while typical photometric metallicity
    uncertainties have a median value of 0.05\,dex, 0.02\,dex, and 0.11\,dex,
    for the above mentioned dSphs, respectively.  If we focus on the
    metallicity range from $-$2\,dex to $-$1.5\,dex, then the agreement
    between the Ca\,T and photometric metallicities seems better, with a
    median $\Delta$[Fe/H] of 0.21\,dex, 0.1\,dex, and 0.13\,dex for Carina,
    Leo\,II and Fornax, respectively, while the range of $\Delta$[Fe/H] is
    equal to 0.37\,dex, 0.99\,dex, and 1.41\,dex, for the above mentioned
    dwarfs, respectively.

  \subsubsection{High--resolution versus photometric metallicities}

    In the case of Fornax, Tafelmeyer et al.~(\cite{sl_tafelmeyer10}) find one
    extremely metal--poor star in common to our photometric sample, shown with
    the red asterisk in the CMD of Fornax (Fig.~\ref{sl_cmds};
    Frx\,05--42). This extremely metal--poor star has a high--resolution Fe\,I
    abundance of $-$3.66\,dex, while it is assigned a photometric metallicity
    of $-$2.98\,dex. This photometric metallicity is again rejected since it
    is an extrapolated value outside the color and metallicity range used for
    the isochrone interpolation. Nevertheless, photometrically this star is
    correctly identified as a very metal--poor candidate.

    Within our photometric metallicities, we identify several stars that
    photometrically have the right color--magnitude position to potentially be
    very metal--poor stars but which are not retained within our analysis
    since their photometric metallicities are extrapolated values. The
    majority of these stars, when we compare them with all the available
    spectroscopic metallicities, show indications of an age--metallicity
    degeneracy, in the sense that their photometric metallicities appear to be
    too metal--poor as compared to the spectroscopic metallicities. Certainly,
    these photometric metallicities are extrapolated values and in some cases
    with large photometric errors, in order to be able to make any solid
    argument.

  \subsubsection{MRS versus photometric metallicities}

    The same trends are observed when we compare the MRS metallicities with
    the photometric metallicities as shown in Fig.~\ref{sl_kirby1}. The
    photometric and MRS spectroscopic metallicities become similarly
    discrepant as in the case of the Ca\,T metallicities. The median of the
    differences between the photometric and MRS metallicities is 0.45\,dex for
    Fornax and 0.33\,dex for Leo\,II, while the full range of the differences
    is 2.56\,dex for Fornax and 2.11\,dex for Leo\,II. The relative
    differences are 47\% for Fornax and 24\% for Leo\,II. The slopes of the
    difference in metallicities $\Delta$[Fe/H] versus the MRS metallicity for
    Fornax and Leo\,II are listed in Table~\ref{table3b}. The typical MRS
    metallicity uncertainties have a median of 0.1\,dex, and 0.11\,dex for
    Fornax and Leo\,II, respectively. 

  \subsubsection{Ca\,T versus MRS metallicities}
    
    The lower left and right panels of Fig.~\ref{sl_kirby2} show the Ca\,T
    versus the MRS metallicities for Fornax and Leo\,II. Both these figures
    show a similarly large scatter as in the case of Sculptor and
    Sextans. Shetrone et al.~(\cite{sl_shetrone09}) compare the Ca\,T--based
    and MRS metallicities for faint Leo\,II stars and also find them to be
    discrepant in a similar way as we find them here. In our study, the median
    of the differences between the Ca\,T and MRS metallicities is 0.24\,dex
    and 0.02\,dex, respectively for Fornax and Leo\,II, while the full range
    of the metallicity differences is 1.12\,dex and 1\,dex,
    respectively. Typical metallicity uncertainties have a median of 0.14\,dex
    (Ca\,T) and 0.11\,dex (MRS) for Fornax, while for Leo\,II these are
    0.16\,dex (Ca\,T) and 0.11\,dex (MRS).  
  
    In the case of Leo\,II and when we compare $\Delta$[Fe/H] in the Ca\,T
    metallicity range from $-$2\,dex to $-$1.5\,dex, the median
    $\Delta$[Fe/H] is 0.07\,dex while its range is 0.88\,dex. Based on these
    values, the agreement between the Ca\,T and MRS metallicities for Leo\,II
    is slightly better than in the Ca\,T on the isoscale versus photometric
    metallicities case. For Fornax, there are not any stars in common for
    metallicities less than $-$1.5\,dex.

\subsection{$\Delta$[Fe/H] dependence on the dSph's SFH}

%
%%%%% FIGURE 12 - ALL %%%%%%%%%%%%% One column figure (place early!) %%%%%%
 \begin{figure}
    \centering
       \includegraphics[width=2.7cm,clip]{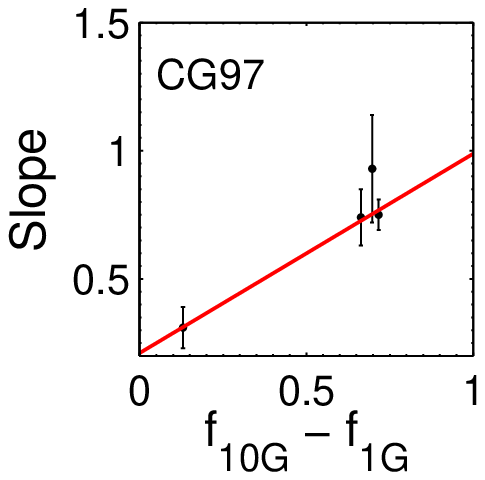}
       \includegraphics[width=2.7cm,clip]{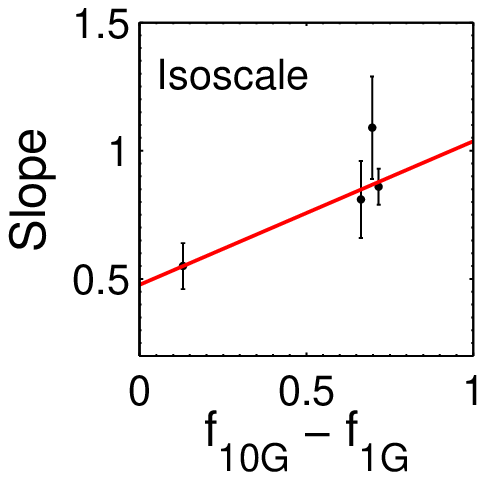}
       \includegraphics[width=2.7cm,clip]{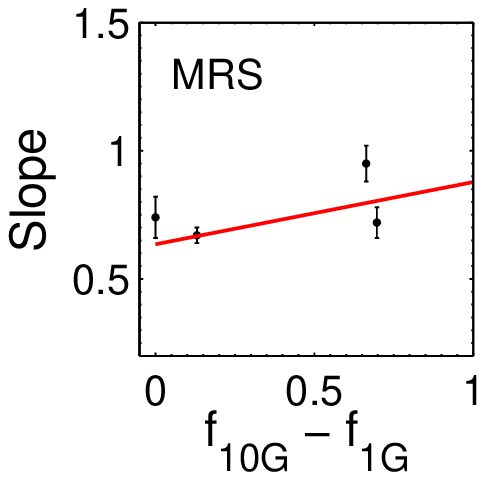}
       \includegraphics[width=2.7cm,clip]{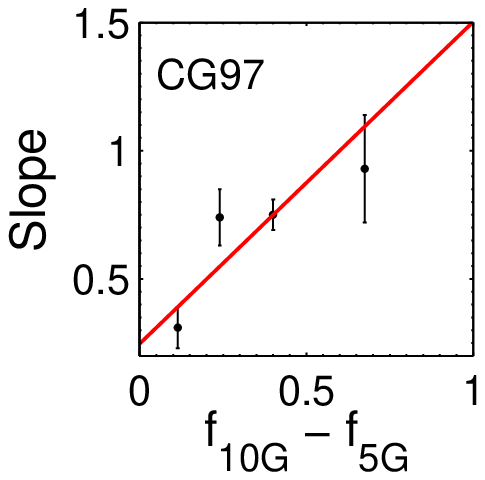}
       \includegraphics[width=2.7cm,clip]{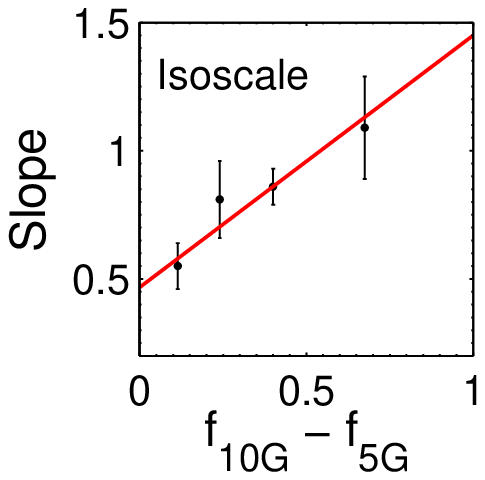}
       \includegraphics[width=2.7cm,clip]{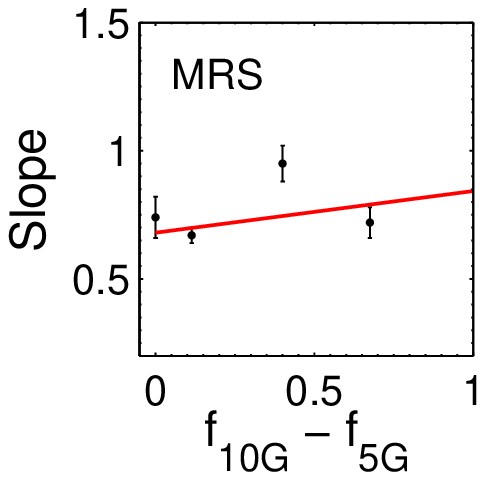}
       \includegraphics[width=2.7cm,clip]{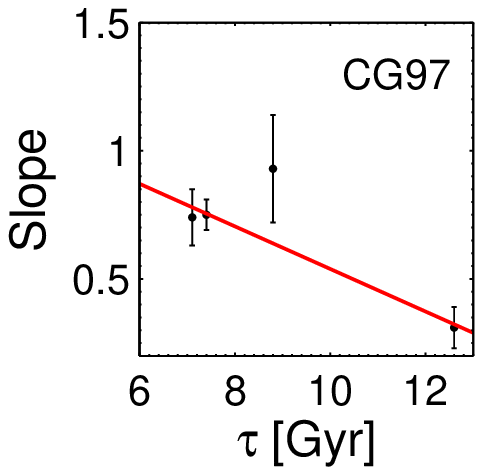}
       \includegraphics[width=2.7cm,clip]{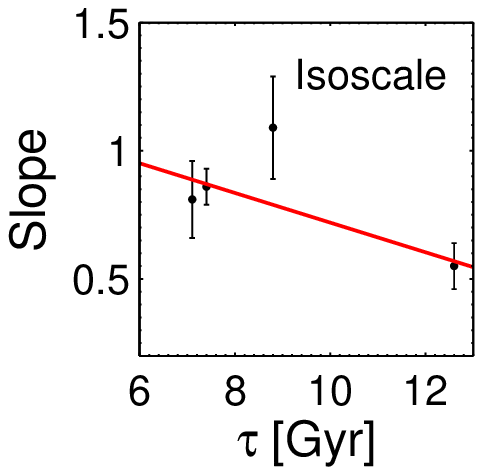}
       \includegraphics[width=2.7cm,clip]{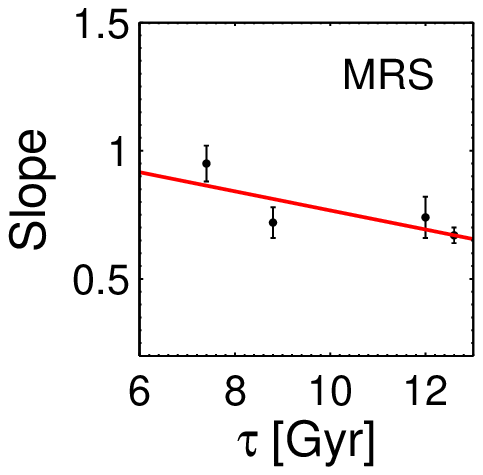}
       \caption{Slopes of the error--weighted linear least squares fit to the
         datapoints of the lower panels of Fig.~\ref{sl_diffeh} and
         Fig.~\ref{sl_kirby1} as a function of the intermediate--age fractions
         (\textit{f}$_{10G}-$\textit{f}$_{1G}$, upper panels;
         \textit{f}$_{10G}-$\textit{f}$_{5G}$; middle panels), as well as a
         function of the mass--weighted mean age $\tau$ (lower panels),
         Sculptor, Sextans, for Carina, Fornax and Leo\,II. The red solid line
         corresponds to an error-weighted linear least squares fit to the
         data. The error bars correspond to the error of the coefficients of
         the fit to the datapoints of Fig.~\ref{sl_diffeh} and Fig.~\ref{sl_kirby1}.}
       \label{sl_slopes}%
\end{figure}
%%%%%%%%%%%%%%%%%%%%%%%%%%%%%%%%%%%%%%%%%%%%%%%%%%%%%%%%%%%%%%%%%%%%%%%%%%
%
    An error--weighted linear least squares fit to the datapoints of the lower
    panels of Fig.~\ref{sl_diffeh} and Fig.~\ref{sl_kirby1}, which show
    $\Delta$[Fe/H] as a function of the spectroscopic metallicities, results
    in the slopes listed in Table~\ref{table3b} for the CG97 metallicity
    scale, the isoscale, and the MRS metallicity scale. In the case of a
    purely old population, one would expect that the slope is zero, as a
    result of ideally zero differences between the photometric and
    spectroscopic metallicities. The non--zero slopes of the $\Delta$[Fe/H] as
    a function of the spectroscopic [Fe/H] are shown in Fig.~\ref{sl_slopes}
    which are now plotted against the \textit{f}$_{10G}-$\textit{f}$_{1G}$
    (upper panels), the \textit{f}$_{10G}-$\textit{f}$_{5G}$ (middle panels),
    as well as against the mass--weighted mean age, $\tau$, of each dSph
    (lower panels), adopted from Orban et al.~(\cite{sl_orban08};
    \textit{f}$_{5G}$ is the fraction of stars formed within the last
    5\,Gyr). The values of \textit{f}$_{10G}$, \textit{f}$_{5G}$,
    \textit{f}$_{1G}$, and $\tau$ are reproduced in Table~\ref{table3b}. 

    In the upper panels of Fig.~\ref{sl_slopes}, there is a tendency of
    increasing the intermediate--age fraction to increase the slope of the
    $\Delta$[Fe/H] as a function of the [Fe/H]$_{spec}$. This is demonstrated
    by the red solid line which is an error--weighted linear least squares fit
    to the data points in Fig.~\ref{sl_slopes}. The error bars correspond to
    the errors of the coefficients of the fit to the datapoints of the lower
    panels of Fig.~\ref{sl_diffeh} and Fig.~\ref{sl_kirby1}. The upper left
    panel of Fig.~\ref{sl_slopes} suggests that the Ca\,T metallicities on the
    CG97 metallicity scale versus the photometric metallicities show the most
    pronounced dependence on increased intermediate--age fractions of stars
    whereas the discrepancies between MRS and photometric metallicities remain
    relatively low regardless of the admixture of younger populations. The
    Pearson correlation coefficients are 0.95, 0.85, and 0.54 for the CG97,
    isoscale, and MRS cases respectively, while it is significant only in the
    case of the CG97 metallicity scale within the 85\% confidence level. 

    In the middle panels of Fig.~\ref{sl_slopes}, the same tendency is
    observed when we plot the slopes of Fig.~\ref{sl_diffeh} and
    Fig.~\ref{sl_kirby1} against the \textit{f}$_{10G}-$\textit{f}$_{5G}$
    fractions. Here, the trend is significant within the 98\% confidence level
    only in the case of the isoscale, with a Pearson correlation coefficients
    of 0.96, but it is not significant in the remaining cases of the CG97 and
    MRS metallicity scales with a Pearson correlation coefficient of 0.87 and
    0.24, respectively.

    In the lower panels of Fig.~\ref{sl_slopes}, there is a much less
    significant trend of a decreasing mass--weighted mean age with an
    increasing slope of $\Delta$[Fe/H] as a function of
    [Fe/H]$_{spec}$. Again, the red solid line is an error--weighted linear
    least squares fit to the data points in Fig.~\ref{sl_slopes}, while the
    error bars correspond to the errors of the coefficients of the fit to the
    datapoints of the lower panels of Fig.~\ref{sl_diffeh} and
    Fig.~\ref{sl_kirby1}. The lower panels of Fig.~\ref{sl_slopes} suggest
    that the trend is insignificant within the 99\% confidence level, with
    Pearson correlation coefficients of 0.81, 0.64, and 0.78 for the CG97,
    isoscale, and MRS cases respectively.

    \subsubsection{Applicability of the photometric metallicity method} 

    The presence of an intermediate--age population in a dSph leads to a
    metal--poor bias in the photometric metallicities, in the sense that the
    stars are assigned with too metal--poor photometric metallicities compared 
    to their spectroscopic values. In the case of the Galactic dSphs studied
    here, where some of them have a pronounced or even dominant
    intermediate--age population, the individual stellar differences of the
    spectroscopic minus the photometric metallicities can reach a range in
    metallicity spanning up to 2.37\,dex in the case of Fornax which has the
    most extended star formation and chemical evolution history. In practice,
    the photometric metallicities become more metal--poor as compared to the
    Ca\,T spectroscopic or MRS metallicities when intermediate--age
    populations contribute. In dSphs where the fraction of the
    intermediate--age population is small, the assumption of a single old age
    when deriving photometric metallicities appears to yield relatively good
    results. In the case of Sculptor, there is a systematic trend of
    increasing $\Delta$[Fe/H] with increasing Ca\,T [Fe/H] that mimics the
    same trend observed in dSphs with a substantial fraction of
    intermediate--age stars present, consistent with the recent findings of de
    Boer et al.~(\cite{sl_deboer11}) and Menzies et al.~(\cite{sl_menzies11})
    regarding the range of ages of the stellar content of Sculptor. We find
    that 24\% of intermediate--age stars with an age of 7\,Gyr are needed in
    order to account for the difference of the median spectroscopic
    metallicity with the median photometric metallicity.

%
%%%%% TABLE 5 %%%%%%%%%%%%%%%%%%%%%%%%%%%%%%%%%%%%%%%%%%%%%%%%%%%%
\begin{table*}
\caption[]{Slopes
  of the $\Delta$[Fe/H] versus the spectroscopic [Fe/H], numbers of
  stars used in our comparisons (N), and population fractions, adopted
  from Orban et al.~(\cite{sl_orban08}).} 
\label{table3b}
\centering
\begin{tabular}{l c c c c c}

\hline\hline
  Galaxy                 &Carina             &Fornax             &Leo\,II              &Sculptor        &Sextans\\
                                                                                           
\hline                                                          
                                                                
$\tau$\,(Gyr)            &7.1                &7.4                &8.8                  &12.6            &12.0  \\

\textit{f}$_{1G}$        &0.0065             &0.013              &0.0028               &0.010           &0     \\

\textit{f}$_{5G}$        &0.43               &0.33               &0.025                &0.025           &0     \\

\textit{f}$_{10G}$       &0.67               &0.73               &0.70                 &0.14            &0     \\
                                                                
N$_{CG97}$               &24                 &131                &25                   &60              &...   \\
                                                                 
N$_{isoscale}$           &19                 &114                &17                   &47              &...   \\

N$_{MRS}$                &...                &90                &127                  &131              &31     \\
                                                                
Slope$_{CG97}$           &$0.74\pm0.11$      &$0.75\pm0.06$     &$0.93\pm0.21$        &$0.31\pm0.08$    &...   \\
                                                                
Slope$_{Isoscale}$       &$0.81\pm0.15$       &$0.86\pm0.07$     &$1.09\pm0.20$       &$0.55\pm0.09$    &...  \\

Slope$_{MRS}$            &...                &$0.72\pm0.12$     &$0.77\pm0.11$       &$0.67\pm0.03$   &$0.74\pm0.08$  \\

\hline                                                                                                 
                                                                                                       
\end{tabular} 
\end{table*}
%%%%%%%%%%%%%%%%%%%%%%%%%%%%%%%%%%%%%%%%%%%%%%%%%%%%%%%%%%%%%%%%%%%%%%%%%%%%%%%%%%%%
%
    In more distant dSphs where the use of the Ca\,T method (or MRS) to derive
    spectroscopic metallicities is not available due to the faintness of the
    stars to be targeted, one has to rely on the photometric method in order
    to have an estimate of their metallicity. The {\em mean} metallicities
    derived from the photometric method in the case of dSphs dominated by old
    stars are biased towards the metal--poor end by approximately 0.08\,dex as
    compared to the Ca\,T metallicities placed on the isoscale (for
    Sculptor). The intrinsic scatter in this case is 0.16\,dex for the
    photometric metallicities and 0.20\,dex for the spectroscopic
    metallicities placed on the isoscale, which leads to an intrinsic scatter
    of their difference of approximately 0.26\,dex. The intrinsic scatter of
    the difference of the spectroscopic metallicity, on the isoscale, minus
    the photometric metallicity is 0.09\,dex, 0.32\,dex, and 0.24\,dex for
    Carina, Fornax, and Leo\,II, respectively, which are less or comparable
    with the median of the $\Delta$[Fe/H]. Thus, depending on the size of the
    fractions of intermediate--age stars, using the photometric method may
    underestimate the mean metallicity by a few tenths of dex in [Fe/H]. 

    Given the fraction of the intermediate--age populations in a dSph, one can
    derive an estimate of how much offset the photometric metallicities may be
    as compared to the spectroscopic metallicities. In more distant dSphs, the
    ability of deriving accurate SFHs is hampered by the same age--metallicity
    degeneracy examined here (Gallart, Zoccali \& Aparicio
    \cite{sl_gallart05}) on the RGB as well as by our inability to obtain CMDs
    that reach the old main--sequence turn--offs. Therefore, one has to rely
    on the  presence of luminous AGB stars as a probe of the presence of
    intermediate age populations. In a study of MDFs of nine dSphs in the M81
    group of galaxies, we detected luminous AGB stars in all of them, with
    fractions ranging from 3\% to 14\% (Lianou et al.~\cite{sl_lianou10}; see
    also Caldwell et al.~\cite{sl_caldwell98}, Da Costa
    \cite{sl_dacosta04}). Similarly, luminous AGB stars were detected in
    early--type dwarfs in other groups of galaxies (Rejkuba et
    al.~\cite{sl_rejkuba06}, Girardi et al.~\cite{sl_girardi10}, Crnojevic et
    al.~\cite{sl_crnojevic11}), as is also the case for Local Group dwarf
    galaxies (e.g., Battinelli \& Demers \cite{sl_battinelli04}; Davidge
    \cite{sl_davidge05}; Groenewegen \cite{sl_groenewegen07}; Whitelock, et
    al.~\cite{sl_whitelock09}; and references therein).

 \section{Summary and Conclusions}

    We test the validity of the photometrically derived stellar metallicities
    generally used under the assumption of a single old age, and we explore
    the effect of the presence of intermediate--age stellar populations on
    photometrically derived stellar metallicities. We choose five Galactic
    dSphs, namely Sculptor, Sextans, Carina, Fornax, and Leo\,II, which have
    different SFHs and contain a different fraction of intermediate--age
    stars, ranging from old ages in Sextans to very prominent
    intermediate--ages in Fornax. We use their resolved RGBs and we derive
    their photometric metallicities using a linear interpolation method
    assuming a constant old age for the theoretical isochrones and a wide
    range in metallicities, from $-$2.5\,dex to $-$0.3\,dex. We compare the
    photometric metallicities with Ca\,T--based spectroscopic metallicities,
    with MRS metallicities, and with high--resolution spectroscopic
    metallicities from the literature in several ways in order to examine the
    effect of the presence of intermediate--age stellar populations on the
    derivation of photometric metallicities. The comparison between the
    photometric and spectroscopic metallicities is performed both on the CG97
    metallicity scale and on the metallicity scale defined by the Dartmouth
    isochrones in the case of the Ca\,T--based metallicities. Moreover, we
    simulate the effect of intermediate--age populations on the photometric
    metallicities via isochrone models of different ages. 

    The comparison of the {\em mean}  photometric metallicity properties with
    the {\em mean} spectroscopic ones shows that we can safely trust the
    photometrically derived stellar metallicities in the case of
    old--dominated systems such as Sculptor and Sextans, where the comparison
    of the photometrically and spectroscopically derived median metallicity
    gives a difference of 0.08\,dex. In systems such as Fornax, which has the
    most extended star formation and chemical enrichment history, the
    comparison between the {\em mean} metallicity properties derived from
    different methods gives highly discrepant results that amount to 0.51\,dex
    in the case of the median MRS metallicity versus the median photometric
    metallicity. In order to account for a difference of 0.43\,dex between the
    median photometric metallicity and the median spectroscopic metallicity
    on the isoscale, as observed in Fornax, it would require that a fraction
    of stars between 100\% to 55\% on the RGB formed from 4 to 2\,Gyr ago, a
    finding that is also supported by Coleman \& de Jong
    (\cite{sl_coleman08}).

    For those stars that are in common in the spectroscopic and photometric
    samples and for galaxies that formed the majority of their stellar
    populations within the last 10\,Gyr, we find  the maximum difference
    between the median Ca\,T metallicity and the median photometric
    metallicity, amounting to 0.52\,dex for Fornax, as well as the maximum
    range of the differences between the MRS and photometric metallicities
    (amounting to 2.56\,dex again for Fornax). These differences become very
    small for almost purely old stellar populations, of the order of less than
    0.1\,dex for Sculptor and Sextans when comparing both MRS and Ca\,T
    metallicities with photometric metallicities. 

    There is the trend of the differences between the individual stellar
    metallicities derived from all methods to increase towards positive
    $\Delta$[Fe/H], and this systematic deviation strongly depends on the
    particular SFH of each studied dSph. As compared to Ca\,T--based
    metallicities, the photometric metallicities seem to show the best
    agreement in the metallicty range from around $-$2 to $-$1.5\,dex,
    independent of each particular SFH. It is interesting that the effect of
    age on the differences between MRS and photometric metallicities is less
    pronounced, regardless of the intermediate--age stellar mass fraction,
    while for Ca\,T metallicities versus photometric metallicities there is a
    stronger manifestation of the age--metallicity degeneracy. 

    Each spectroscopic method yields different results. Our comparison between
    metallicities from different spectroscopic methods shows differences of a
    similar size as the comparison of metallicities between spectroscopic
    methods and photometry in the case of the old--age dominated dSphs
    Sculptor and Sextans. Such differences are of the order of 0.1\,dex. In
    the case of Leo\,II and Fornax, the comparison between different
    spectroscopic methods show differences smaller than those when comparing
    spectroscopic with photometric metallicities. As expected, we do find
    effects of the age--metallicity degeneracy for galaxies with high
    fractions of intermediate--age stellar populations. Therefore, we find
    that we are justified to use the photometric method of deriving stellar
    metallicities in the case of old or intermediate--age dominated dSphs when
    we focus on the metallicity range from $-$2\,dex to $-$1.5\,dex where
    Ca\,T metallicities and photometric metallicities agree the best
    independent of each dSph's SFH. Furthermore, we are justified to use the
    photometric method in a wider metallicity range only when the dSph is
    old--age dominated, since then we get the least discrepant results between
    photometric and spectroscopic metallicities, although we note that the
    same discrepancy is observed between spectroscopic methods. Therefore, an
    estimate of the intermediate--age stars present in a dSph is important for
    stellar metallicities studies. 

\begin{acknowledgements}
      The authors thank an anonymous referee for the thoughtfull comments. We
      kindly thank Giuseppina Battaglia and Matthew Walker, the former for
      sharing with us the full photometric datasets of Sculptor and the Ca\,T
      spectroscopic dataset of Sextans, and the latter for sharing with us the
      full photometric datasets of Sculptor, Fornax, and Carina. We also
      kindly thank Aaron Dotter for extremely useful discussions on Dartmouth
      isochrones. Katrin Jordi and Thorsten Lisker are also acknowledged for
      useful discussions. SL acknowledges an IAU travel grant to participate
      to the XXVII GA, during which this work was motivated to initiate. SL
      and this research were supported within the framework of the Excellence
      Initiative by the German Research Foundation (DFG) via the Heidelberg
      Graduate School of Fundamental Physics (HGSFP) (grant number GSC
      129/1). AK acknowledges support by an STFC postdoctoral fellowship and
      funding by the DFG through Emmy-Noether grant Ko 4161/1.

      This research has made use of the VizieR catalogue access tool, CDS,
      Strasbourg, France. This research has made use of the NASA/IPAC
      Extragalactic Database (NED) which is operated by the Jet Propulsion
      Laboratory, California Institute of Technology, under contract with the
      National Aeronautics and Space Administration. This research has made
      use of NASA's Astrophysics Data System Bibliographic Services.   
\end{acknowledgements}

\end{document}